\documentclass[twocolumn,aps,prl,amsmath,amssymb,superscriptaddress,notitlepage]{revtex4-1} 

\usepackage{graphicx}
\usepackage{amssymb}
\usepackage{amsmath}
\usepackage{dsfont}
\usepackage{bm}
\usepackage{mathrsfs}
\usepackage{times}
\usepackage[utf8]{inputenc} 
\usepackage{gensymb} 
\usepackage{xcolor} 
\usepackage{bbold}

\begin{document}

\title{Non-linear Processes and Stimulated Hawking Radiation in Hydrodynamics\\ for Decelerating Subcritical Free Surface Flows with a Subluminal Dispersion Relation}
\author{L\'{e}o-Paul Euv\'{e}}
\affiliation{Bluerium, Avenue Louis Philibert, 13100 Aix-en-Provence, France.}
\affiliation{Laboratoire de Physique et M\'{e}canique des Milieux H\'{e}t\'{e}rog\`{e}nes, UMR 7636, CNRS, ESPCI, Sorbonne Universit\'e, Universit\'e PSL, 1 rue Jussieu, 75005 Paris, France.}
\author{Germain Rousseaux}
\affiliation{Institut Pprime, UPR 3346, CNRS, Universit\'e de Poitiers, ISAE ENSMA, TSA 51124, 86073 Poitiers Cedex 9, France.}

\begin{abstract}
In \cite{PRL2011}, the authors have {\it ``conducted experiments in order to verify the thermal nature of the stimulated Hawking process at a white hole horizon in a fluid analogue gravity system"} namely the linear mode conversion giving rise to negative energy waves {\it i.e.} the classical ingredient at the root of the Hawking effect in astrophysics. However, here we show that these experiments in Vancouver operated in a weakly non-linear regime that obscure them as was the case for the seminal experiments in Nice within a stronger non-linear regime \cite{NJP2008}. We finally shed some light on these matters by demonstrating that the linear conversion of water waves on a counter-current takes place with or without a dispersive white hole horizon as anticipated in the Nice experiment no matter the frequency is conserved within the entire process provided there is an absence of wave breaking during the wave-current interaction. The main novelty is the role of free (and not forced as usual in non-linear effects) harmonics generation in the interpretation of both the Nice (for at least the stimulating mode) and Vancouver (for the converted modes only) experiments. Unfortunately, the thermality of the spectrum is not demonstrated in the Poitiers reproduction of the Vancouver experiments based on the analysis of the scattering coefficients themselves and not just of their ratio as was done previously.
\end{abstract}

\maketitle

Hawking radiation is a linear scattering process with an amplification of both classical and quantum fields ruled by a dispersion relation featuring a Doppler-shift of the field frequency and a linear relationship between the co-moving frequency and the wavenumber in the long wavelength approximation (see the Appendix). The latter shift is due to either a moving inhomogeneous medium (or a space-changing physical characteristics of the medium) or a gravitational space-time and leads to the existence of negative energy modes whose relative frequency for a co-moving observer is negative (as well as their conserved norm associated to phase invariance). The frequency of some waves or modes may become negative if it is measured in a coordinate system which is moving faster than the phase velocity of dispersion-less waves \cite{Musha1964a, Musha1964b, Sturrock1960, Pierce1961, Dodo1964, Volovik2003}. A resulting amplification of outgoing positive energy modes (the universal Hawking effect) that are propagating away from the horizon towards an asymptotic observer is predicted by the very appearance of the outgoing negative ones inside the horizon of real black holes or their analogues \cite{Volovik2003, BLV, Scott, Como, Barcelo, PRL2020, Royal, PTRSA2020}.

Historically, the quantum vacuum noise, namely quantum field fluctuations seen as pairs of particle-antiparticle, has been predicted by Stephen Hawking in 1974 to be separated by a curved space-time featuring an event horizon (for dispersion-less modes obeying Lorentz invariance): negative norm modes corresponding to antiparticles are absorbed inside the horizon whereas the positive norm modes (the astrophysical Hawking radiation) are amplified and emitted as a quantum field radiation at the horizon towards flat space-time regions \cite{Hawking}. The domain of existence of the negative modes inside the black horizon is dictated by the dispersion-less behavior of the light in the curved space-time of the black hole which corresponds to supercritical (the current speed is superior to the waves speed) flows in analogue systems. It was realized by White in 1973 that acoustic waves feel similarly a moving medium (like wind) as an effective space-time with a corresponding acoustic metric and a Doppler-shifted frequency due to the flow current \cite{White} (see the Appendix for a graphical resolution of the dispersion relation in many settings). Independently, Anderson and Spiegel in 1975 noticed that light waves propagating in a moving medium could be kinematically trapped within an effective optical metric akin to the acoustic metric of White \cite{AS}. In 1981 Unruh, willing to observe in the laboratory such an effect, transposed the reasoning of Hawking to dumb holes featuring a trapping supersonic region due to a trans-sonic flow for non-dispersive acoustics waves propagating in a flowing air and he predicted an analogue acoustic Hawking radiation \cite{Unruh1981}: the Newton escape velocity plays the role of an analogue current flow velocity in a trans-sonic regime that reaches the speed of waves at the horizon \cite{BLV, Scott, Como, Barcelo}. Barcelo \cite{Barcelo} and a special issue after a Royal Society of London Meeting \cite{Royal} have recently summarized the current state of affairs in the growing field of "analogue space-times" or "analogue gravity" and the recent achievements in the comprehension of Hawking radiation seen as a universal phenomenon common to astrophysics (Hawking's original prediction) and condensed matter physics (including classical hydrodynamics as discussed in the review issue included in \cite{PTRSA2020} and our own works \cite{PRL2009, NJP2010, PRE2011, PRD2011, Chaline, faltot, PRD2015, POF2016, PRD2016, PRL2016, PRD2017, viscouspaper, PhD2017}). 

Here, we reveal that weakly non-linear effects (as opposed to catastrophic non-linear regimes like wave breaking) have played a critical role in the Vancouver experiments (like in the Nice experiments \cite{NJP2008, NJP2010} and Chapters 5 and 7 of \cite{Como}) for the very same reasons that were not accounted for in \cite{PRL2011, LWTPU2012, WTPUL2013}. We have reproduced in Poitiers the experimental setup and data analysis of the Vancouver team using exactly the same obstacle geometry, flow parameters and range of stimulating frequencies, but with different wave amplitudes (see details in the Appendix and in \cite{PRD2015, PhD2017}): we used smaller wave amplitudes than in \cite{PRL2011, WTPUL2013} and still got significant nonlinear effects. 

\begin{figure}[!htbp]
\includegraphics[width=8cm,height=5cm]{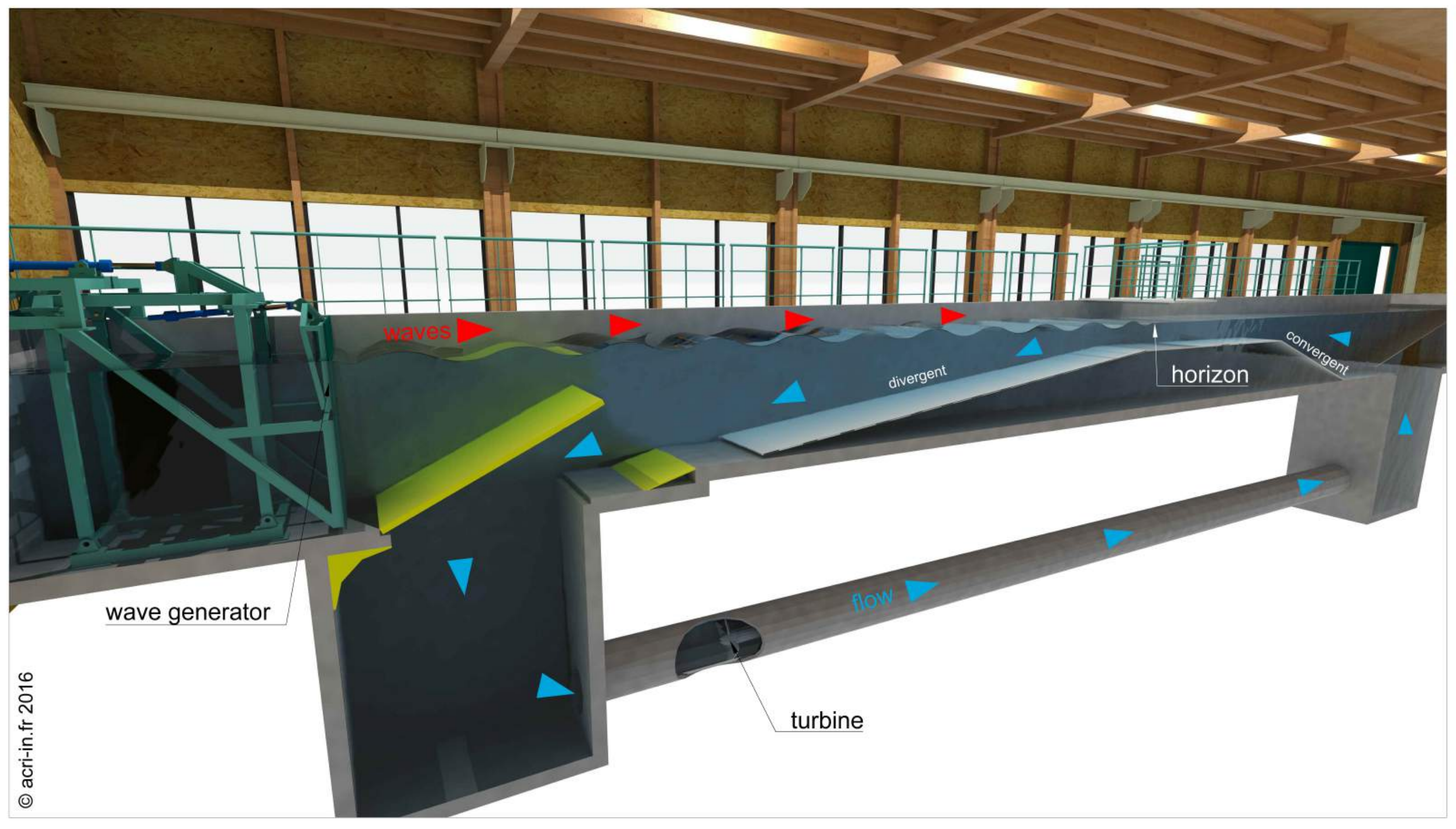}
\includegraphics[width=8cm,height=5cm]{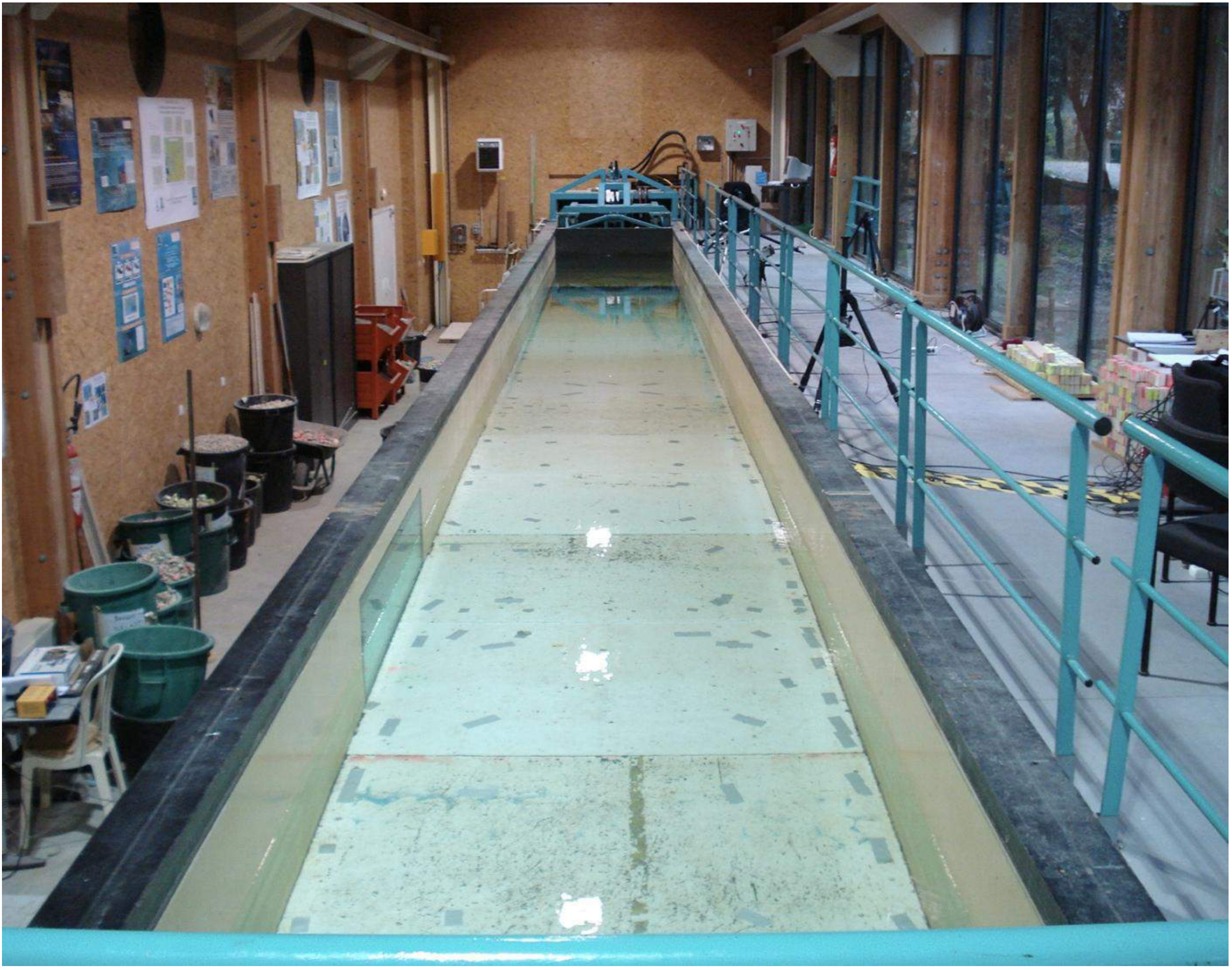}
\caption{(Top) Experimental setup of the 2010 Nice experiments \cite{Chaline} with geometrical parameters: maximum water height $h=1.6m$ downstream and upstream of the bottom obstacle, minimum water height $h=0.5m$ on the flat part of the geometry, ascending slope angle $18.5^\circ$ for the water current, descending slope angle $7.5^\circ$, length of the ascending slope $3.3m$, length of the descending slope $8m$, length of the flat part of the obstacle $4.8m$; (Bottom) A picture of the water channel seen from the center of the tank's width which is $1m80$ looking towards the wave-maker.}
\label{NiceBump}
\end{figure}

\begin{figure}[!htbp]
\includegraphics[width=8cm,height=6cm]{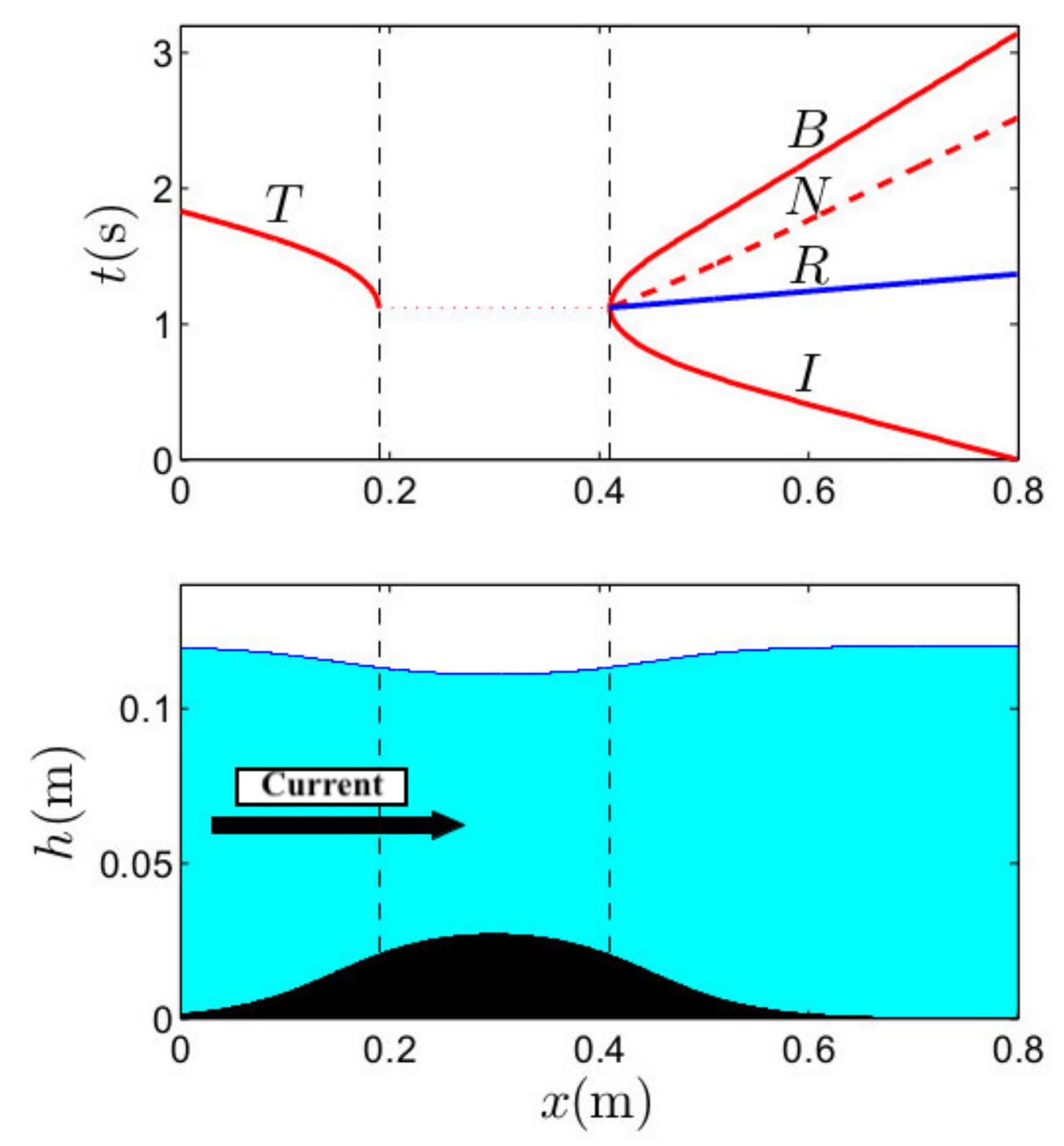}
\caption{(Top) Modes conversion for a wave-current interaction above a bottom obstacle in a water channel \cite{NJP2010}: the so-called stimulated Hawking radiation corresponds usually to the mode conversion at the constant frequency $\omega$ of the wave-maker from the incoming mode $k_I^\omega$ towards a pair of a blue-shifted mode $k_B^\omega$ and a negative mode $k_N^\omega$ without a transmission mode $k_T^\omega$ and retrograde mode $k_R^\omega$ in terms of wavenumber; (Bottom) The theoretical dispersion relation with free harmonics for pure gravity waves on a counter-current \cite{SU, NJP2008, NJP2010, PRL2011, Como} including the transverse modes $k_{t_p}$ whose transverse component is $k_y=p\pi/W$ where p is an integer and W is the width of the channel.}
\label{theory}
\end{figure}

In 2006-2010 \cite{NJP2008, NJP2010, Chaline} (see Fig.~\ref{NiceBump}), one of us belonged to a team who designed a dispersive white hole flow (the time reversed of a black hole flow, featuring a -zero group velocity- dispersive white horizon and a decelerating sub-critical water current) in order to observe the mode conversion of a long stimulating incoming free surface water wave (easily generated but hardly imaged) (see the Appendix for an extensive discussion on the effect of dispersion on mode conversion). The latter denoted as an I(ncident) mode, produced by a downstream wave-maker, should convert into a pair of short (hardly generated but easily imaged) positive and negative energy modes denoted as the B(lue) and N(negative) modes propagating on the counter-current in the water channel with a bottom obstacle and a given initial water depth \cite{NJP2008} (see Fig. \ref{theory}). This wave-current interaction system \cite{SU} was supposed to reproduce part of the Physics of the quantum emission of a white hole horizon in General Relativity by probing the classical counterpart of Hawking prediction and this was formulated at that time by the following expression "the classical ingredient at the root of Hawking radiation" or "the simulated Hawking radiation" \cite{BLV}. The purpose of the experiments was to test the analogy put forward by White-Unruh between the propagation of light waves in curved space-time and the propagation of hydrodynamics waves (sound, free surface waves) in a flowing fluid (air, water) \cite{White, Unruh1981, SU}.

\begin{figure}[!htbp]
\includegraphics[width=8cm]{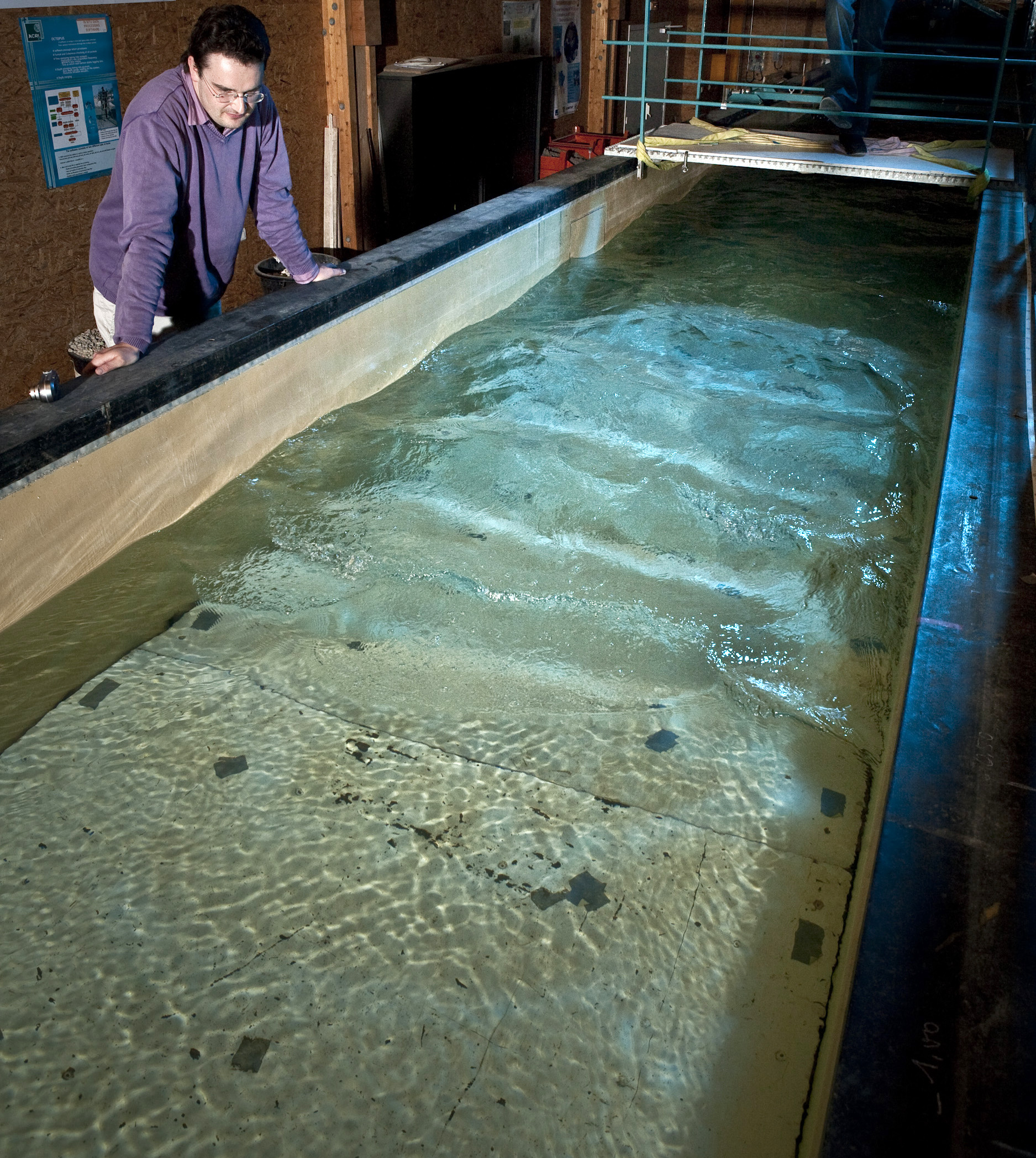}
\caption{Wave blocking at a dispersive group velocity horizon in the 2010 Nice experiments \cite{Chaline} (see also Figure 1 of \cite{NJP2010} reporting a similar picture about the 2008 experiments \cite{NJP2008}): the short waves correspond mainly to the blue-shifted modes $B^\omega $ at the dispersive horizon or wave blocking line (if amplified by the simultaneous appearance of negative modes $N^\omega $, they would correspond to stimulated Hawking radiation) and which are converted from the long incoming modes $I^\omega$. In the near horizon region where $k_I^\omega  \simeq k_B^\omega$ one observe sometimes partial wave breaking and parasitic capillary waves conversion on the front wave. Picture by Olivier Monge, Agence MYOP.}
\label{ACRI}
\end{figure}

The title of the paper "Observation of negative phase velocity waves in a water tank: A classical analogue to the Hawking effect?"  featured a question mark since the understanding of what had been observed was incomplete. In the experiments reported in \cite{NJP2008}, the flow current driven by a pump in a free surface flow above a bottom obstacle was subcritical (the flow speed was inferior to the speed of non-dispersive long gravity waves akin to the light speed). This was considered as a drawback since the White-Unruh analogy maps trans-critical flows with an accelerating speed on one hand and black hole space-times with a horizon where the Newton escape velocity matches the speed of light on the other hand \cite{White, Unruh1981}. White holes (dispersion-less) are considered as unstable solutions in General Relativity \cite{Eardley} so to use them in Fluid Mechanics was already suspicious but subcritical dispersive white holes can be created easily in the experiments as demonstrated for the first time in \cite{NJP2008}. Wave blocking is nevertheless possible but itthus depends on the period of the incoming waves. In the present work, we will try to answer to the following questions:

- Is Hawking prediction robust for these subcritical flows \cite{NJP2008, NJP2010, PRL2011, PRD2015}?

We will show that the sub-luminal dispersion correction in fluid mechanics extends the domain of validity of the original Hawking prediction but Hawking radiation may not reach the asymptotic observer because of harmonics (of a certain kind) generation in some regimes.

- Can we observe positive energy modes and negative energy modes in pairs \cite{SU, NJP2010} or separately \cite{NJP2010}?

Both scenarii are possible because of the subluminal dispersion correction in fluid mechanics. Here, we will create pairs of converted modes from both fondamental and harmonics modes type.

- Are the scattering coefficients the same \cite{MP2014, RFP2016, AS2016, MP2018, MPR2018, CS2018, BCFV2020} as in Hawking semi-classical calculation without dispersion \cite{Hawking}?

The answer is negative because of either dispersive wave transmission or non-linear harmonics generation or the appearance of a dispersive zero frequency solution.

- Is unitarity \`a la Hawking \cite{Hawking} (norm or wave action conservation) checked \cite{MP2014, RFP2016, AS2016, MP2018, MPR2018, CS2018, BCFV2020}?

The answer is also negative because of either dispersive wave transmission or non-linear harmonics generation or the appearance of a dispersive zero frequency solution.

The constraint that the 2006 and 2010 Nice experiments reported in \cite{NJP2008, NJP2010, Chaline} were performed with subcritical flows was dictated by the fact that the free surface over an obstacle remains flat (or features a small depression of the order of the cm) for a large range of flow rate and initial water depth (around $1.4-1.6m$ and a maximum obstacle height of $1.1m$) in the open channel that was used at that times. For higher flow rates and/or small water depth above the maximum height of the obstacle, a turbulent hydraulic jump was observed (see the Appendix for an illustrative picture of the flow regime with the turbulent jump, the J regime of hydraulic flows as reported in \cite{PTRSA2020}): hence, this regime was dismissed since counter-current incoming waves would not have reached the non-dispersive white hole horizon without going through the time-dependent aerated hydraulic jump. Hence, a dispersive and sub-critical white hole flow was used instead with a flat surface (F regime) or a small depression (D regime) of a few cm (depending on the flow rate) above the obstacle for the dynamical depth at a give flow rate and a fixed initial water depth which has the same value on either side of the obstacle in the asymptotic regions. By F, D and J regimes, we refer to the classification of the shape of the free surface reported in the hydraulic phase diagram of our previous work starting from an initial static free surface and by increasing the flow rate to go through the many free surface shapes \cite{PTRSA2020}.

Waves with either positive or negative phase velocity/wavenumbers were definitely observed sometimes (depending on the physical parameters) in the experiments \cite{NJP2008, NJP2010, Chaline} (even in the absence of a white-hole group-velocity horizon) on the top of the long incoming waves as seen by naked eyes and by cameras filming transversally the waves traces on the side wall opposite to the former (see Fig.~\ref{ACRI}). Unfortunately, the origins of those waves were unclear in particular for the cases without wave blocking: because of the sign of the phase velocity, the "positive" waves were considered as an indication of the classical Hawking radiation in the peculiar case of wave blocking (the blue-shifted waves B using the fluid mechanics terminology) however other "positive" waves in absence of wave blocking were also observed and superposed to the incoming stimulating modes I and whose nature was debated (case (b) of Figure $8$ of the  paper \cite{NJP2008}); similarly, "negative" waves (both with and without wave blocking) were also observed and considered as an indication of the negative partners in astrophysics (the negative energy waves using the same terminology) albeit with arguments in favor of a loss of linearity since the period of the waves does not seem to be conserved despite the fact that the period of the short waves propagating backward with respect to the incoming waves I was proportional to the wave-maker period hence of the incoming wave (either linear conversion or harmonics generation at double the frequency is compatible with this observation). In addition, space-time diagrams of the free surface showed a dispersive wave blocking boundary (a line in 2D transverse to the water channel, see Fig.~\ref{ACRI}) depending on the incoming period of the wave-maker (demonstrating the existence of the conversion towards the positive energy mode \`a la Hawking \cite{PRL2009} because of the finite slope of the incoming/blue positive modes at the dispersive group velocity horizon). The space-time diagrams prove also the blue-shifting of the stimulating mode wavelength by the changing slope of the wave crest close to the dispersive group velocity horizon (see Figures 1 and 2 of \cite{NJP2010}). However, the comparison with numerical simulations in addition to the difficulty to test the conservation of the frequency (hence energy in the linear regime only) with only optical visualisations on one side of the wave-tank lead the authors to wonder whether they had observed a parasitic non-linear phenomena blurring the interpretation of the experiments : {\it ``It is conceivable that we have seen a new fluid-mechanics phenomenon that significantly enhances the Hawking effect. Could it be a nonlinear mode conversion, a nonlinear process generating harmonics with negative frequencies? We observed that the incident waves become steeper as they propagate against the current"} \cite{NJP2008}. In the Appendix, we revisit the physical origin of some of the mode seen in absence of wave blocking in the light of the reproduction of the following experiments.

Indeed, another experimental setup in Vancouver was designed shortly after the seminal Nice experiments. According to the Vancouver team, the Nice experiments were plagued with several problems (see the oral presentation at Pirsa in 2009 p. 40/80 of slide show \cite{PIRSA2009}): {\it ``negative frequency waves with wrong frequency (indicating non-linear effects); negative frequency modes with and without acoustic horizons ; separation of flow"}. The new experiments performed in Vancouver lead to the claim that {\it ``Hawking radiation has indeed been measured and shown to possess a thermal spectrum, as predicted"} \cite{PRL2011, PIRSA2011, LWTPU2012, WTPUL2013, Unruh2014}. Noticeably, the flow regime featured a stationary undulation (the so-called U regime of \cite{PTRSA2020} with a depression followed by secondary waves called whelps) namely a zero frequency solution of the dispersion relation in presence of a current, see \cite{Unruh2008, NJP2010, Como, CP2014}. All the scattering experiments in Vancouver were done with an undulation. In a separate paper \cite{PRL2016}, we have shown that the undulation over-amplifies the Hawking scattering effect, a purely dispersive effect due to the sub-luminal character of the water waves dispersion relation \cite{Unruh2008} since the converted modes propagates on the undulation up to the asymptotic observer downstream of the obstacle. For the highest amplitudes used in \cite{NJP2008, NJP2010, Chaline}, partial wave breaking was reported in particular in presence of wave blocking with potential parasitic capillary ripples decorating the white hole horizon (see Fig. \ref{ACRI}). The Vancouver team \cite{PRL2011} set up a flow with an undulation at the limit of wave breaking \cite{PTRSA2020}. Therefore, the incoming amplitude superposed to the undulation must not combined so as to reach a too high camber $k\times a$ otherwise wave breaking will occur like in some of the Nice experiments (mainly in presence of wave blocking) \cite{NJP2008, PhD2017}.

Hence, the results reported in \cite{NJP2008, NJP2010, Chaline} met strong criticisms since the experimental configuration departs from the expected transposition from astrophysics \cite{SU}: in  particular, the size, geometry and physical parameters retained by the Nice group did not allow to create an analogue trans-critical flow featuring a dispersion-less horizon but an inhomogeneous flows driven by the obstacle geometry (and not dissipation for instance \cite{viscouspaper}) with a zero-group velocity real horizon or without wave blocking (a complex horizon \cite{CS2018} may albeit be present) in a dispersive white hole configuration (a loose version of the time reversed non-dispersive black hole studied by Hawking). This choice was explained theoretically in \cite{PRL2009} since it was shown that the dispersion relation of pure gravity waves allows a partial conversion without wave blocking for any velocity. It was subsequently tempered by the inclusion of the effect of surface tension which leads to a threshold in speed (the so-called Landau speed related to the Landau-Cerenkov wake, see Chapter 6 of \cite{Como}) for the mode conversion into both positive and energy modes in addition to the appearance of an undulation.

Criticisms of the Vancouver experiments have already been formulated: the dispersive white hole group velocity horizon (blocking point) was not present in the Vancouver experiments for most of the stimulated frequencies, as checked both experimentally and numerically in \cite{PRD2015}: only four experimental points out of nine involved a blocking point, a fact not reported by the Vancouver team, because the stimulated frequency of the incoming mode $\omega_I > \omega_ {\ min} = 2.4 \rm {Hz}$ was superior to the minimum blocking frequency (computed theoretically from the dispersion relation) in this four regimes only with the threshold $ | \tilde{A} | ^ 2 <$ 1/16 for the transmission coefficient $\tilde{A}$: the so-called unitarity condition (norm conservation \cite{PRL2011}) was modified accordingly to include a scattering coefficient associated to transmission. In addition, the linearity with the frequency of the measured natural log ratio of positive to negative norm components by the Vancouver team was shown to be a consequence of the obstacle geometry using numerical simulations in the {\it linear} regime only \cite{MP2014, MP2018}. The present work introduces more concerns about the observations reported in \cite{PRL2011} since we will take into account transmission, harmonics generation without filtering our results at the wave-maker frequency as was done in \cite{PRL2011} to dismiss spurious effects like free surface noise due turbulence for instance.

\begin{figure}[!htbp]
\includegraphics[width=8cm,height=4cm]{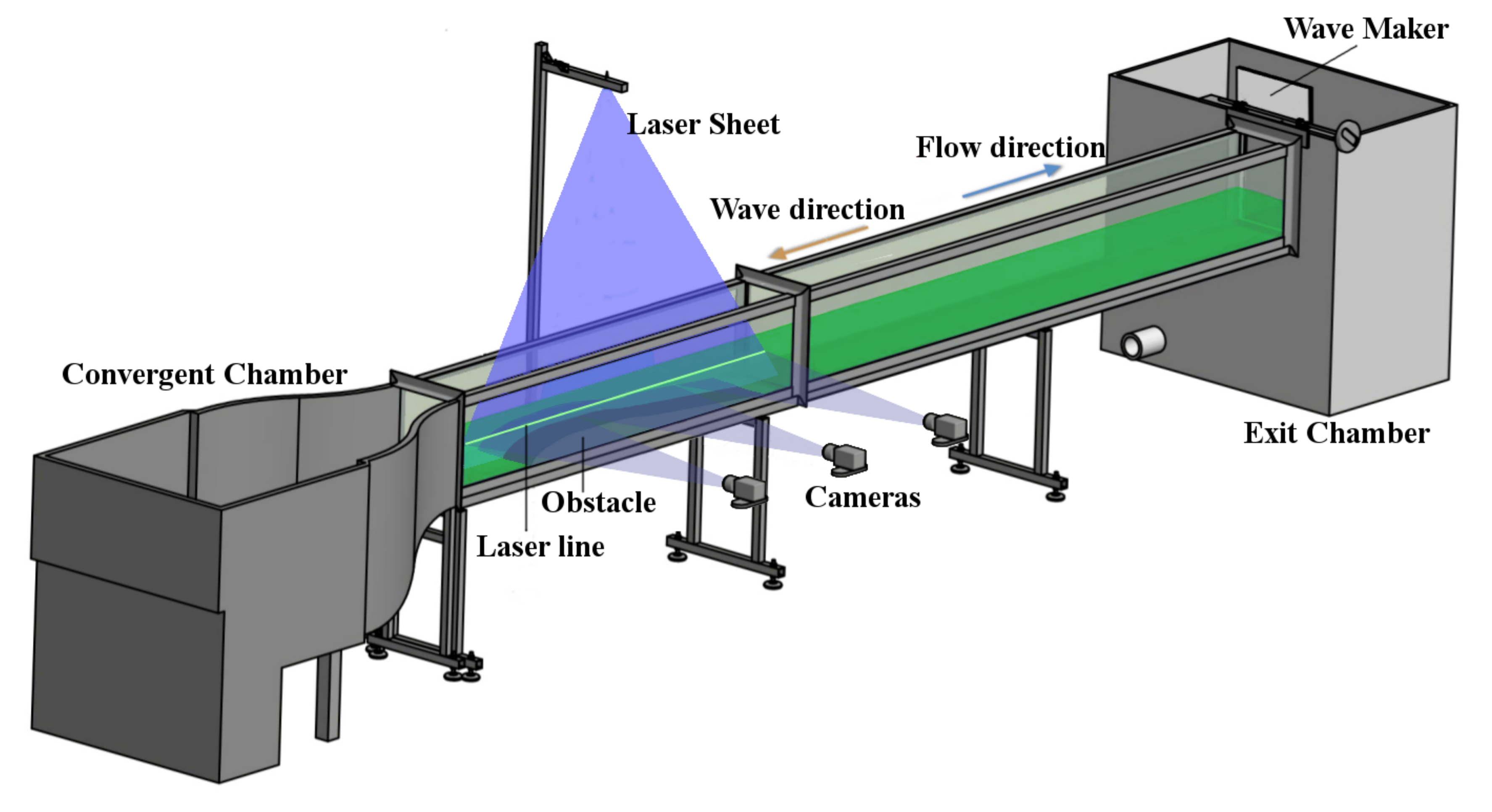}
\includegraphics[width=8cm,height=4cm]{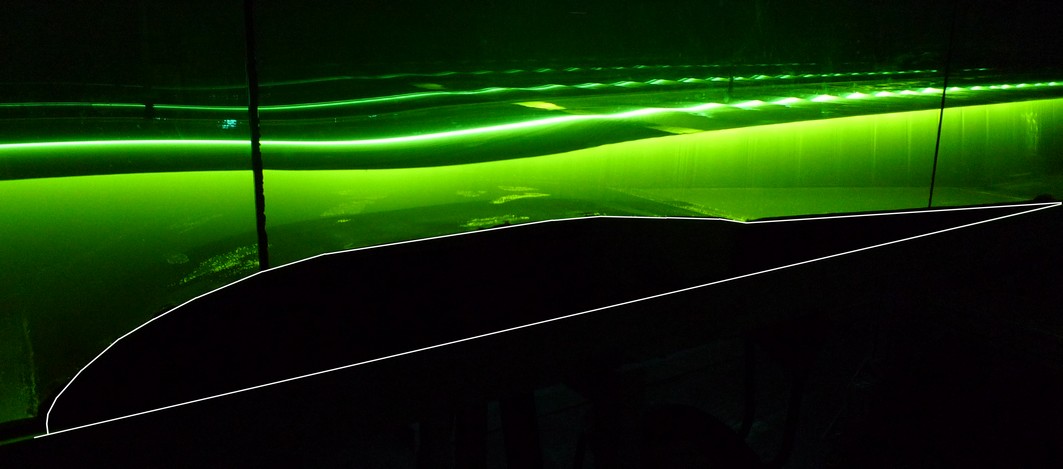}
\includegraphics[width=8cm,height=4cm]{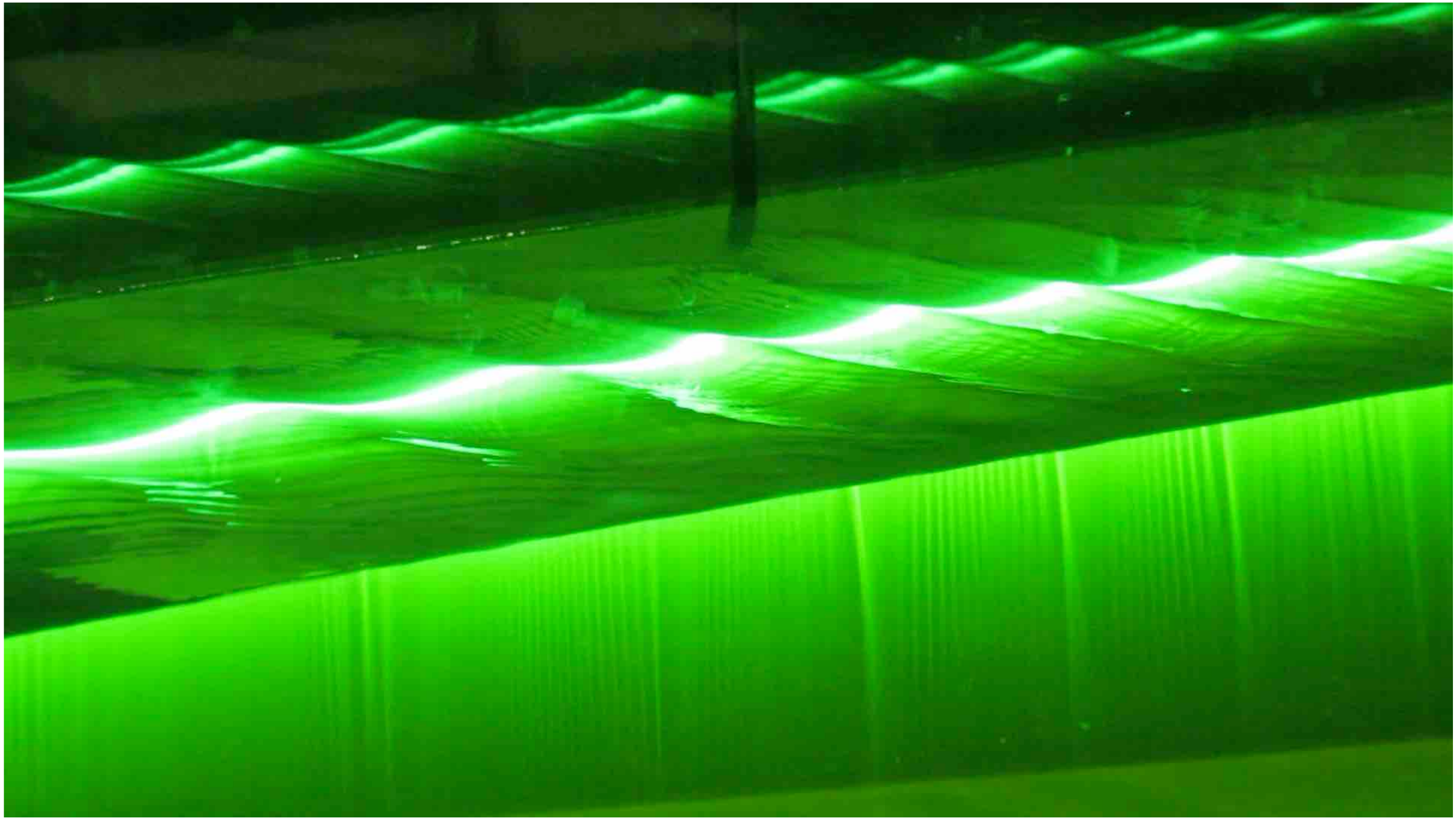}
\caption{(Top) Poitiers experimental setup: the length of the water channel is 7m long and its width is $W=0.39m$; (Middle) The stationary zero mode in the Vancouver flow regime; (Bottom) A zoom on the stationary undulation where both medium gravity caustics and small capillary caustics in-between appear beneath the free surface.}
\label{setup}
\end{figure}

In \cite{PRL2011, WTPUL2013}, the amplitudes of the long waves, generated against the current in the region downstream of the obstacle, were reported to be around $1-2 mm$ without a possible dependance on the wave-maker frequency whose mathematical expression is the so-called transfert function relating the imposed mechanical amplitude to the resulting hydrodynamical amplitude. Not knowing if this is the sine wave amplitude ($ a $) or peak-to-trough amplitude ($2a$), we decided to investigate amplitudes less than $1 mm$. As we seek to verify the presence or not of non-linearities, several incident wave amplitudes are then studied: $0.05, 0.15, 0.35, 0.8 mm$ corresponding to a mechanical displacement of the wave-maker of $0.25, 0.50, 1$ and $2 mm$ assuming small variations from one wave-maker frequency to the other for the range probed by the wave-maker in order to measure the mode mixing and the associated scattering coefficients. Since the transmission at the lowest frequencies have already been studied thoroughly \cite{PRD2016} (see the Appendix for complementary results on the envelope shape when varying the frequency for a given flow rate), we decided to focus on a blocking case ($\omega_I = 3,14 \rm {Hz}> \omega_ {min}=2.4 \rm {Hz}$) in the linear regime (wave transmission across the obstacle may occur simply by increasing the asymptotic incoming amplitude).

\begin{figure}[!htbp]
\includegraphics[width=9cm,height=4cm]{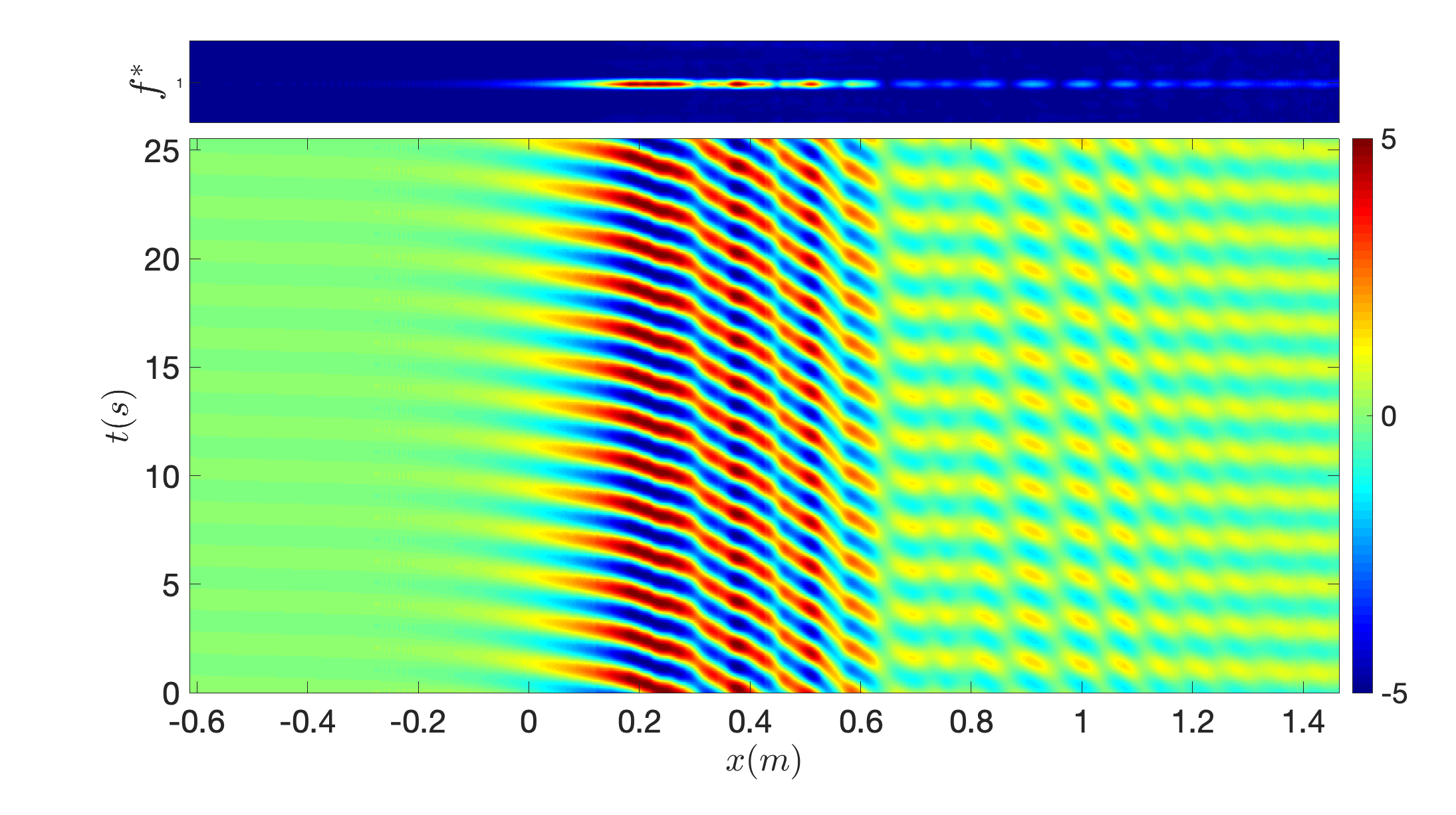}
\includegraphics[width=9cm,height=4cm]{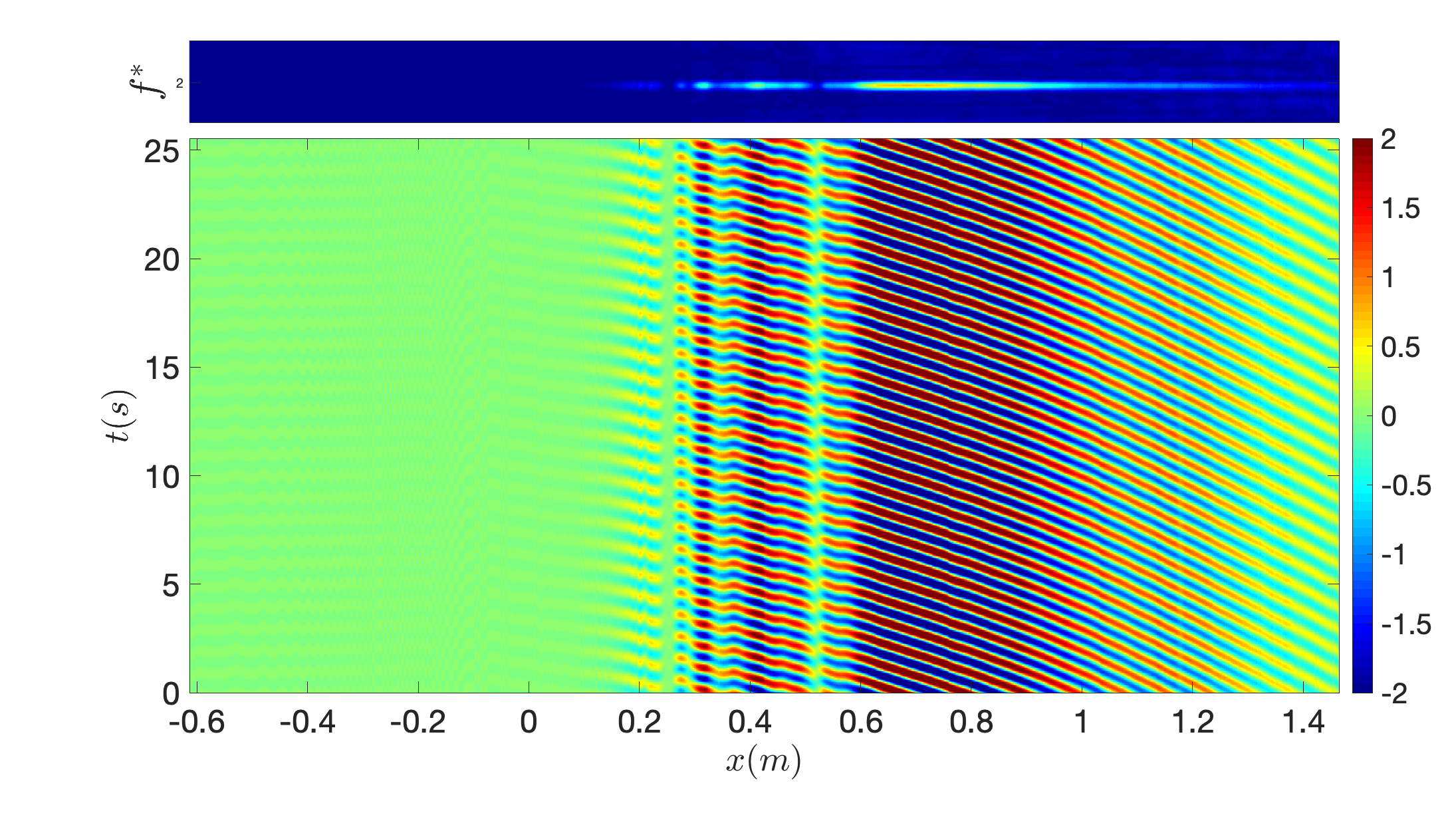}
\includegraphics[width=9cm,height=4cm]{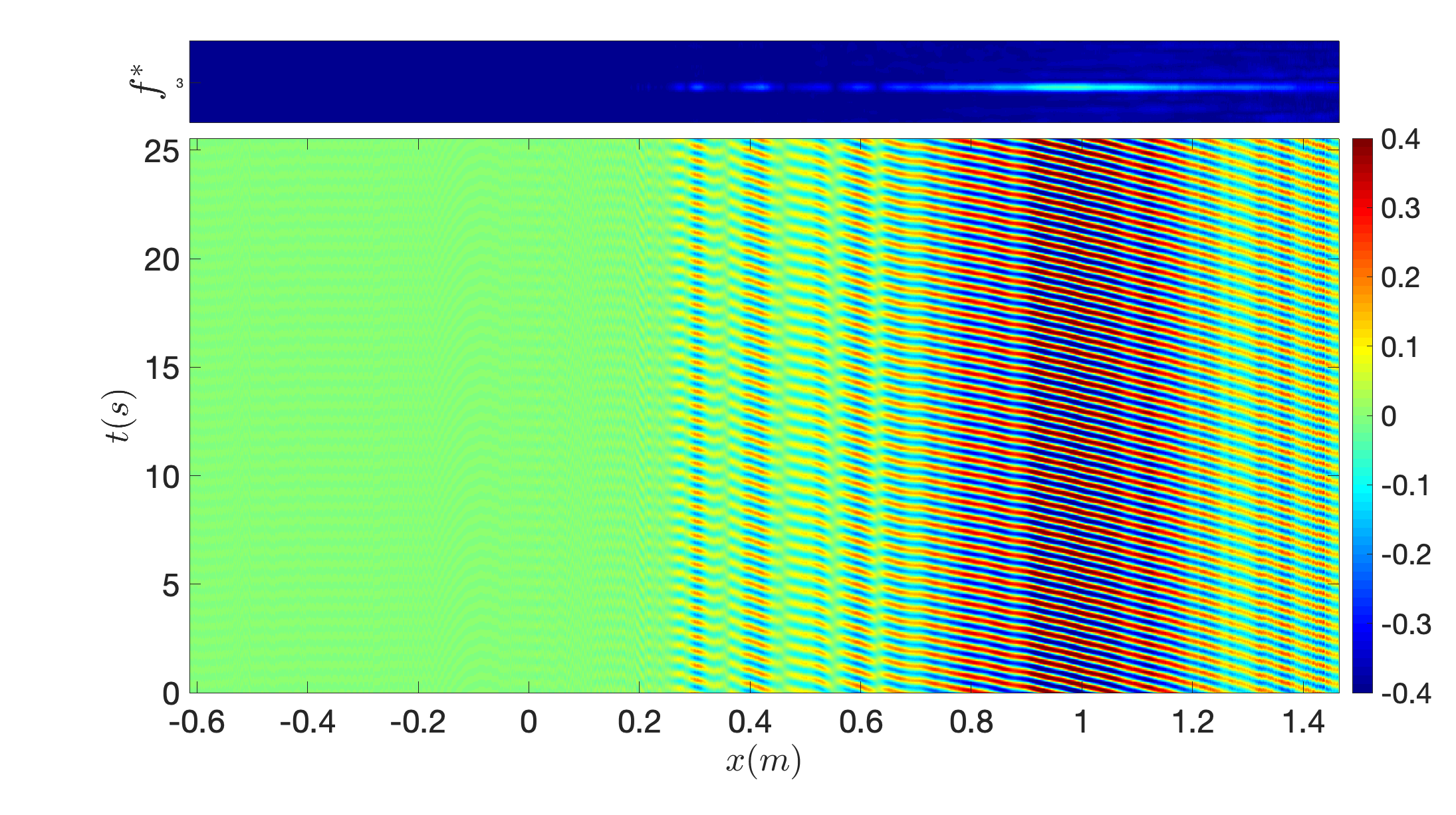}
\caption{Space-time diagrams of the free surface fluctuations filtered at different frequencies for a given asymptotic incoming wave amplitude $a_I=0.8mm$ where the zero mode has been removed (see the Appendix for details): (Top) the interferences pattern of the modes (incoming mode $k_I^\omega $, blue-shifted mode $k_B^\omega$ and negative mode $k_N^\omega$) corresponding to the Vancouver filtering at the wave-maker frequency $\omega_I  = 3.14 \rm {Hz}$  \cite{PRL2011} ; (Middle) The interferences pattern of the modes at $2\omega _I$; (Bottom) The interferences pattern of the modes at $3\omega _I$. The upper part of each block corresponds to the spatial spectrogram with a concentration of the signal at integer values of the dimensionless frequency $f^*=f/f_I=\omega/\omega_I$ namely 1, 2 or 3. The color bars for the water fluctuations are in mm.}
\label{xvst3}
\end{figure}

The free surface is illuminated by a laser sheet and a resulting green fluorescence trace imaged by three side cameras (whose pictures are calibrated and glued together) is recorded at the acquisition rate $f_{acqui}=20 Hz$ (Fig.~\ref{setup}). Then, we use the same method of extraction of the free surface with sub-pixel accuracy as in references \cite{PRL2011, faltot, PRL2016, PRD2017, PRL2020}. In the Vancouver flow regime \cite{PRL2011} with a flow rate per unit width $q=0.045 m^2/s$ and an asymptotic dynamical water depth $h_{upstream}=19.4cm$ controlled by a downstream double-weir (see the Appendix for a detailed description of the wave-maker), one observes Hawking radiation at zero frequency \cite{Unruh2008, NJP2010, Como, CP2014} for subcritical flows with a subluminal dispersive behavior of the dispersion relation: the noise due to the bulk turbulence excites a free surface wave turbulence that stimulates with infinitely long period modes the dispersive white hole horizon that emits a stationary undulation. The downstream wave-maker generates water waves that propagate on the counter-current and then reach the bottom obstacle region which does feature this zero mode solution or undulation displayed in the Fig.~\ref{setup}. Then, one must be very careful in using filters (see Fig.~\ref{xvst3}) to single out the many modes that are generated in the wave-current interaction process and not just dismiss the harmonics contribution as was done in \cite{PRL2011} (see also the Appendix for the several steps of the filtering procedure including the zero mode).

\begin{figure}[!htbp]
\includegraphics[width=8cm,height=5cm]{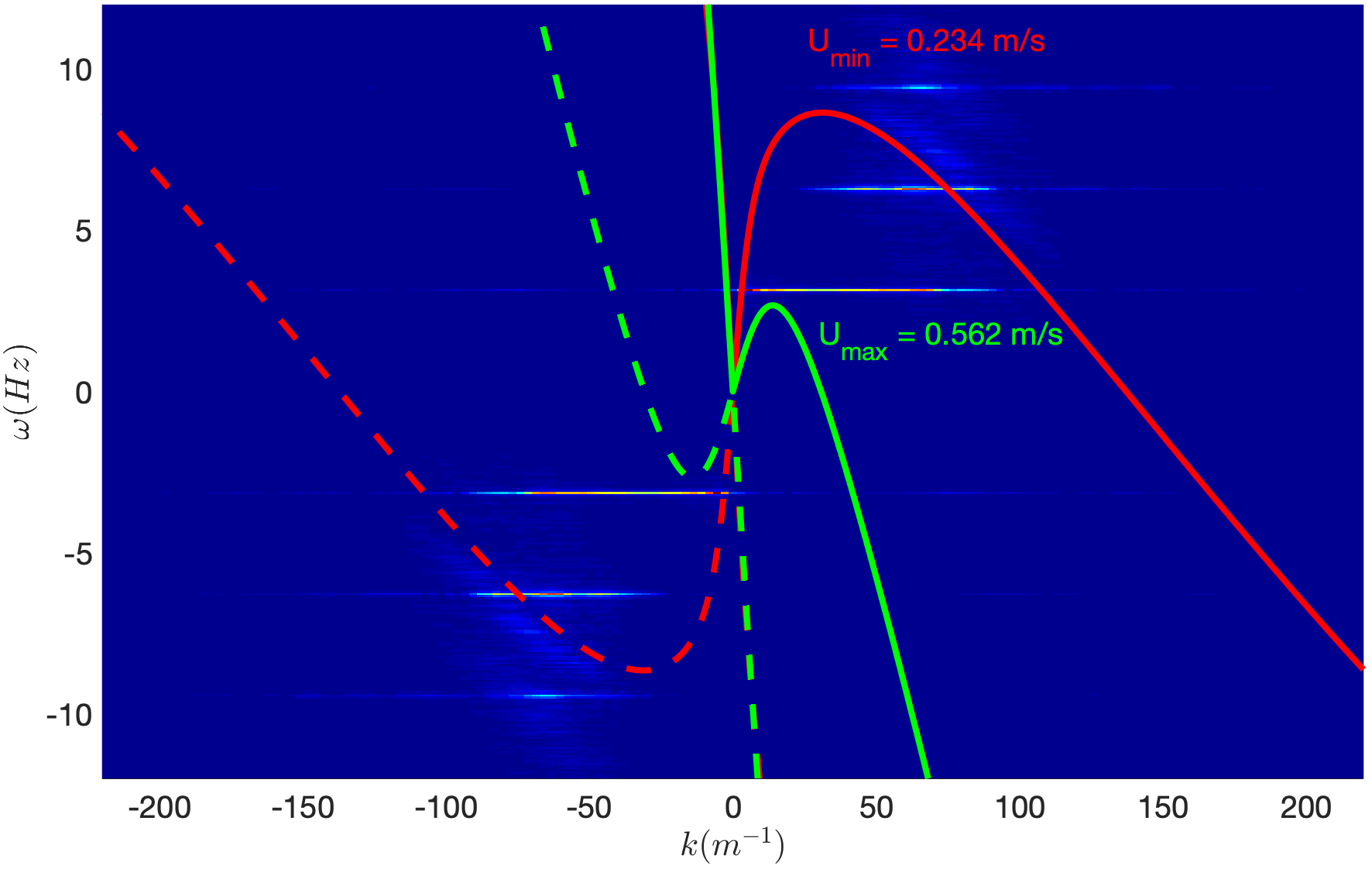}
\includegraphics[width=8cm,height=5cm]{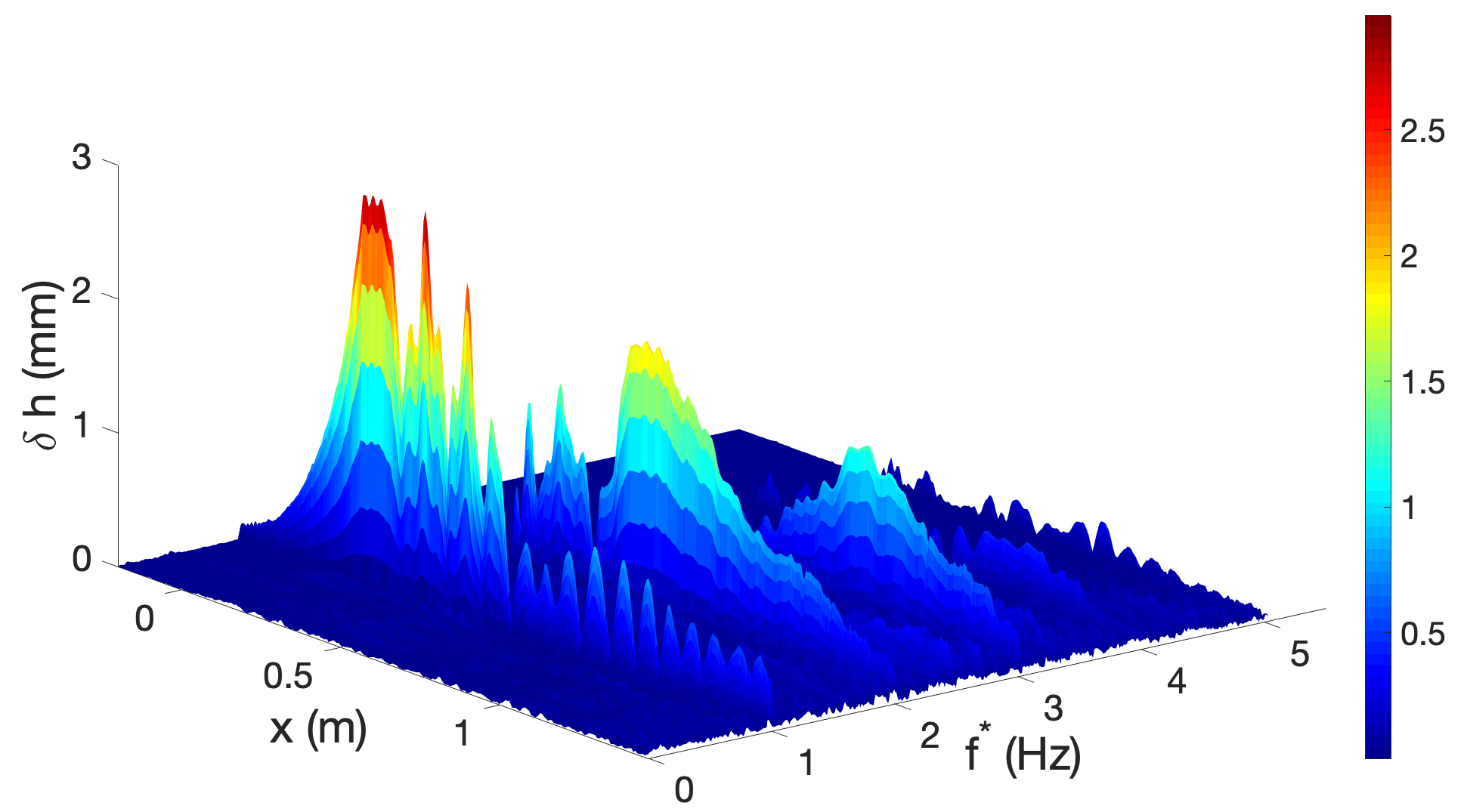}
\caption{(Top) 2D Fourier transform of free surface fluctuations $|\delta \hat{h}(k, \omega) |$ calculated over the entire viewing window of the cameras in the Poitiers experiments, for an incident frequency $\omega_I  = 3.14  Hz$ with an asymptotic wave amplitude $a_I=0.8 mm$. The red lines indicate the dispersion relation in the region downstream of the obstacle, those in green lines correspond to the dispersion relation above the obstacle where the Froude is maximum (the continuous/dotted lines pertain to the positive/negative branch); (Bottom) A 3D plot of the wave amplitude envelope $\delta h$ (in mm) as a 2D function of both the longitudinal position $x$ along the water channel and the dimensionless frequency $f^*=f/f_I=\omega /\omega_I$.}
\label{DispersionRelation}
\end{figure}

\begin{figure}[!htbp]
\includegraphics[width=8cm,height=6cm]{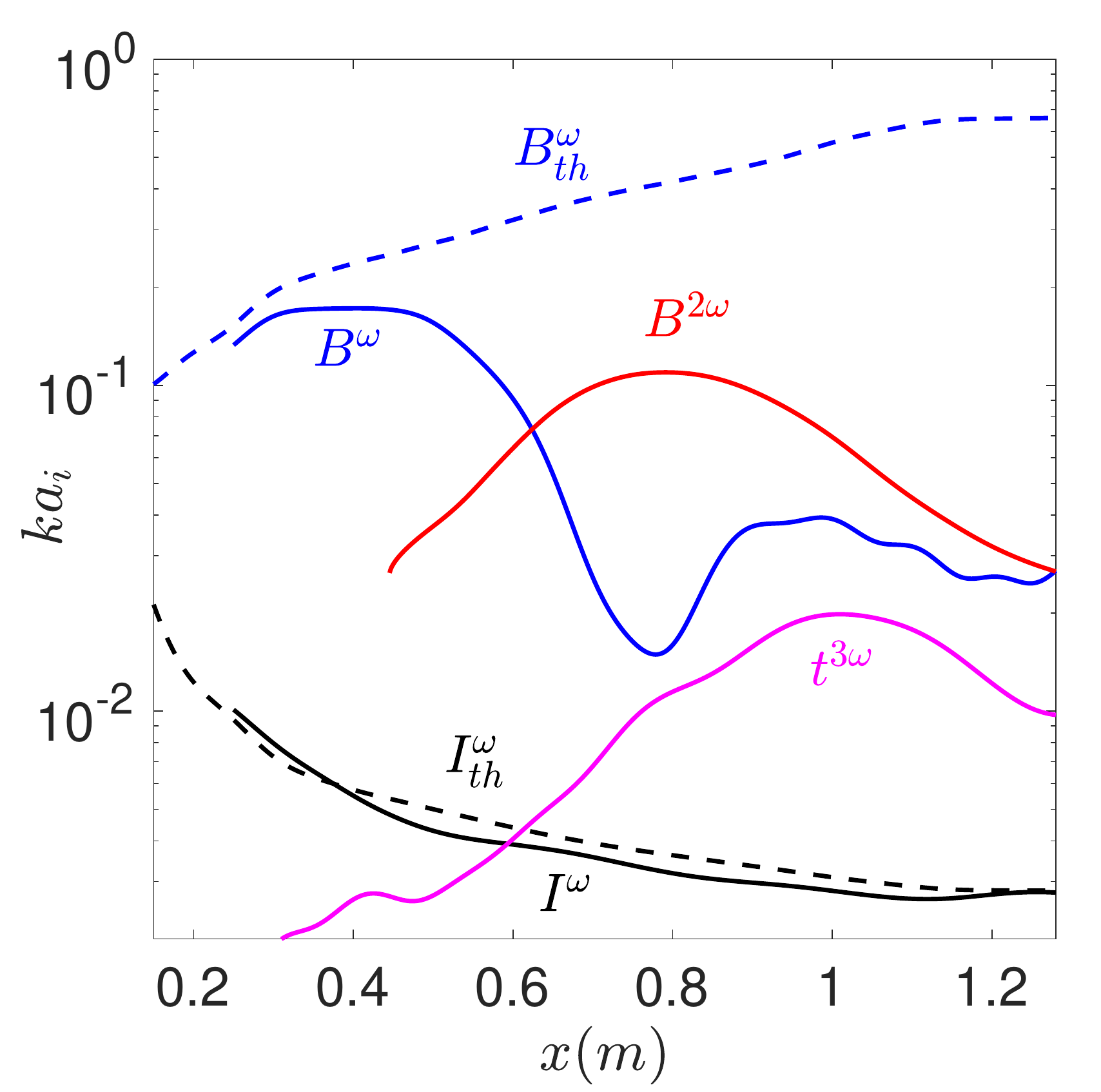}
\caption{Spatial evolution of the camber $k\times a_i$ (where $a_i$ is the wave amplitude of the mode i) of the several modes: fundamental $I^{\omega}$ + converted $B^{\omega}$ + harmonics $B^{2\omega}$ and $t^{3\omega}$ (here positive to simplify) for a wave-maker angular frequency $\omega_I  = 3.14 \rm {Hz}$ and an asymptotic incoming wave amplitude $a_I=0.8mm$. The predicted curves for the linear theory were added on the basis of the wave action conservation \cite{PhD2017, PRD2017} demonstrating the disappearance of the converted modes $B^{\omega}$ downstream of the obstacle to the detriment of free harmonics including longitudinal converted modes $B^{2\omega}$ as well as transverse modes $t^{3\omega}$ \cite{PRD2011}.}
\label{kaivsx}
\end{figure}

From the measurement of the free surface fluctuations $\delta h(x,t)$, a Fourier transform (in space and time $\delta \hat{h} (k,\omega)$ or just in time $\tilde{\delta h}(x,\omega)$) permits to observe the presence of non-linearities for high wave amplitudes. As can be seen in the Fig.~\ref{DispersionRelation}, there is no modification of the fundamental frequency which remains constant (there is no Benjamin-Feir type instability with a frequency downshifting for instance \cite{Benjamin1967, Shugan2014}) but we observe an important distribution of amplitude on the harmonic angular frequencies ($ 2 \omega_I $, $ 3 \omega_I $, etc...). The frequency of the stimulating wave is somehow conserved but new modes of the harmonics type appear (we can infer that the energy is distributing in all these new channels): so-called  {\it free harmonics} are created during the converted waves propagation \cite{Qin1997, Gutierrez2017, Ning2014, Ning2015} as can be checked by looking to the peaks position in the Fourier space \cite{PhD2017} and by checking that the free harmonics are solutions of the dispersion relation to the contrary of bounded harmonics waves. The non-linearities of the positive norm modes of the Hawking pairs $k_B$ are very peculiar with $k_B^{2\omega}<k_B^{\omega}$ which makes this nonlinear channel more effective in the process of reducing the waves camber (see the Fig.~\ref{DispersionRelation}): the system "prefers" to excite free harmonics instead of inducing wave-breaking when amplitude hence energy accumulates too much on a given mode. On the contrary and for completeness, the negative norm modes are such that $k_N^{2\omega}>k_N^{\omega}$ hence negative harmonics mode are less excited than positive harmonics mode. If the wave-current interaction is too strong, forced harmonics and/or wave breaking will relax the energy. The incoming wave is such that its frequency is conserved but the converted modes corresponding to the positive and negative norm modes (analogous to the particle-antiparticle pair in astrophysics) are distributing their amplitude hence their energy into harmonic modes at twice, three times {\it etc.}\ the value of the incoming angular frequency $\omega_I$ (see Fig.~\ref{kaivsx}).

\begin{figure}[!htbp]
\includegraphics[width=8cm,height=8cm]{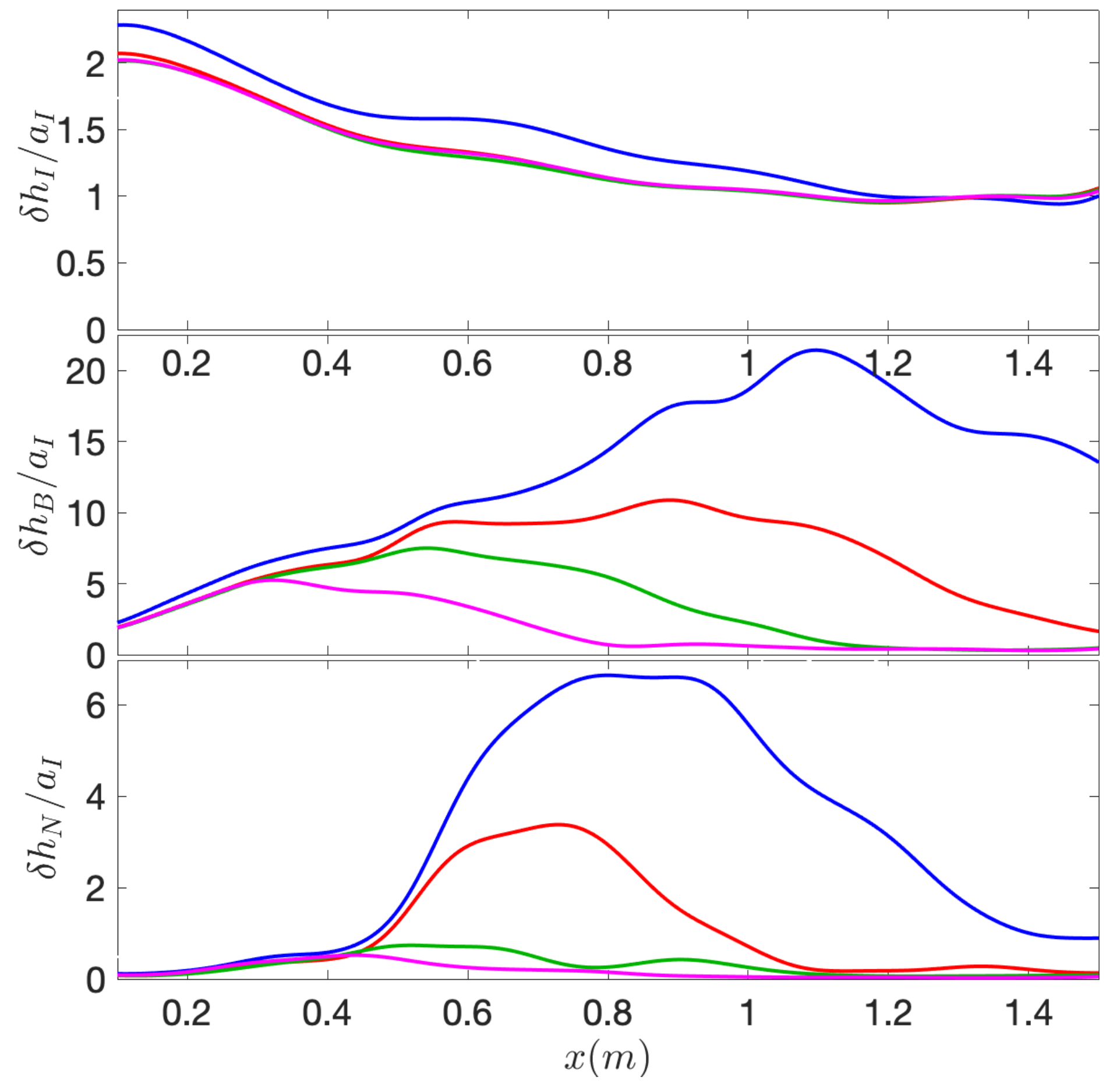}
\caption{Normalized amplitudes of the incident and converted modes as a function of space in the downstream region of the obstacle at a fixed wave-maker angular frequency $\omega_I  = 3.14 \rm {Hz}$ for several asymptotic incoming amplitudes  $a_I$ (0.05 mm (blue), 0.15 mm (red), 0.35 mm (green), 0.8 mm (purple)).}
\label{normalized}
\end{figure}

As these free harmonics are still solutions of the dispersion relation \cite{Massel1983, Brossard2009, LiTing2012, Kuznetsov2021}, they propagate at different speeds than the corresponding fundamental mode contrary to the more usual {\it bounded harmonics} \`a la Stokes that propagate at the very same velocity and which are no more solutions of the dispersion relation \cite{Qin1997, Gutierrez2017, Ning2014, Ning2015}. Hence, the norm conservation \cite{PRL2011, Como} associated to the phase invariance of the wave equation in the linear regime ($\vert \alpha \vert ^2 - \vert \beta \vert ^2=1$ where $\alpha /\beta $ pertain to the positive/negative norm modes akin to the analogue pair of particle/antiparticle) must be generalized since new channels for different harmonic frequencies are opened and populated by the redistribution of the norm between both free and bounded harmonics in general. Here, mainly the free harmonics channels are involved, as was the case for transmission at small frequencies \cite{PRD2015}, and this is another strong departure from the divergence behavior of the scattering coefficients $\alpha (\omega)$ and $\beta (\omega)$ as a function of the frequency for the small values of the latter based on Hawking's prediction \cite{Hawking}. Depending on the asymptotic value of the incoming wave amplitude $a_I$, the evolving amplitude of the incoming and converted modes are not the same when plotted as a function of space for a given wave-maker frequency  \cite{PhD2017} (see Fig.~\ref{normalized} and the Appendix).

\begin{figure}[!htbp]
\includegraphics[width=8cm,height=6cm]{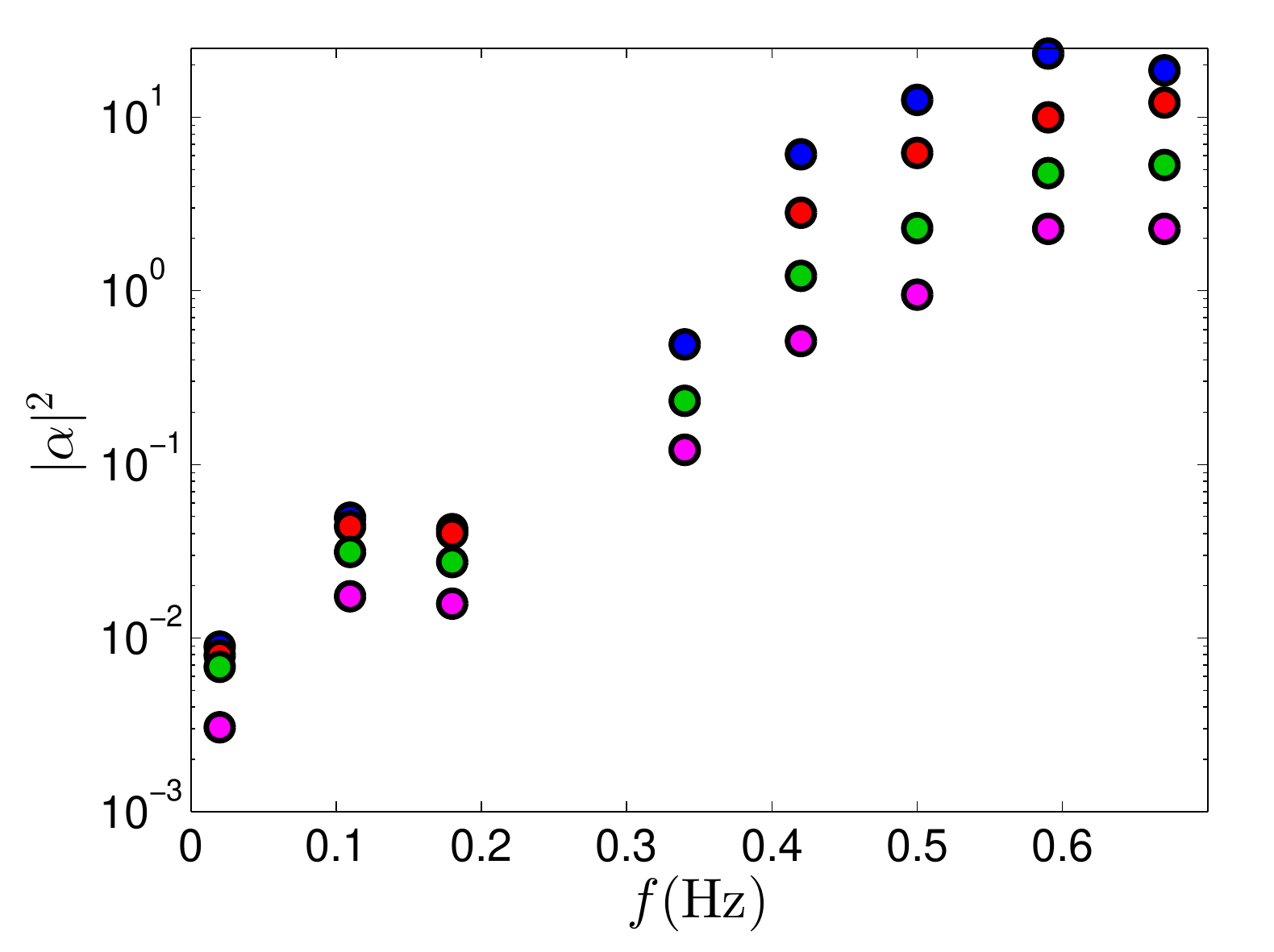}
\includegraphics[width=8cm,height=6cm]{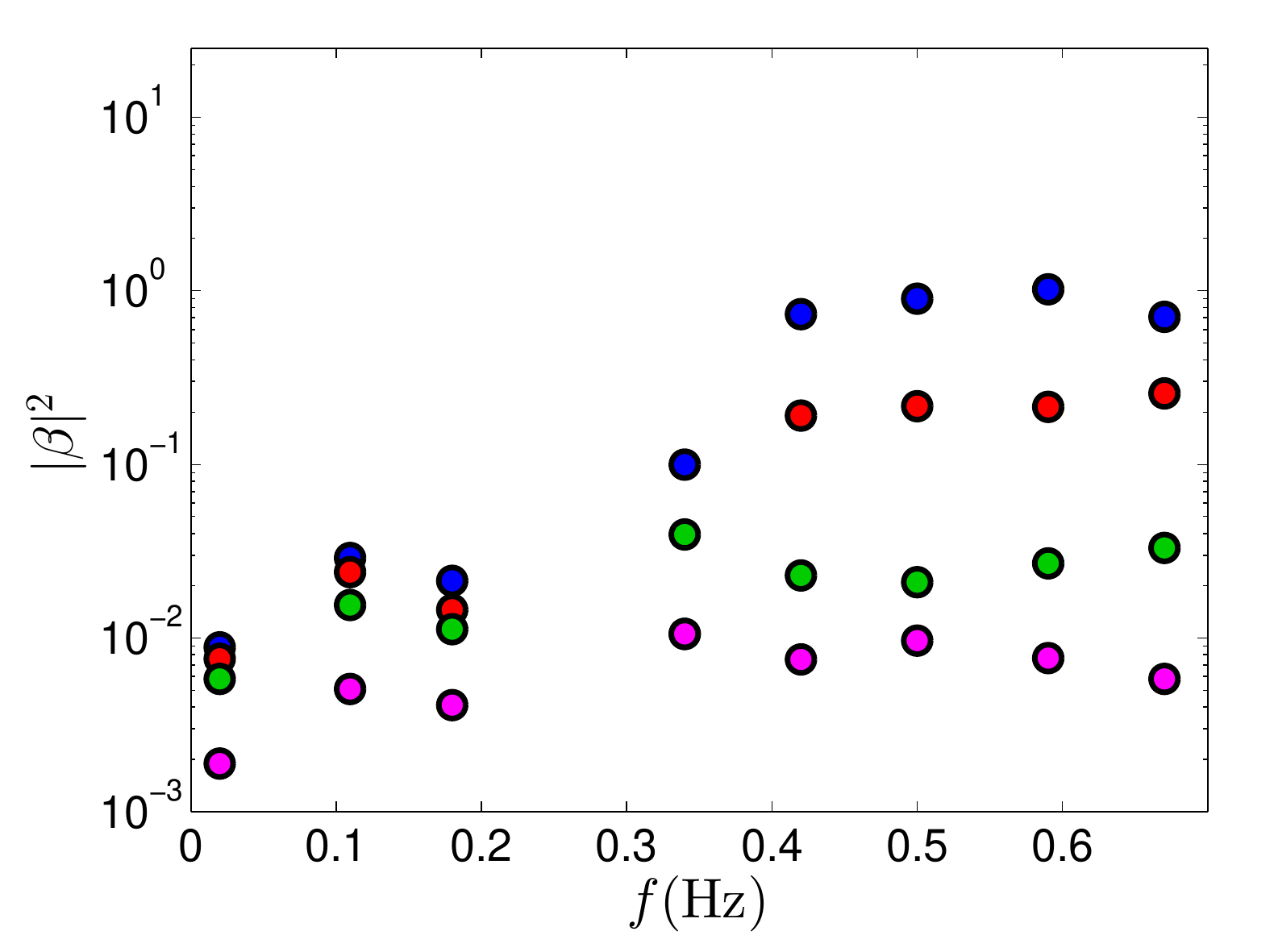}
\caption{Scattering coefficients in the downstream region of the obstacle for several asymptotic incoming wave amplitudes (0.05 mm (blue), 0.15 mm (red), 0.35 mm (green), 0.8 mm (purple)) as a function of the wave-maker frequency $f=\omega/2\pi$ for the positive norm mode \`a la Hawking (Top) and for the negative norm mode akin to the partner (Bottom).}
\label{Scattering}
\end{figure}

In Fig.~\ref{Scattering} we display the most salient features, namely the plots of the scattering coefficients $\vert \alpha \vert ^2$ and $\vert \beta \vert ^2$ as a function of the frequency $f$ of the wave-maker. The scattering coefficients are obtained from the ratio between the norm of both converted modes to the incoming wave as discussed in \cite{PRL2016, PhD2017}. So if the incoming wave amplitude varies with the frequency of oscillation for a given mechanical stroke of the wave-maker, the fact of taking the ratio dismissed this effect and an eventual frequency dependence. Clearly, the scattering coefficients do not diverge as the frequency cancels. Hence, the expected thermal spectrum for trans-critical flow is not observed, since the scattering coefficients vanish when the angular frequency goes to zero (in accordance with the numerical and theoretical predictions for subcritical flows \cite{MP2014, RFP2016, AS2016, MP2018, MPR2018, CS2018, BCFV2020}).

\begin{figure}[!htbp]
\includegraphics[width=8cm,height=6cm]{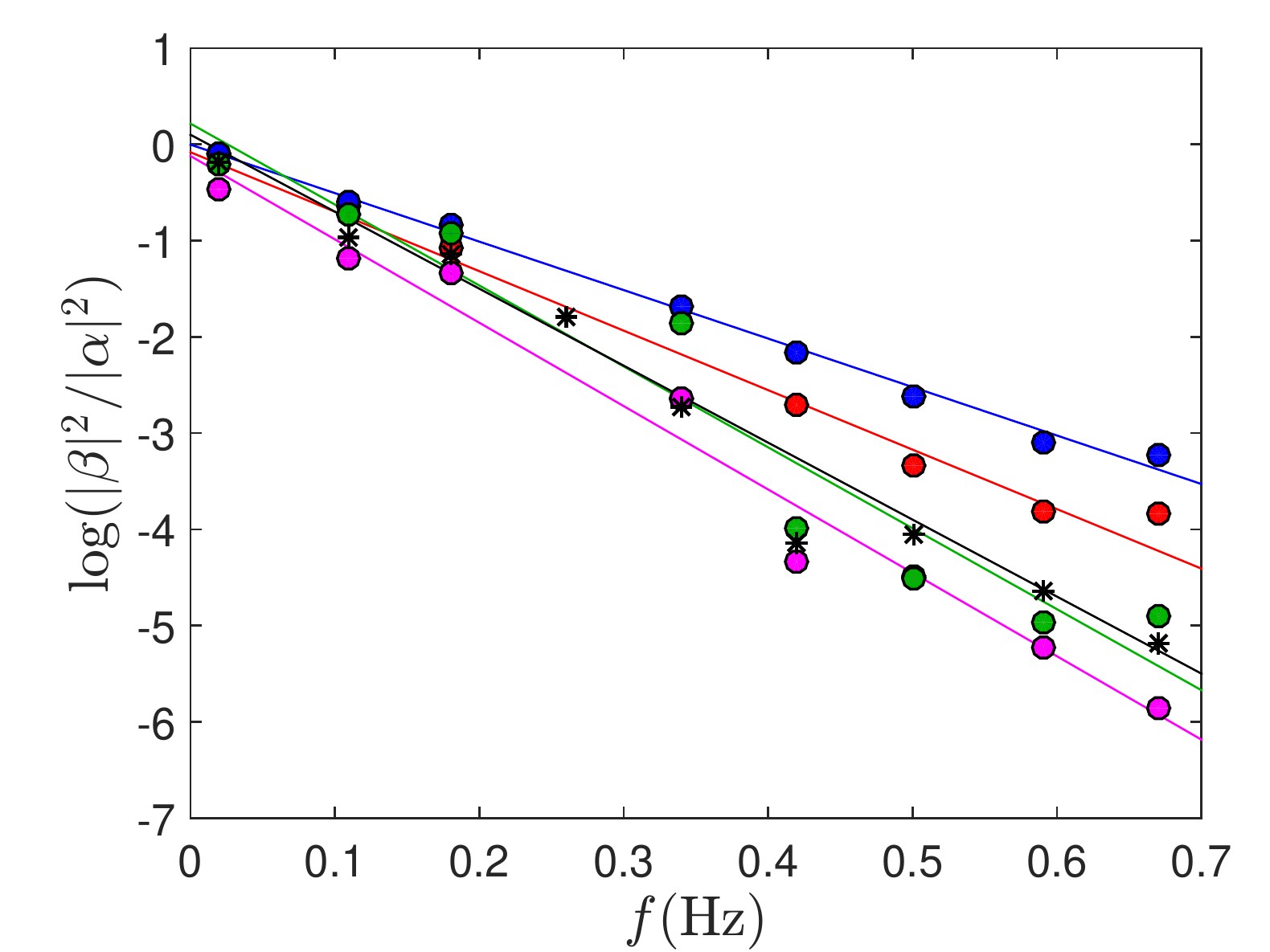}
\caption{Natural log of the ratio $R = |\beta ^2|/|\alpha ^2|$ of positive to negative norm components in the downstream region of the obstacle for several asymptotic incoming wave amplitudes and linear least-squares fit (continuous lines) as a function of the wave-maker frequency $f=\omega/2\pi$: 0.05 mm (blue), 0.15 mm (red), 0.35 mm (green), 0.8 mm (purple). The black stars correspond to the Vancouver experiments \cite{PRL2011} with an asymptotic incoming wave amplitude $a_I$ around $1-2mm$ \cite{PRL2011}.}
\label{Spectrum}
\end{figure}

As a major criticism of \cite{PRL2011}, the slope of the natural log ratio of positive to negative norm components $\log (R)=\log \left(\vert \beta \vert ^2/\vert \alpha \vert ^2\right)$ is not the same for several asymptotic incoming amplitudes (even if it appears like a linear Boltzmann law as a function of frequency), a fact not captured by the linear treatment discussed in \cite{MP2014, RFP2016, AS2016, MP2018, MPR2018, CS2018, BCFV2020}. Because of the disappearance of the converted modes at $\omega _I$ downstream of the obstacle, the Vancouver team was forced to measure the ratio of the norm in the vicinity of the obstacle where the flow is inhomogeneous by introducing a change of variable (equation 2 of the \cite{PRL2011}), a procedure that we have reproduced carefully by taking also the same window size. We notice that the blue-shifted and negative modes (at the wave-maker frequency) propagate over a shorter distance, before seeing their amplitudes transferred to the harmonic frequencies, when the amplitude of the incident wave becomes larger. These converted modes are then diffused more weakly by the undulation because they propagate on it over a shorter distance. Whereas, when the amplitude of the incident wave is lower, the converted modes propagate over a very large part of the zero mode, they then undergo a larger wave scattering by the undulation \cite{PRL2016} and the ratio $R = |\beta ^2|/|\alpha ^2|$ increases (see Fig.~\ref{Spectrum}). The Vancouver conversion was probably linear but the propagation was non-linear and does not allow an observation downstream from the obstacle of the stimulated Hawking radiation since the converted modes were transferred towards free harmonics for the range of amplitudes that was probed. On the contrary, the Poitiers reproduction has shown that to decrease the amplitudes of the incoming mode allows to avoid harmonics generation to the detriment of an extra scattering of the linear converted modes by the undulation since the converted modes propagate downstream across the undulation before reaching the asymptotic observer (as already shown in \cite{PRL2016}). To use as in \cite{PRL2011} the same surface gravity $dU/dx(x_{horizon})$ whatever the frequency for a dispersive flow (with or without blocking) to infer an analogue Hawking temperature is puzzling for the present authors since the position of the group velocity horizon (whenever a wave blocking line is present) changes with the incoming frequency. Moreover, the expression of the surface gravity takes into account only the fact that the speed of the flow varies with the position but not the speed of the non-dispersive long waves $c(x)=\sqrt{gh(x)}$ (see the generalized formula by Visser in \cite{BLV}) which is another oddity for the present authors which do consider that the measurements of the scattering coefficients should be done in a downstream homogeneous region...

Finally, the Poitiers reproduction of the 2011 Vancouver experiments demonstrates the linear mode conversion both close to a dispersive horizon in case of wave blocking or in a conversion band in absence of wave blocking (as shown by the interferences in the space-time diagram filtered at the fundamental frequency) but then shows subsequent non-linear conversions into free harmonics (keeping all the frequencies without filtering them at the wave-maker frequency), during the downstream propagation of the converted modes away from the linear interference region: a classical analogue of the Zitterbewegung region \`a la Schroedinger is seen where positive and negative norm modes are mixed \cite{Schro1930, Schro1931}. An asymptotic observer cannot measure Hawking radiation in the weakly non-linear regime because of free harmonics generation that localized the non-linear converted modes spatially to the detriment of the linear converted modes that disappear, a scenario not envisaged in Analogue Gravity until now. Regrettably, the Vancouver team missed the harmonics generation by filtering their results at the wave-maker frequency. Unfortunately, the thermal Hawking spectrum was not verified in any fluid mechanics experiments so far. As far as the Nice seminal experiments are concerned, we have shown that they operated in a stronger non-linear regime (because of higher wave cambers) with respect to the weakly non-linear experiment in Vancouver in the sense that incoming modes generated free harmonics in some of the Nice experiments in agreement with the observations reported in 2008 but only converted modes generated free harmonics in the Vancouver experiments and not the incoming modes. 

As a conclusion, we plan to come back to a modern data analysis of the old Nice experiments of 2006-2010 looking to the traces of the negative modes and of a new scenario featuring non-blocking of the incoming mode at $\omega _I$ but the production of its harmonics and a following linear conversion with a Hawking process of this latter harmonics waves which may be blocked or not depending on the frequency with a total or partial conversion. These possibilities combining weak non-linearity and dispersion were not anticipating in astrophysics (Analogue Gravity is somehow more general than General Relativity...) and may finally explain completely the Nice observations in addition to the important step reported in this work. We plan to study in the future the effect of the Landau speed threshold on the robustness of stimulated Hawking radiation which limits its propagation towards the asymptotic observer but not its production: this is another scenario not envisaged in astrophysics due to a dispersive microscopic scale, namely the capillary length, akin to the Planck scale in Quantum Gravity where Einstein's equations would not be valid anymore. As Carl Sagan used to say "{\it extraordinary claims require extraordinary evidence}", the analogue Hawking radiation and its astrophysical prediction are still keeping some of their secrets...\\

\newpage

{\bf Acknowledgements}

This research was supported by the University of Poitiers (ACI UP on Wave-Current Interactions 2013-2014), by the Interdisciplinary Mission of CNRS (PEPS PTI 2014 DEMRATNOS), by the University of Tours in a joint grant with the University of Poitiers (ARC Poitiers-Tours 2014-2015), by the French national research agency (ANR) through the grant HARALAB (ANR-15-CE30-0017-04). This work pertains to the French government program "Investissements d'Avenir" (LABEX INTERACTIFS, reference ANR-11-LABX-0017-01 and EUR INTREE, reference ANR-18-EURE-0010). The authors would like to thank Romain Bellanger, Patrick Braud, Laurent Dupuis and Jean-Marc Mougenot for their help with respect to the experimental aspects of this work.

\section{Appendix}

\subsection{Theoretical section}

Richard White in 1973 formulated a ray theory which describes the propagation of sound in moving fluids which are inhomogeneous and inviscid \cite{White}. The dispersion relation of sound pulses can be written in a compact and 1+3 form within geometrical acoustics $g^{\mu \nu}k_\mu k_\nu =0$ with the 4D wave-number $k_\mu = (-\frac{\omega}{c}, {\bf k})$ and the so-called acoustic metric $g^{\mu \nu}= \begin{bmatrix}
1 & {\bf U}^\intercal /c\\
{\bf U} /c& {\bf U}\otimes {\bf U}/c^2 -\mathbb{1}
\end{bmatrix}$. The same metric appears in the propagation equation (also derived by White) $\frac{1}{\sqrt{-g}}\partial _\mu \left[\sqrt{-g}g^{\mu \nu} \partial _\nu \phi =0\right]$
with $g=det(g^{\mu \nu})$ where $\phi$ is the velocity potential for sound waves \cite{White}. The speed of acoustic waves $c$ becomes the speed of long gravity waves $\sqrt{gh}$ in hydraulics for free surface flows as derived by Unruh and Sch\"{u}tzhold in 2002 with $h$ the water depth and $g$ the gravity field (not to be mixed with the determinant of the metric) \cite{SU}. The resulting dispersion relation writes $(\omega - {\bf U.k})^2=c^2k^2$ that is $\omega - {\bf U.k}=\pm c \lvert k\rvert$ with $k=\sqrt{k_x^2+k_y^2}$ in the experiments where ${\bf U}=U{\bf e_x}$ and $U<0$ for a black/white hole flow with a(n) acceleration/deceleration in the rest of the paper. The flow is directed along the $x$ direction and the water waves propagate in the plane $(x,y)$ where $y$ is the transverse direction to the flow current and the channel.

\begin{figure}[!htbp]
\includegraphics[width=8cm,height=4cm]{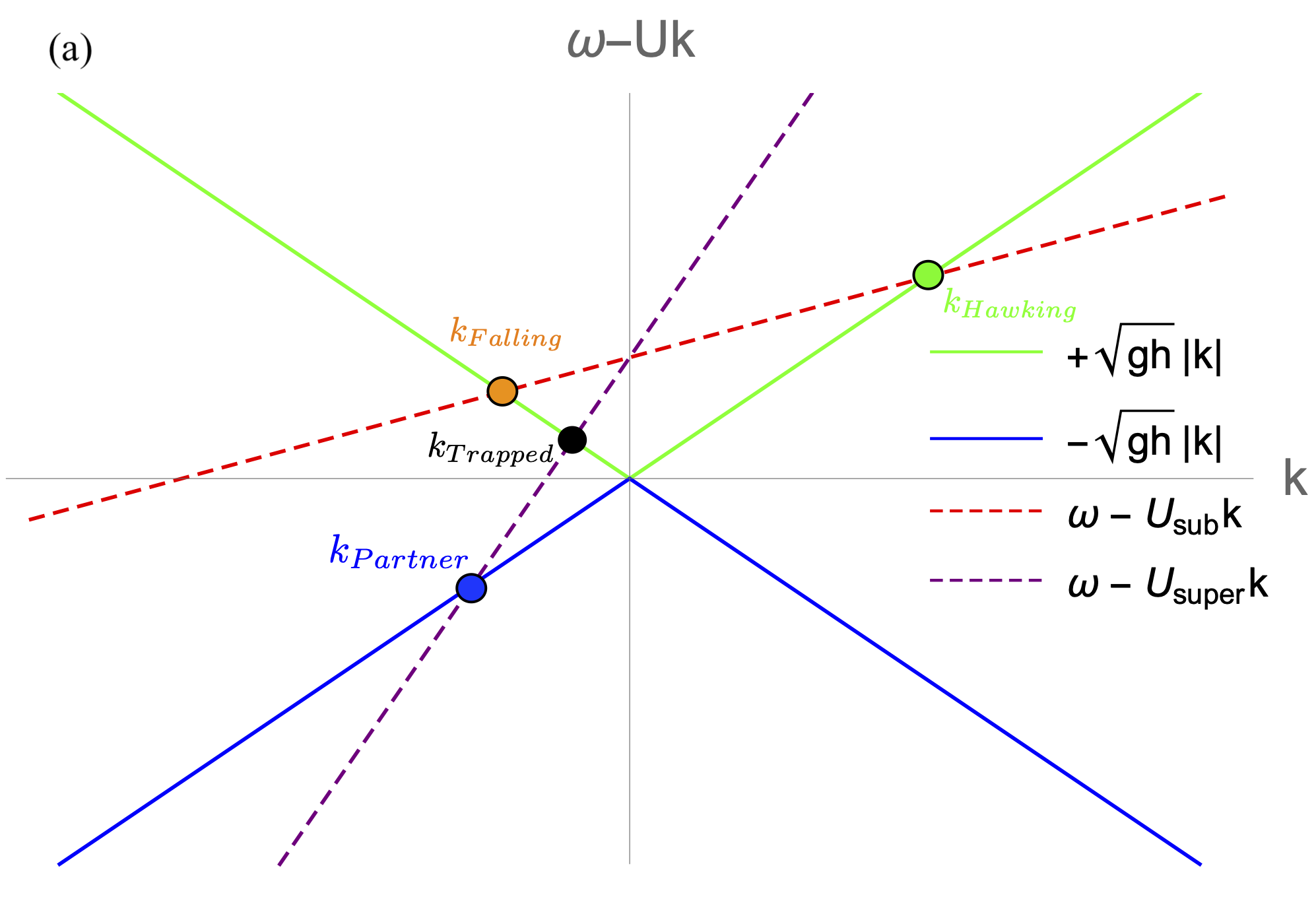}
\includegraphics[width=8cm,height=4cm]{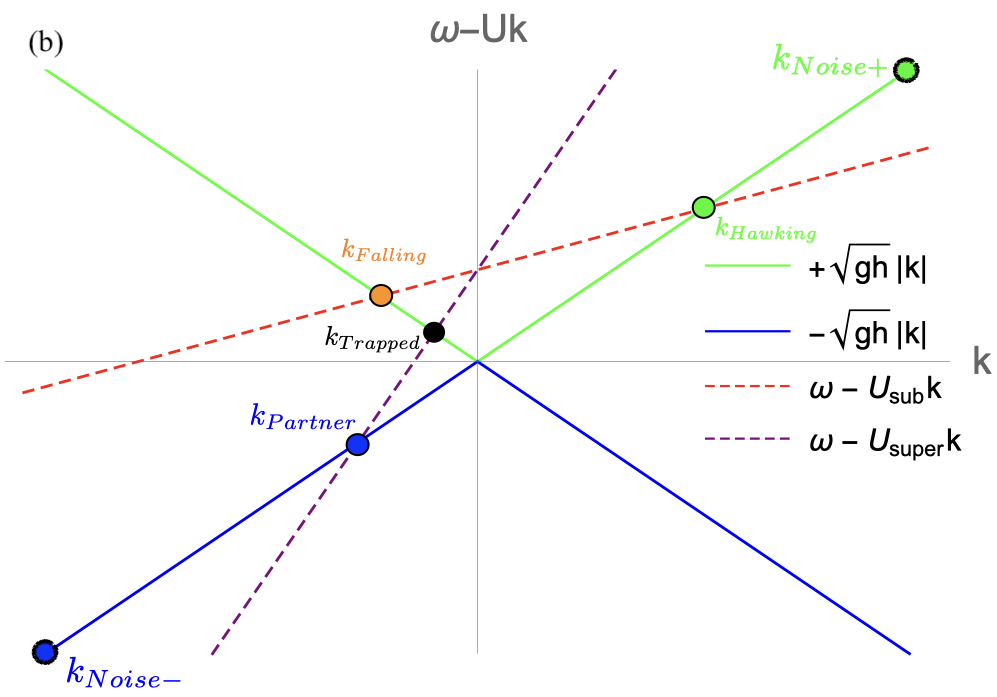}
\caption{Mode mixing in the dispersion-less limit ($kh<<1$) by a (a) conformal coupling when the wave $c=c(x)$ varies with space  corresponding to the extreme shallow water limit ($k_{Falling}=>k_{Trapped}+k_{Hawking}+k_{Partner}$) as reported in \cite{PRL2020} for a black hole flow OR by a (b) "Spontaneous" Hawking Process ($k_{Noise^+}=>k_{Hawking}$ AND $k_{Noise^-}=>k_{Partner}$) with a seed of classical (mechanical/thermal noise in aquatic analogues) or quantum (vacuum noise in astrophysics/Hawking's original prediction) origin for a transcritical flow with a Froude number $Fr=U/c$ crossing one. If $Fr<1$, then $U=U_{sub}$ and if $Fr>1$, then $U=U_{super}$ in the graphics.} 
\label{RD1}
\end{figure}

A falling ingoing modes (water or acoustic waves or light) has negative wavenumber and lives outside of a black hole horizon. When going through the trans-critical boundary or horizon in the dispersion-less limit, the falling mode becomes a trapped mode which cannot escape from the  interior of the horizon. If the speed $c$ of waves does not vary with space, the solution $k_{Falling}$ is totally converted into $k_{Trapped}$ with obvious names for waves interacting with a black hole horizon. However, when the wave speed varies with the position according to $c=c(x)$, then a partial conversion called "conformal coupling" towards the pair of outgoing modes occurs ($k_{Hawking}$ and $k_{Partner}$) but we insist on the fact that this not the Hawking process which should occurs at the horizon (whether its spontaneous or stimulated exemplifications): we have been testing experimentally this scenario recently \cite{PRL2020}. $k_{Hawking}$/$k_{Partner}$ is a mode which is propagating in the counter-current with a speed superior/inferior to it. The genuine Hawking process corresponds to the conversion where the outgoing modes origin corresponds to transplanckian modes for which the wavelength goes to zero at the horizon in the dispersion-less limit: we denote them as $k_{Noise\pm}$ depending on the sign of their norm. $k_{Noise^+}$/$k_{Noise^-}$ leaves in the subcritical/supercritical region of the black hole flows such that the speed is $U_{sub}$/$U_{super}$ (see Fig. \ref{RD1}).

\begin{figure}[!htbp]
\includegraphics[width=8cm,height=4cm]{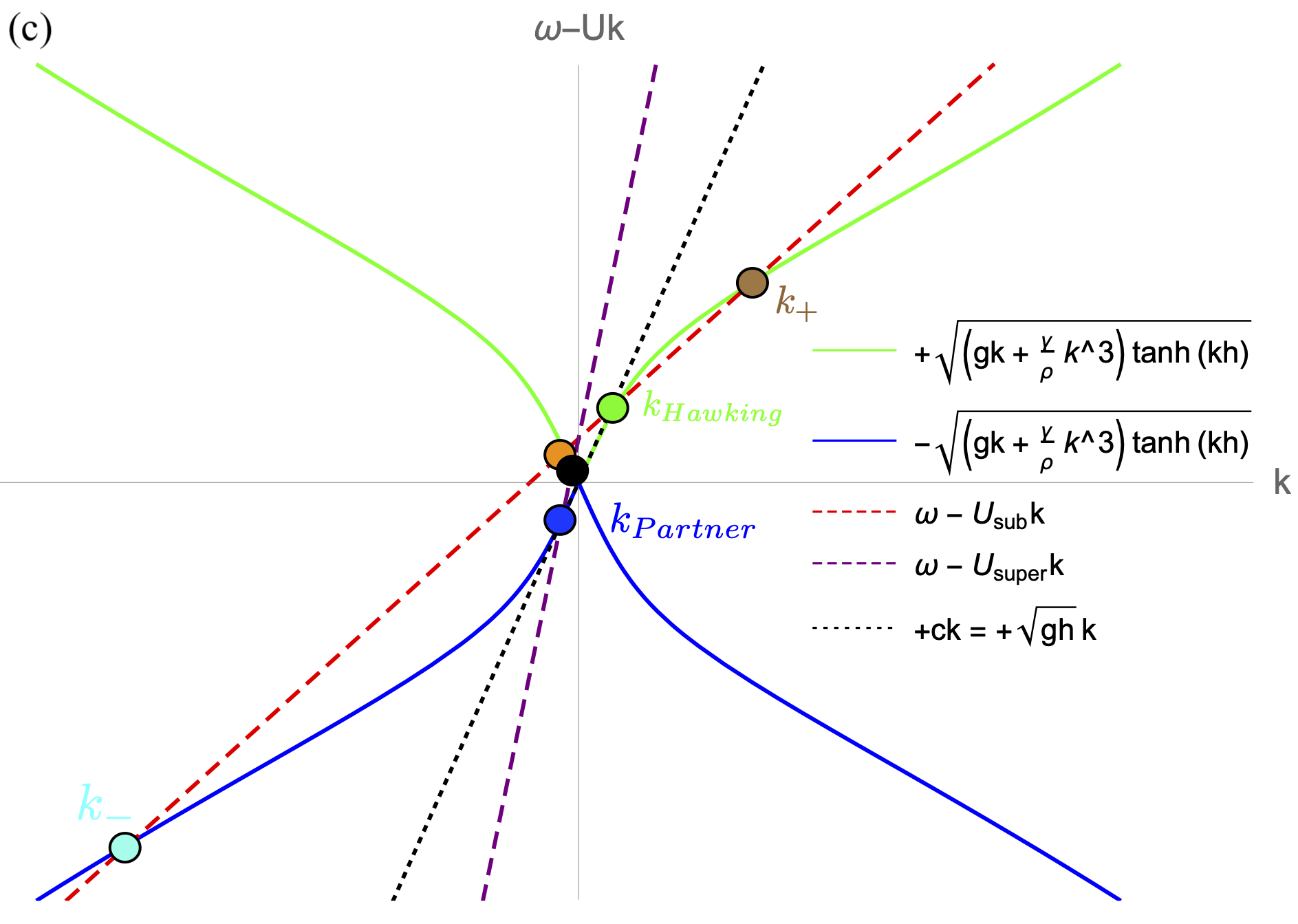}
\includegraphics[width=8cm,height=4cm]{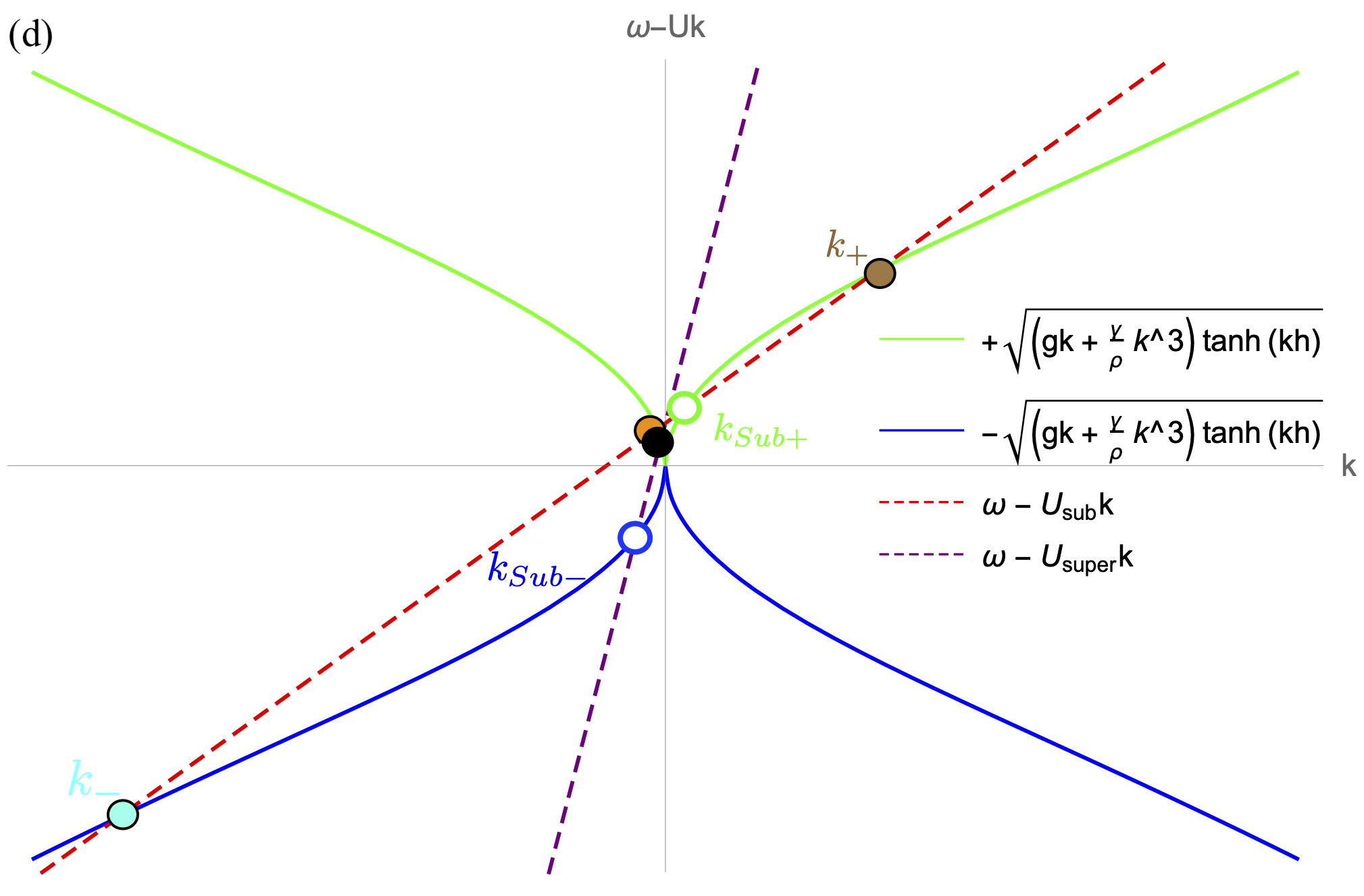}
\caption{Stimulated Hawking Radiation from initial short gravity waves in a subcritical black hole flow ($k_{^+}=>k_{Hawking}$ and $k_{^-}=>k_{Partner}$) or from long gravity waves in a white hole flow \cite{PRL2009} ($k_{Sub^+}=>k_{^+}+k_{^-}$ in the subcritical region AND $k_{^-}=>k_{Sub^-}$ in the supercritical region from the group velocity point of view in this dispersive regime) with a Froude number inferior to one but a subluminal dispersive correction followed by a superluminal dispersive correction. The incoming short gravity modes originate from outside the dispersive horizon for example when looking at a white hole flow as reported in the main text: (c) neglecting surface tension whatever the water depth and (d) neglecting surface tension in the deep water limit $kh>>1$.}
\label{RD2}
\end{figure}

When including dispersion, two behaviors (see Figure \ref{RD2}) are possible according to the sub-luminal (the group velocity $\partial \omega /\partial k$ decreases with $k$) or sub-super-luminal (the group velocity decreases then increases when including microscopic Physics, here surface tension effect with a characteristic scale, the capillary length, playing the role of the Planck scale) regimes of the dispersive correction \cite{BLV, Scott, Como}. Another regime (see Chapter 7 of \cite{Como}) with both shallow water and surface tension effect with a pure superluminal corrections akin to the one occurring in a Bose-Einstein condensate has been dismissed since the water depth is of the order of a few centimeters (it would have been relevant for the case of a circular jump akin to a white fountain \cite{PRE2011}).

To simplify the notations for a dispersive subcritical white hole flow in the deep water gravity regime, the incoming mode from outside the white hole and propagating on the counter-current is denoted by $k_I$ in the main text ($k_{Sub^+}$ in the Figure \ref{RD2}), the corresponding retrograde mode is $k_R$ propagating in co-current and the pair of converted modes are the blue-shifted one $k_B$ and the negative one $k_N$ \cite{PRL2009} ($k_{^+}$ and $k_{^-}$ in the Figure \ref{RD2}).

\begin{figure}[!htbp]
\includegraphics[width=8cm,height=4cm]{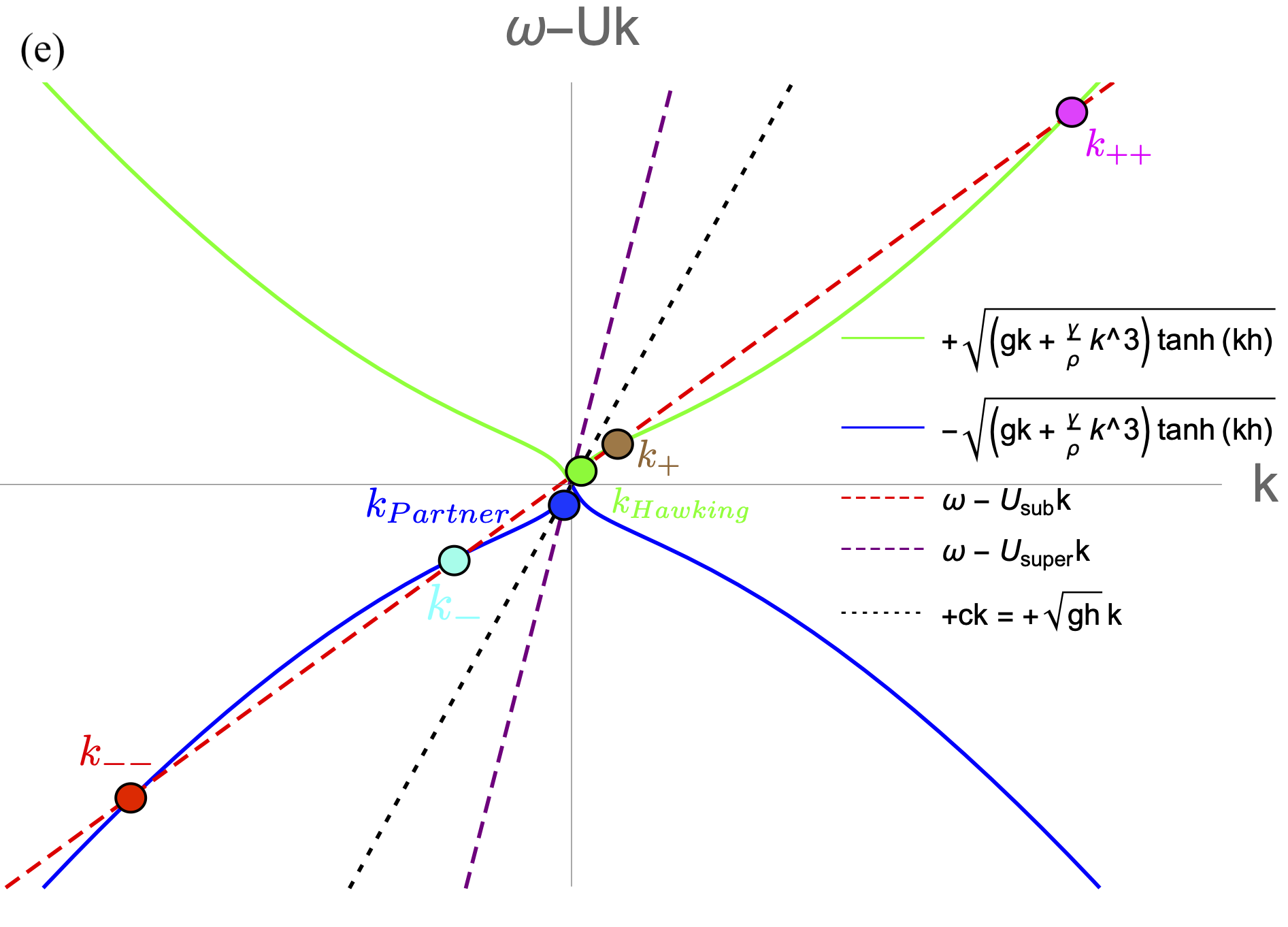}
\includegraphics[width=8cm,height=4cm]{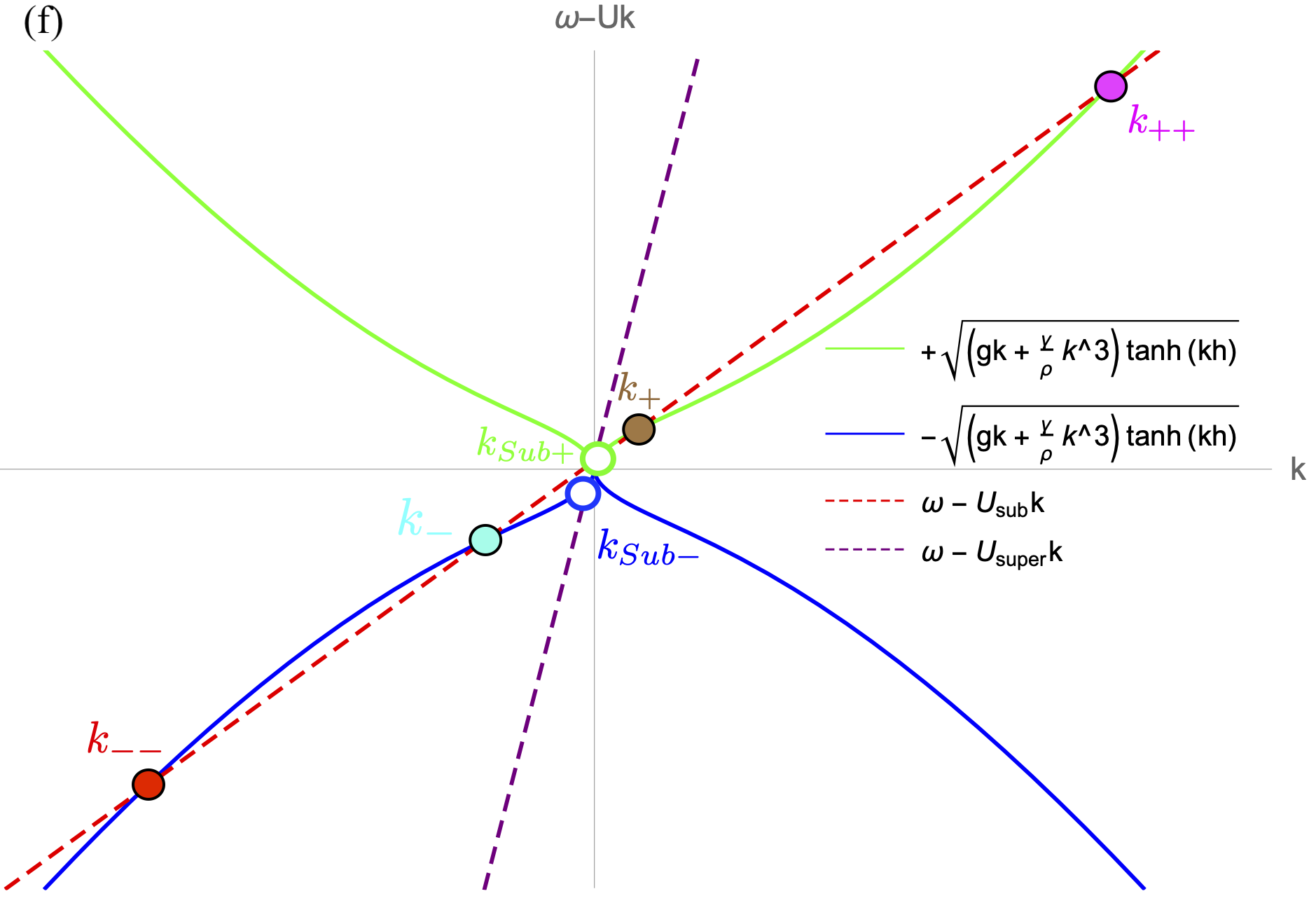}
\caption{Stimulated Hawking Radiation from initial short capillary waves in a subcritical black hole flow ($k_{^{++}}=>k_{^+}=>k_{Hawking}$ and $k_{^{--}}=>k_{^-}=>k_{Partner}$) or from short gravity waves in a white hole flow ($k_{Sub^+}=>k_{^+}+k_{^-}$ and $k_{^+}=>k_{^{++}}$ and $k_{^-}=>k_{^{--}}$ in the subcritical region AND $k_{^-}=>k_{Sub^-}$ in the supercritical region from the point of view of the group velocity) with a Froude number inferior to one but a subluminal dispersive correction followed by a superluminal dispersive correction. The incoming capillary modes originate from inside the dispersive horizon for example when looking at a white hole flow as reported in \cite{PRD2017}: (e) including surface tension whatever the water depth and (f) including surface tension in the deep water limit $kh>>1$.}
\label{RD3}
\end{figure}

In the current work, the effect of capillarity induces a threshold for the appearance of either mode mixing \`a la Hawking with positive and negative norm modes $k_B$ and $k_N$ in th main text (the so-called Landau speed at 23.1cm/s for long period \cite{NJP2010}) or the appearance of an undulation (the zero frequency solution $\pm k_z=k_B(\omega=0)=-k_N(\omega=0)$ of the dispersion relation in presence of a current as discussed in \cite{Unruh2008, NJP2010, Como, CP2014}). The extra modes due to capillarity (see Fig. \ref{RD3}) will not studied here but the interested reader shall have a look to the following references \cite{NJP2010, Como, PRD2017}.

\subsection{Experimental section}

\subsection{The Poitiers reproduction of the Vancouver experiments}

\begin{figure}[!htbp]
\includegraphics[width=8cm,height=4cm]{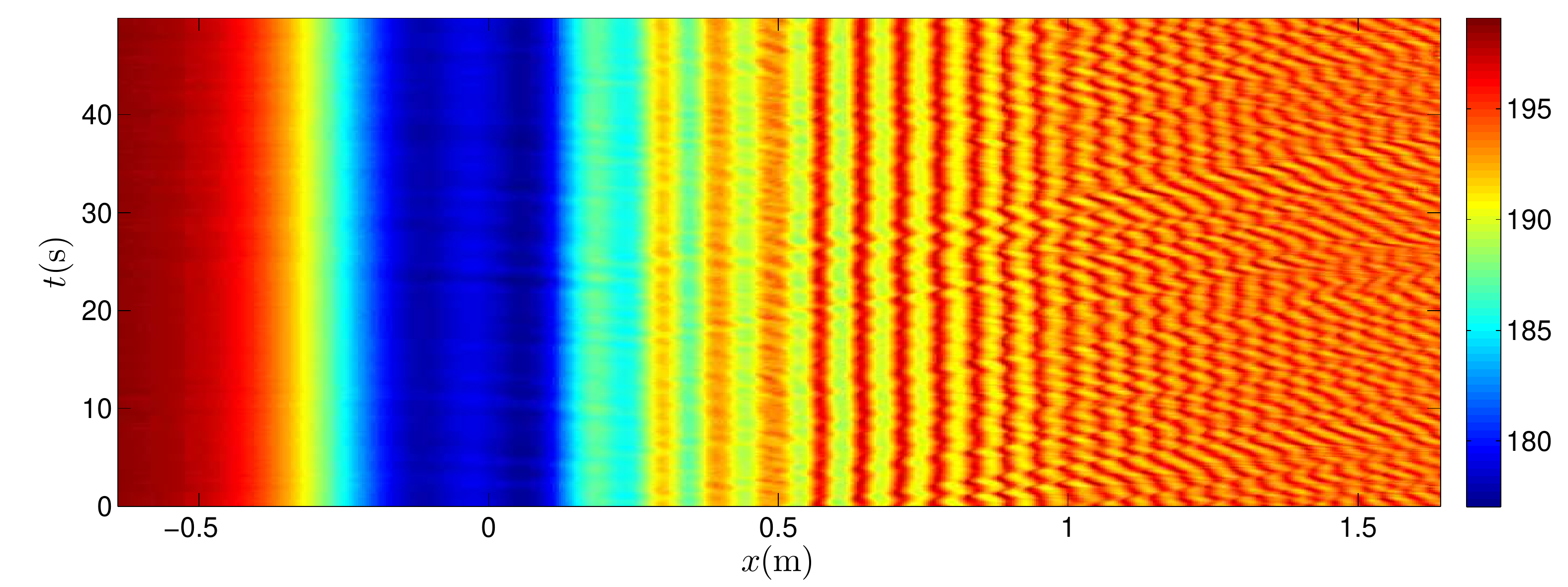}
\includegraphics[width=8cm,height=4cm]{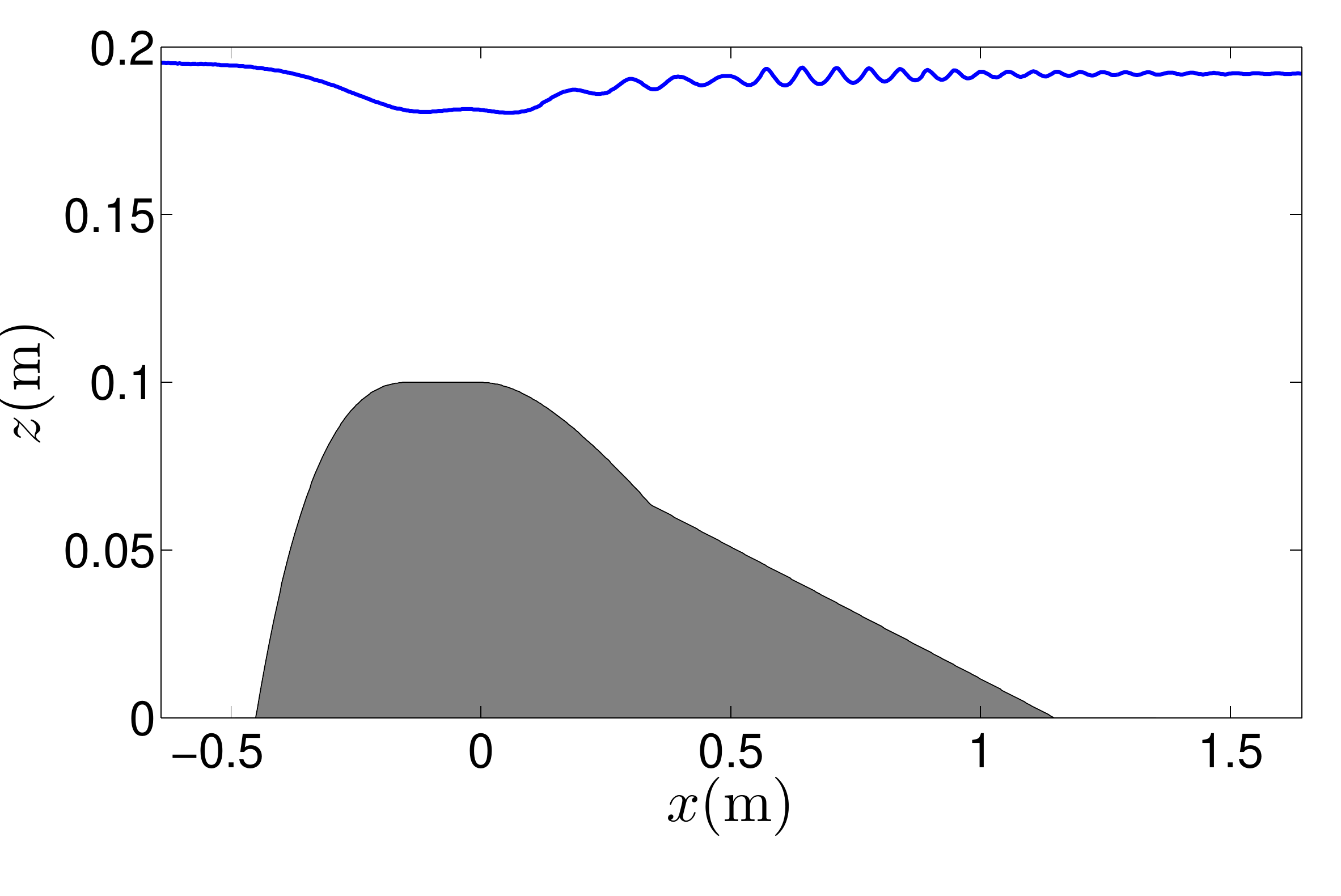}
\caption{(Top) Space-time diagram of the water height $h (x, t)$ (in mm) of the permanent regime used in the Vancouver experiments \cite{PRL2011} without incoming water waves except some random noise; (Bottom) the average undulation in time featuring a depression (above the flat part of the geometry where the speed is higher) followed by secondary waves or whelps $\left \langle h (x, t) \right \rangle$ in blue on the descending part of the bottom obstacle in grey.}
\label{Undulation}
\end{figure}

The geometry of the Vancouver obstacle \cite{PRL2011} is given by (see Fig.~\ref{Undulation}):
\begin{eqnarray}
	o(x) = \left\{
    \begin{array}{c}
        2a(1-(x-x_1)-e^{-b(x-x_1)})\\
        0,1\\
   		2a(1-(x-(x_1-x_2))-e^{-b(x-(x_1-x_2))})\\
   		o(x_4)-(x-x_4)\tan(4,5^\circ)
    \end{array}    
    \begin{array}{c}
        x\in[x_1, x_2] \\
   		x\in[x_2, x_3]  \\
   		x\in[x_3, x_4] \\
   		x\in[x_4, x_5]
    \end{array}
\right.
\label{eq:reg:obstacle}
\end{eqnarray}
with $x_1$=-0,45~m, $x_2$=-0,15~m, $x_3$=0~m, $x_4$=0,34~m, $x_5$=1,15~m, $a$=0,094, $b$=5,94.

Let us first characterized the mean free surface over the obstacle for the Vancouver obstacle in the Fig.~\ref{Undulation} and parameters which are: flow rate $Q=17.5 l/s$ ; flow rate per unit width $q=0.045 m^2/s$ since the channel width is $W=0.39m$; asymptotic upstream dynamical water depth $h_{upstream} = 194 mm$. The free surface measurements (optical measurement with the laser line using fluoresceine) show that the maximum Froude number observed above the obstacle is $Fr = 0.67$ (for $ h _ {\rm {min}}= 77.5 mm$ corresponding to the depression in front of the secondary waves of the undulation) whereas the upstream Froude number is 0.22. The undulation reaches a maximum amplitude of about $4 mm$ and its wavelength varies from $50 mm$ to $120 mm$ (see Fig.~\ref{Undulation}). The zero mode is at the verge of wave breaking in order to have a Froude number as large as possible and as close to 1 that would correspond to a non-dispersive white hole horizon. In other words, by slightly increasing the flow rate or by reducing the overall dynamical height with the downstream weir then the undulation would  certainly break. The wave numbers $k_B$ and $k_N$ are close to that of the zero mode $k_Z$ or the zero mode is nothing else than the blue shifted $k_B(\omega=0)=+k_Z$ and negative $k_N(\omega =0)=-k_Z$ modes at zero frequency. Hence, the fact of superimposing both blue-shifted and negative solutions (even of small amplitude) "on the top" of the zero mode which is itself at the limit of the breaking limit makes the blue-shifted and negative modes non-linear. Their wave numbers are very close to that of the zero mode, and even assuming that they are equal (although not exactly but almost), the maximum curvature would be equal to the sum of the curvatures; that of the zero mode being already at the limit of the surge, a slight addition generates the non-linearities which are observed.

\begin{figure}[!htbp]
\includegraphics[width=8cm,height=6cm]{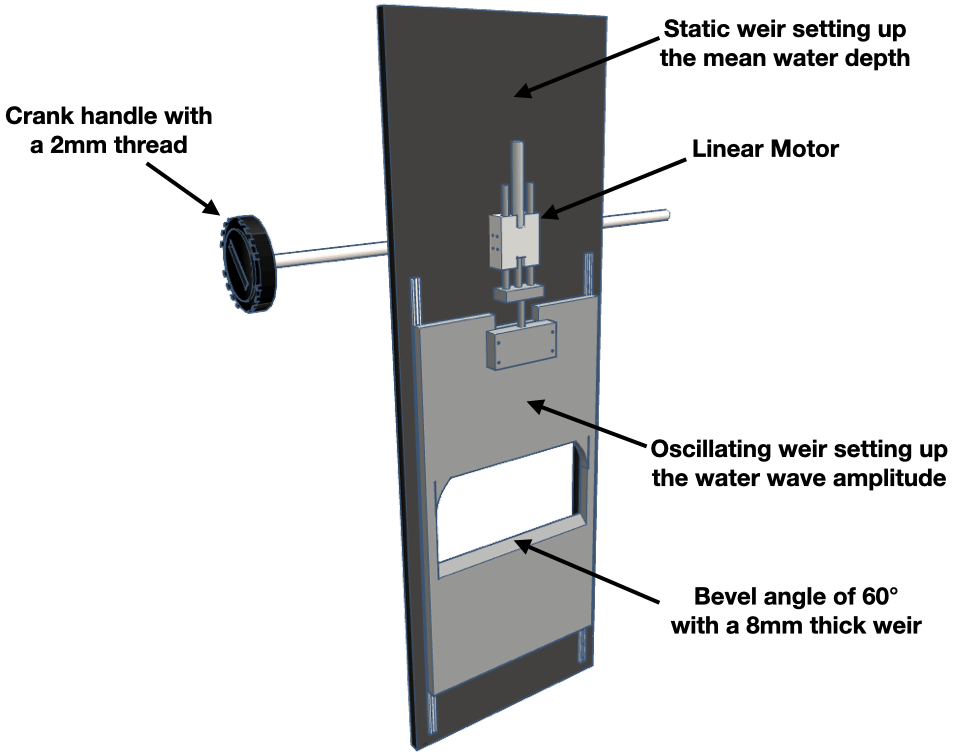}
\caption{Oscillating double-weir wave-maker used to produced the counter-current water waves downstream of the obstacle (Picture courtesy of Johan Foudrinoy, see \cite{JGCGC2018}).}
\label{Weir}
\end{figure}

Water waves are produced downstream of the bottom obstacle by a double-weir system (see Fig.~\ref{Weir}). The first weir is a usual one in open channel flows and serves to fix the stationary dynamical water depth as a function of the flow rate and connects the channel to the exit reservoir. Mounted on the first weir with vertical and lateral sliders, a second weir is driven by a linear motor Linmot from TransTechnik ($2PS01-23x160H-HP-R$ with a standard stroke of 60mm namely a maximum wave-maker amplitude $\pm a_m=30mm$ with a maximum speed of $5.3m/s$). A detailed study of the transfert function (wave amplitude as a function of the stroke displacement for several frequencies) of the wave-maker was undertaken elsewhere \cite{JGCGC2018}.

\begin{figure}[!htbp]
\includegraphics[width=8cm,height=4cm]{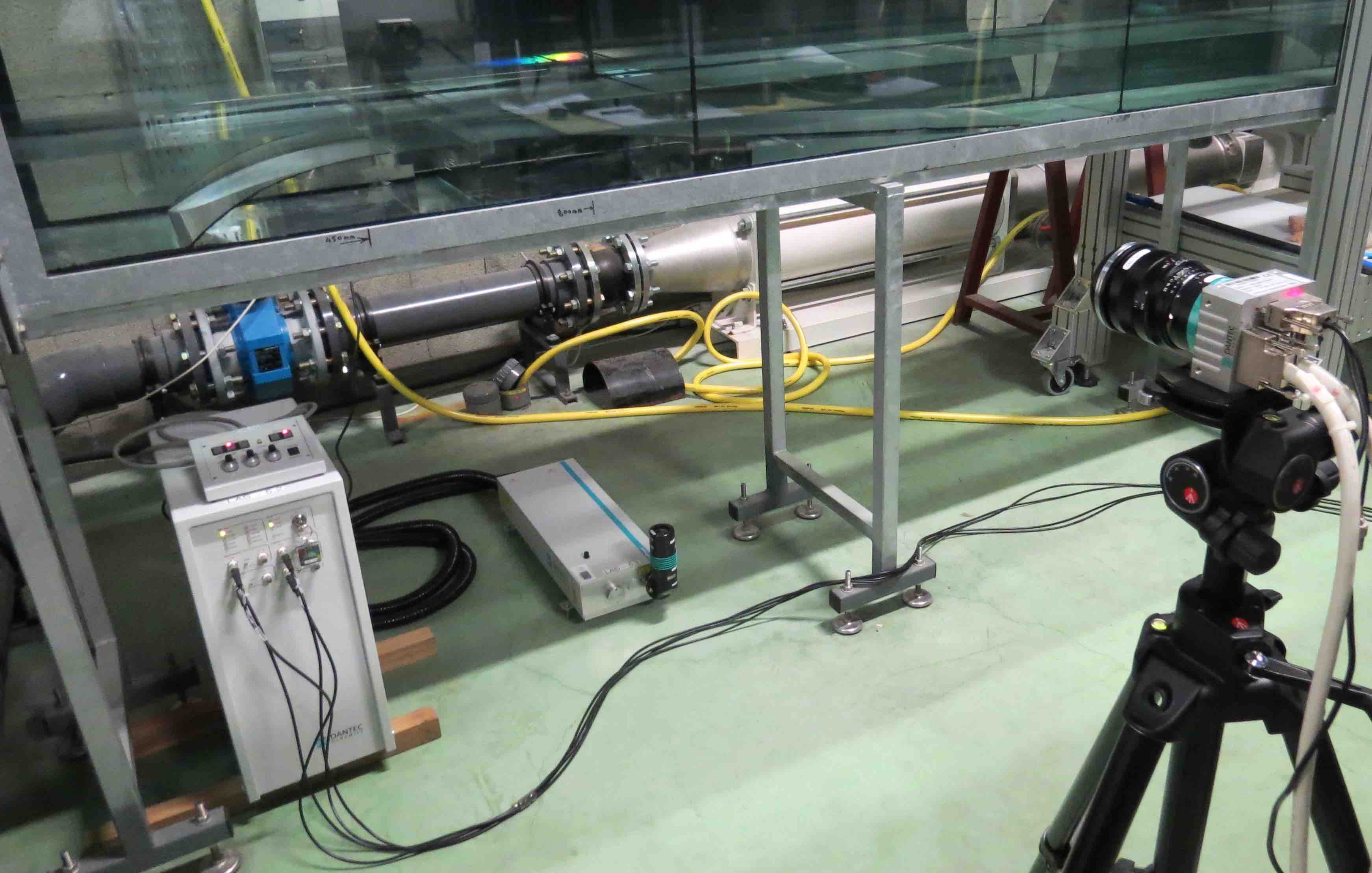}
\includegraphics[width=8cm,height=4cm]{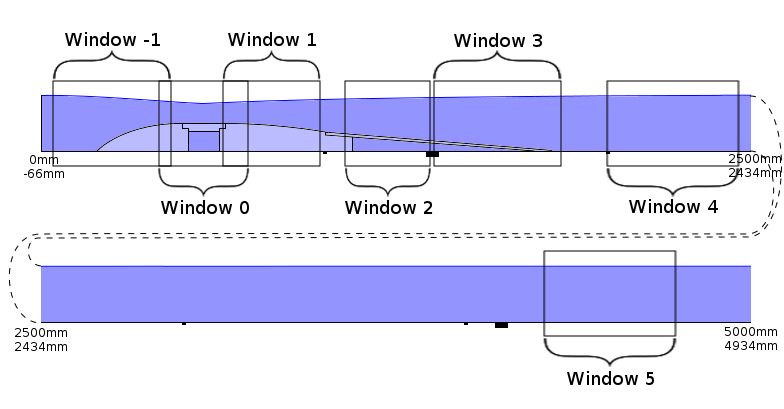}
\includegraphics[width=8cm,height=4cm]{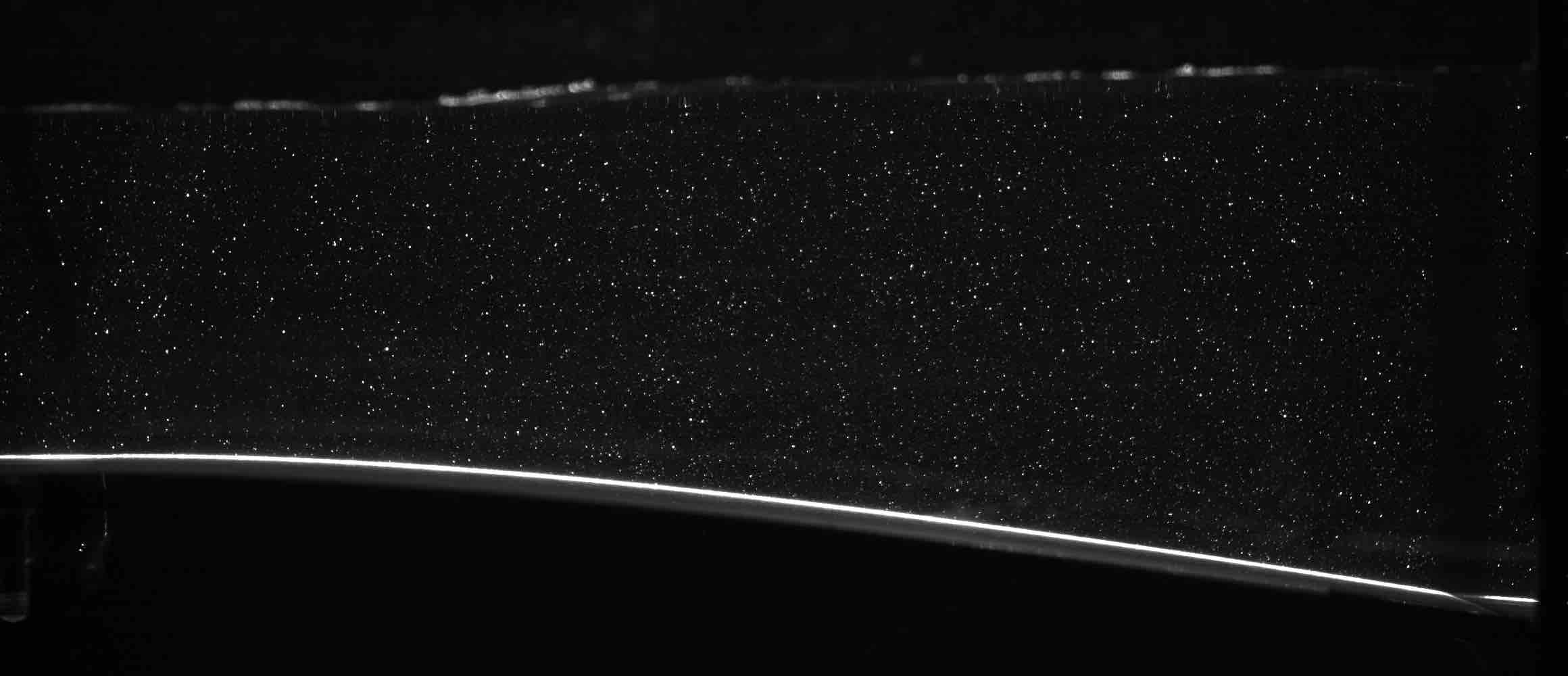}
\caption{(Top) Experimental setup of the Particle Image Velocimetry (PIV) measurements with the LASER placed below the bottom window of the water channel and the side CMOS camera; (Middle) The positions of the numbered windows used to reconstruct the longitudinal velocity profile; (Bottom) A picture of the seeded particles in the flow with the laser line on the obstacle corresponding to Window $1$.}
\label{PIV}
\end{figure}

\begin{figure}[!htbp]
\includegraphics[width=9cm,height=4cm]{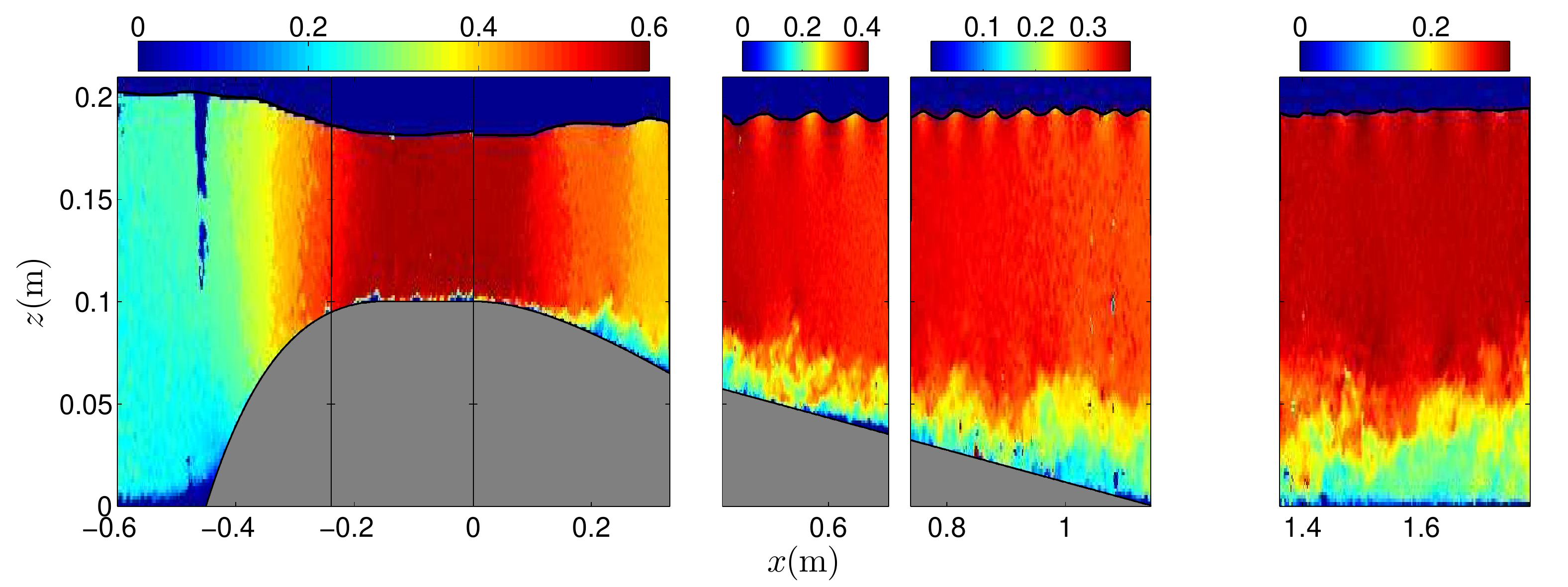}
\includegraphics[width=9cm,height=4cm]{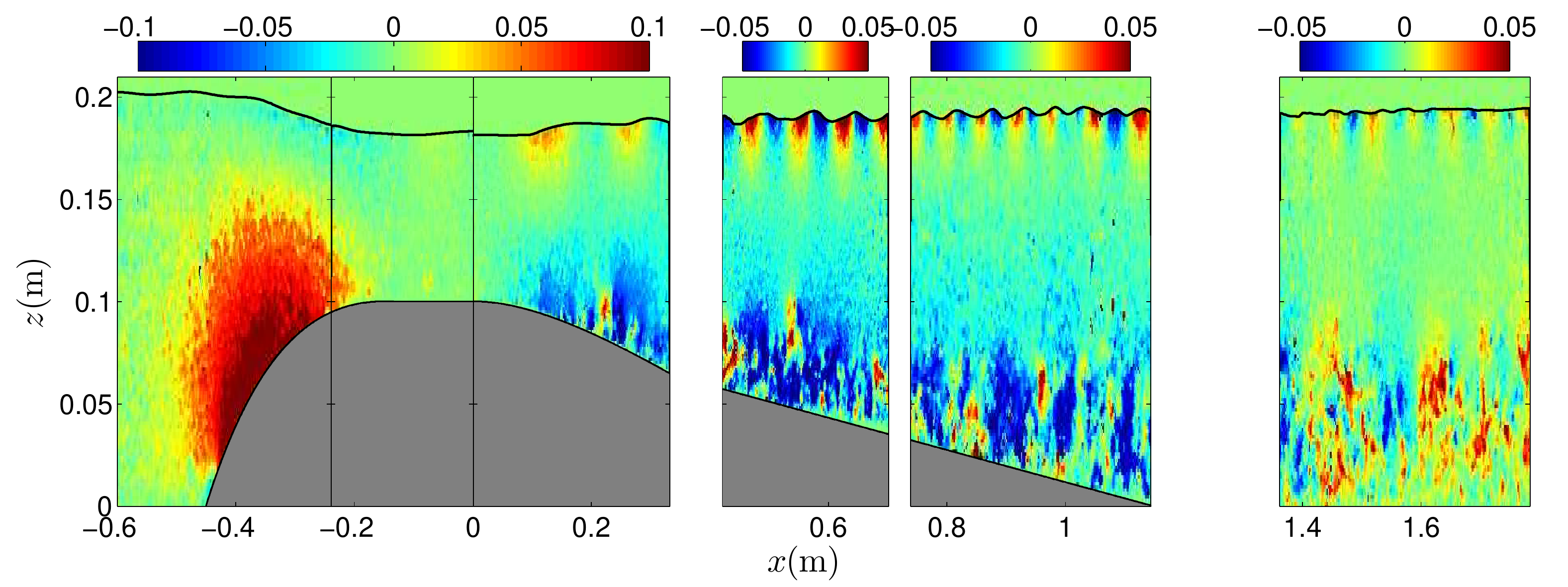}
\caption{(Top) Horizontal component of the instantaneous velocity field $ U_x (x, z, t) $ (m / s) of the flow above the Vancouver obstacle; (Bottom) Vertical component of the instantaneous velocity field $ U_z (x, z, t) $ (m / s) of the flow above the same obstacle.}
\label{Uvsx}
\end{figure}

\begin{figure}[!htbp]
\includegraphics[width=9cm,height=4cm]{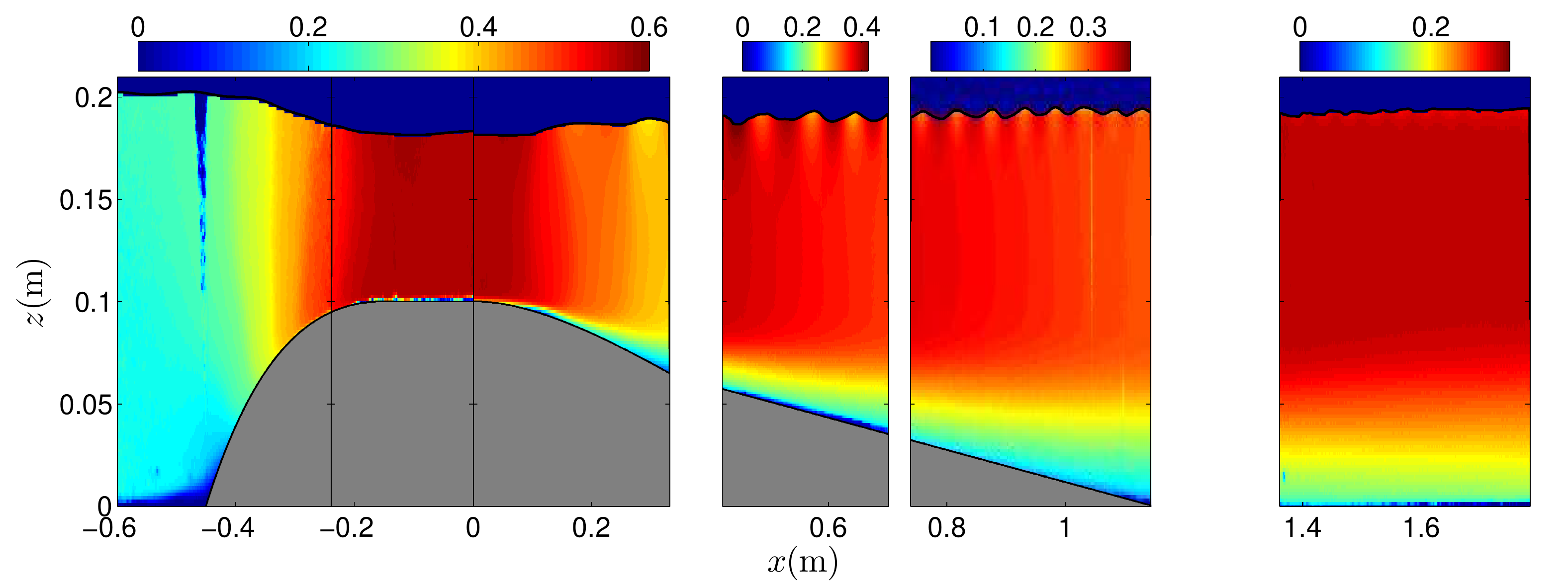}
\includegraphics[width=9cm,height=4cm]{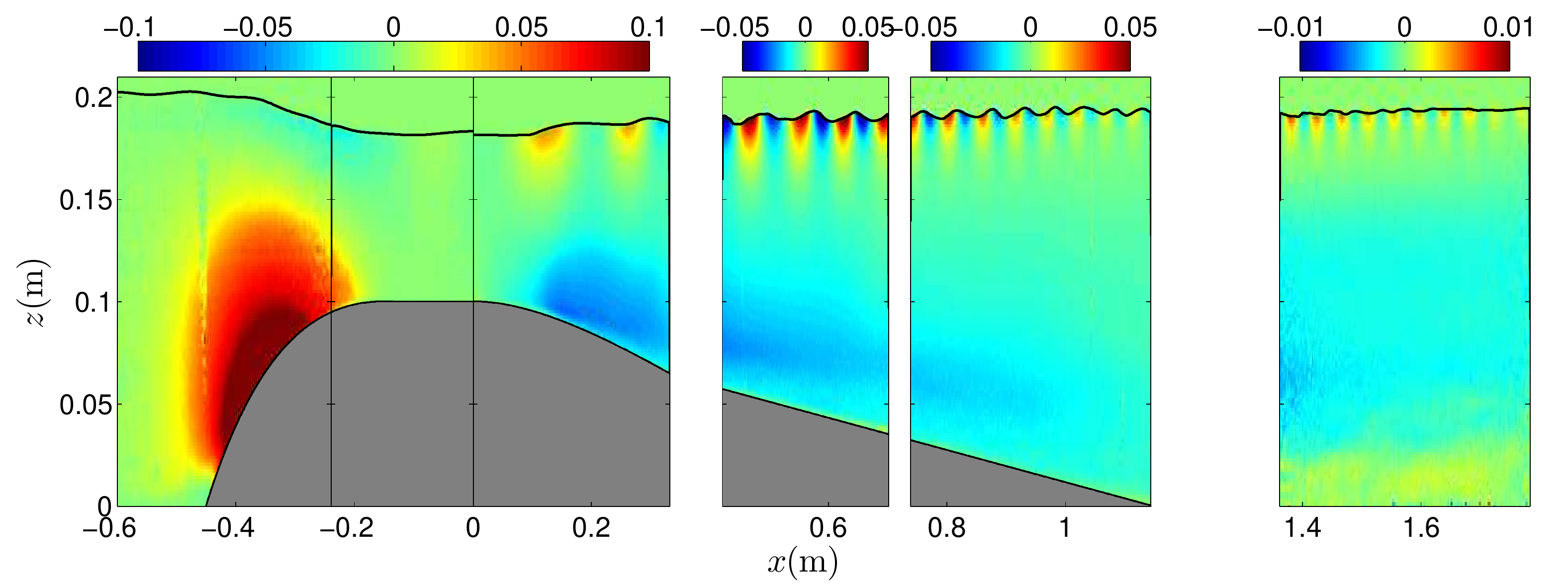}
\includegraphics[width=9cm,height=4cm]{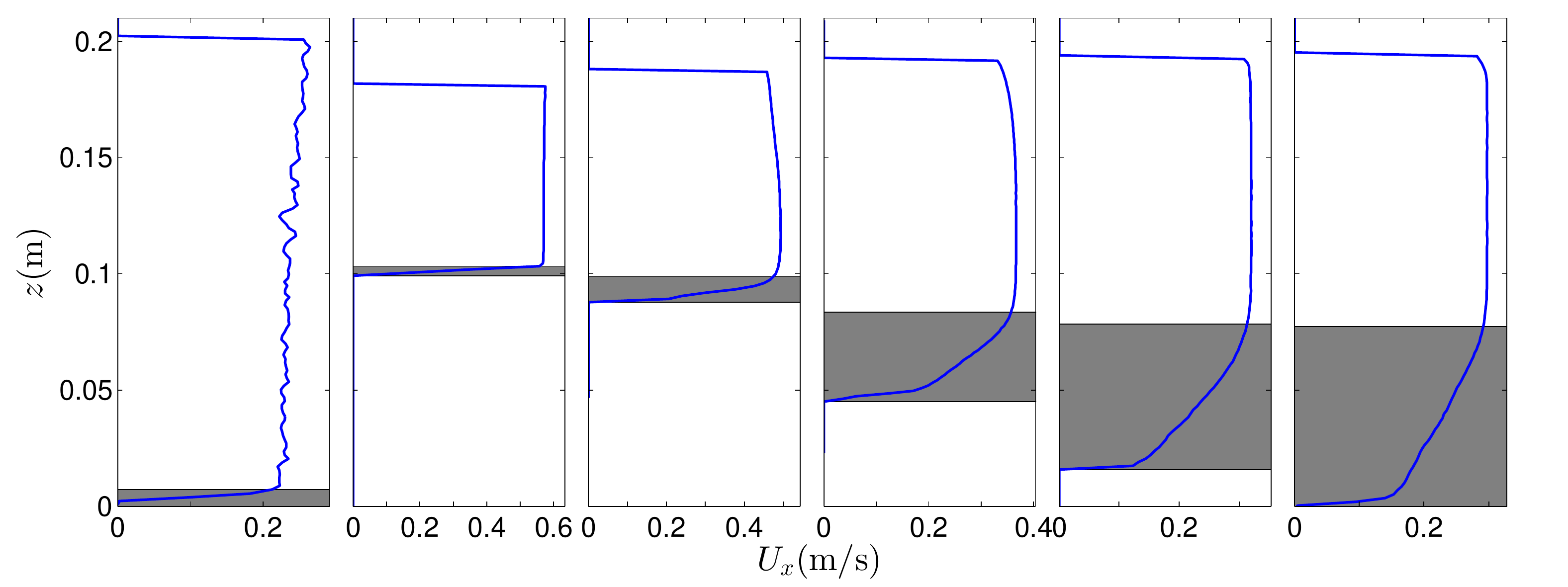}
\caption{(Top) Horizontal component of the time-average velocity field $ \left \langle U_x \right \rangle $ (m / s) of the flow above the Vancouver obstacle ; (Middle) Vertical component of the time-average velocity field $ \left \langle U_z \right \rangle $ (m / s) of the flow above the same obstacle; (Bottom) Time-average vertical profile of the horizontal component $ U_x (z) $ (m / s) for different positions along the water channel. From left to right: $ x $ = - 0.580 m, -0.120  m, 0.166  m, 0.561  m, 0.941  m and 1.572  m (in the center of each window except for the one upstream). The shaded areas represent the boundary layer thickness.}
\label{Uvsz}
\end{figure}

In order to measure the velocity fields under the free surface for this regime, a particle image velocimetry (PIV) method was set up (see Fig.~\ref{PIV}, \ref{Uvsx} and \ref{Uvsz}). The flow is seeded with polyamide particles (Vestosint 2158) whose small size and density close to that of water ($d_{50} = 20 \rm{\mu}m$, $ \rho $ = 1016 ~ $ \rm{kg / m ^ 3} $) allow to follow the displacements of the fluid. A dual-cavity pulsed Nd: Yag (\textit{Dantec Dynamics Dual Power} 2x50 mJ) LASER is used to generate two Laser layers separated by a time interval $ \Delta t $ (depending on the average speed of the flow in order to to obtain an average displacement of 8 pixels between two images), in our case $\Delta t $ is between 2 ~ ms and $5 ms$ depending on the region of the flow to be studied. A CMOS camera (SpeedSense 1040 from Dantec Dynamics - 2320x1726) synchronized with the Laser then captures 4000 pairs of images at an acquisition rate of $40 Hz$. The measurements are made in the plane ($ xOz $), with $ \vec {z} $ the vertical and $ \vec {x} $ the longitudinal direction of the flow, and in the center of the channel. As the Laser sheet cannot be projected from above because of the deformations that the free surface would induce, a transparent PMMA obstacle had to be created so that the sheet, which is emitted by below, can pass through it. In order to have a good enough resolution, the camera is placed so that it observes a field of $400 mm$ wide. It is then displaced to observe different regions of the flow (see Fig.~\ref{PIV}): some of the latter were reduced because of shadow zones due to the reinforcement posts under the channel or due to the junctions of the different parts of the obstacle.

\begin{figure}[!htbp]
\includegraphics[width=8cm,height=4cm]{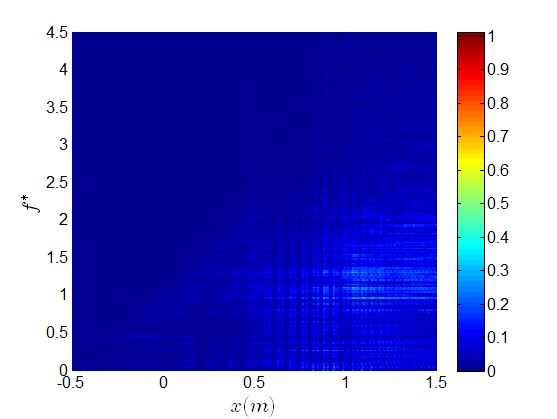}
\includegraphics[width=8cm,height=4cm]{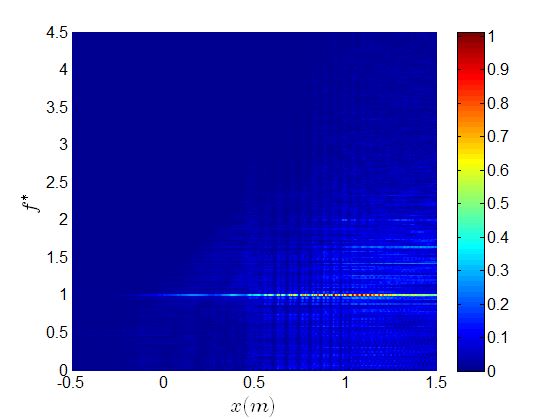}
\includegraphics[width=8cm,height=4cm]{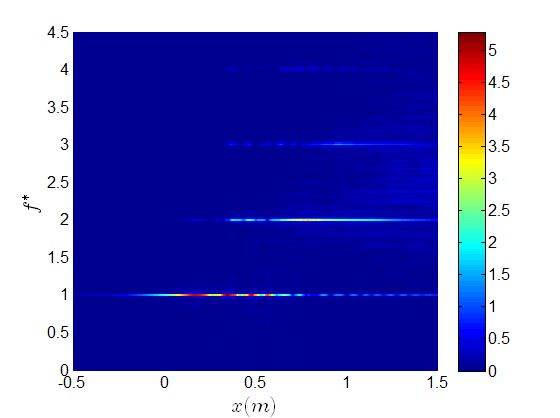}
\caption{Spatial spectrograms $\tilde{\delta h}(x,\omega)$ with the dimensionless frequency $f^*=f/f_I=\omega /\omega_I$ in meter ($x=0m$ is the flat part right extremity of the obstacle). Parameters: (Top) free surface noise only; (Middle) wave-maker frequency = incoming wave angular frequency $ \omega_I = 3.14 \rm {Hz}$ and an asymptotic incoming wave amplitude $a_I=0.05mm$ without free harmonics generation; (Bottom) $ \omega_I = 3.14 \rm {Hz}$ and $a_I= 0.8 mm$ with free harmonics generation. See the additional videos.}
\label{Spatial}
\end{figure}

The results of the velocimetry (see Fig.~\ref{Uvsx} and \ref{Uvsz}) show the generation of a boundary layer in the downstream part of the obstacle. We observe that the upper limit of the boundary layer is approximately equal to the maximum height of the obstacle. Note that the minimum speed $U_{plug}=0.232 m/s$ (assuming a plug flow) is greater than $ U_{\gamma} =0.231 m/s$, the Landau threshold \cite{NJP2010} for this speed range. In practice, the measured speed at the free surface is even higher $U_{real}=0.295 m/s$ (only the waves with frequency $2.4 \rm {Hz}<\omega <8.3 \rm {Hz}$ will be blocked because either the geometry does not accelerate sufficiently the flow speed with a maximum value or because the flow rate sets up a minimum speed compatible with the wave blocking line obtained by canceling the group speed \cite{NJP2010}). Hence, the hypothesis of a plug flow is invalidated in our reproduction of the Vancouver experiments. It is the measured downstream surface speed that must be used to fit the experimental dispersion relation downstream of the obstacle because of the thickening of the boundary layer as reported elsewhere \cite{PhD2017}. Therefore, the disappearance of the converted modes cannot be associated to the Landau speed threshold downstream of the obstacle but to the appearance of free harmonics as demonstrated for the first time in this work by changing the asymptotic amplitude of the incoming modes (see the spatial spectrograms of Fig.~\ref{Spatial} and the corresponding videos).

\begin{figure}[!htbp]
\includegraphics[width=8cm,height=4cm]{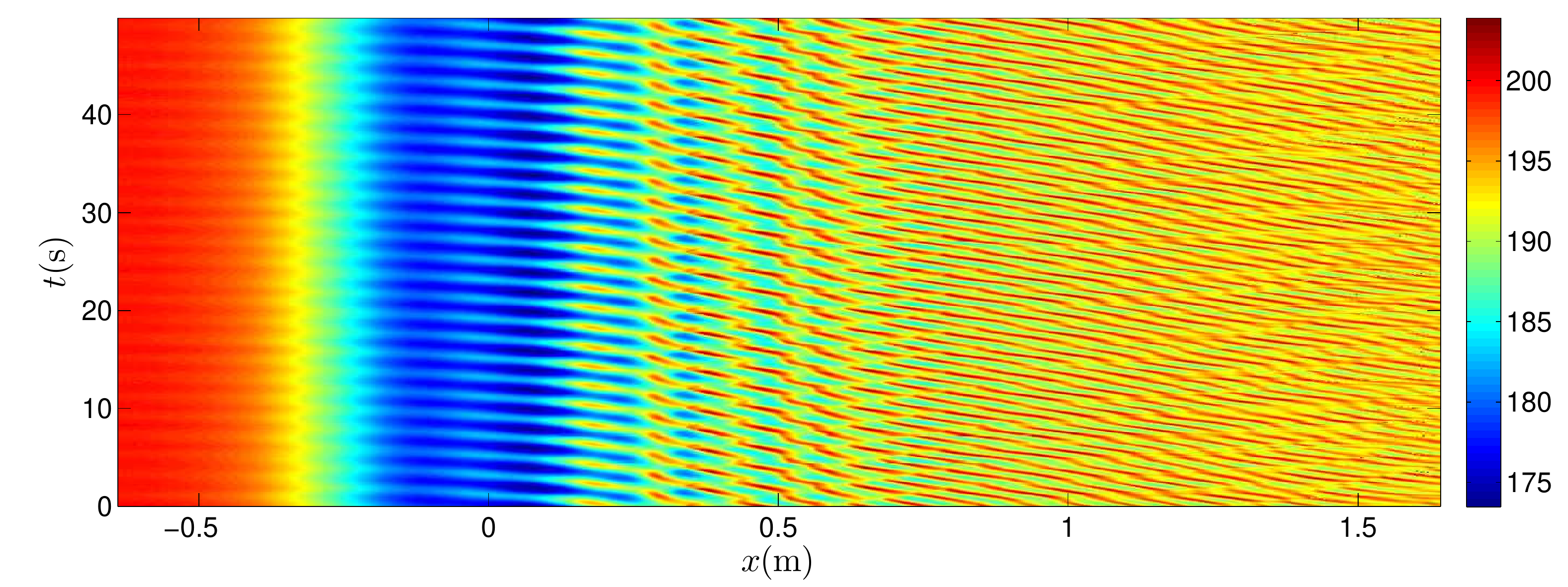}
\includegraphics[width=8cm,height=4cm]{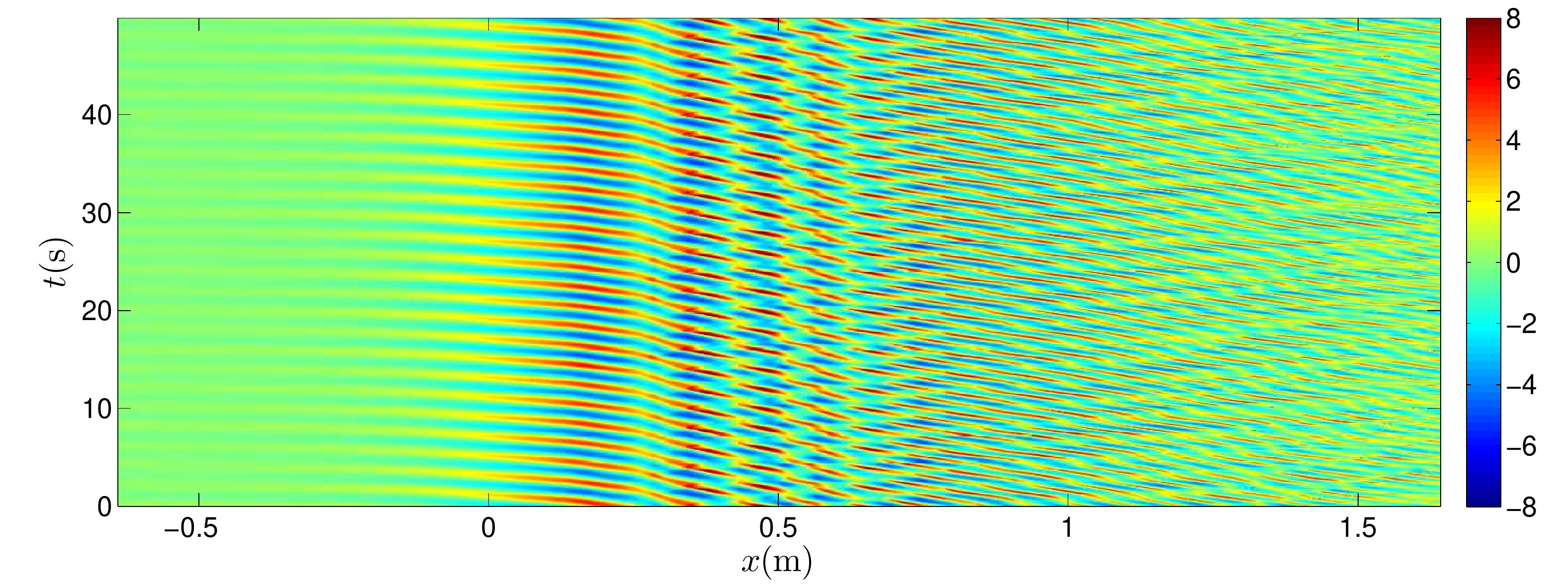}
\includegraphics[width=8cm,height=4cm]{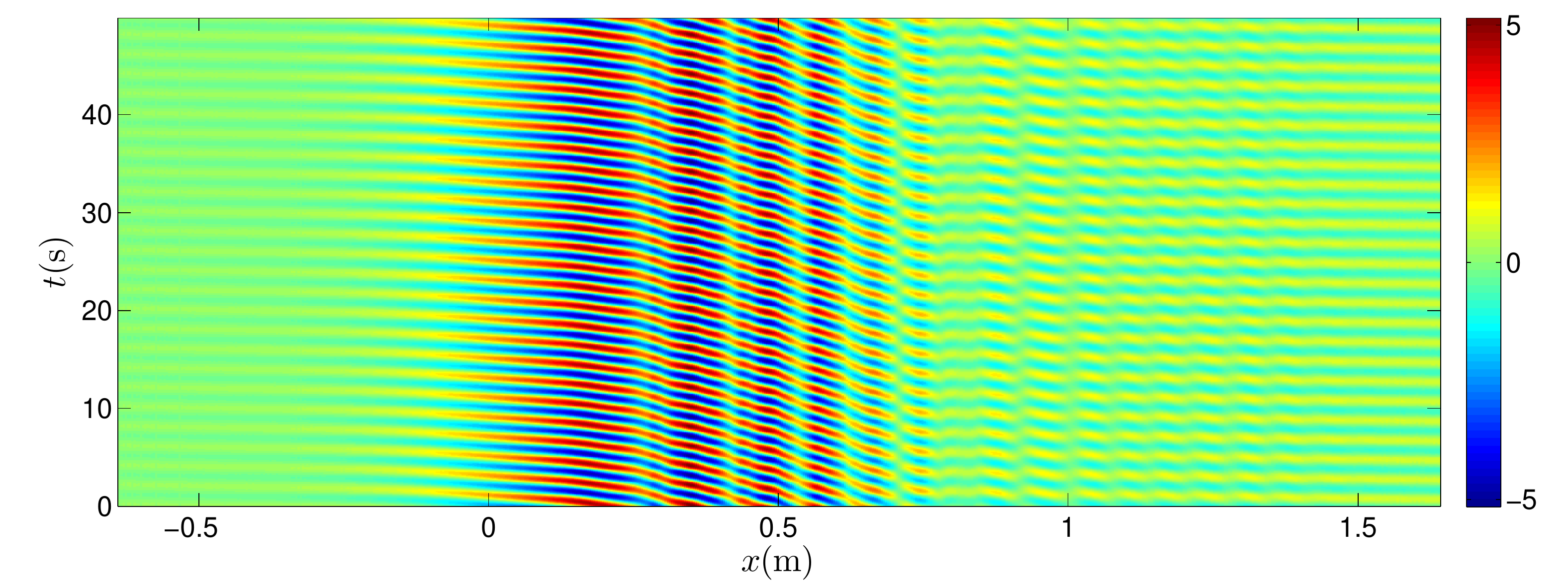}
\caption{Space-time diagrams of the free surface deformations (in mm) for an incident angular frequency $ \omega_I = 3.14 \rm {Hz}$ sent with an asymptotic incoming wave amplitude $a_I=0.8mm$: (Top) undulation + noise + waves namely the total depth $h(x, t)$ without the obstacle height; (Middle) by subtracting the stationary undulation $ \delta h (x, t) $; (Bottom) by filtering at the incident angular frequency which removes both the noise and harmonics $ \delta h (\omega_I, x, t) $ as in the Vancouver experiments \cite{PRL2011, WTPUL2013}.}
\label{xvst050}
\end{figure}

\begin{figure}[!htbp]
\includegraphics[width=9cm,height=12cm]{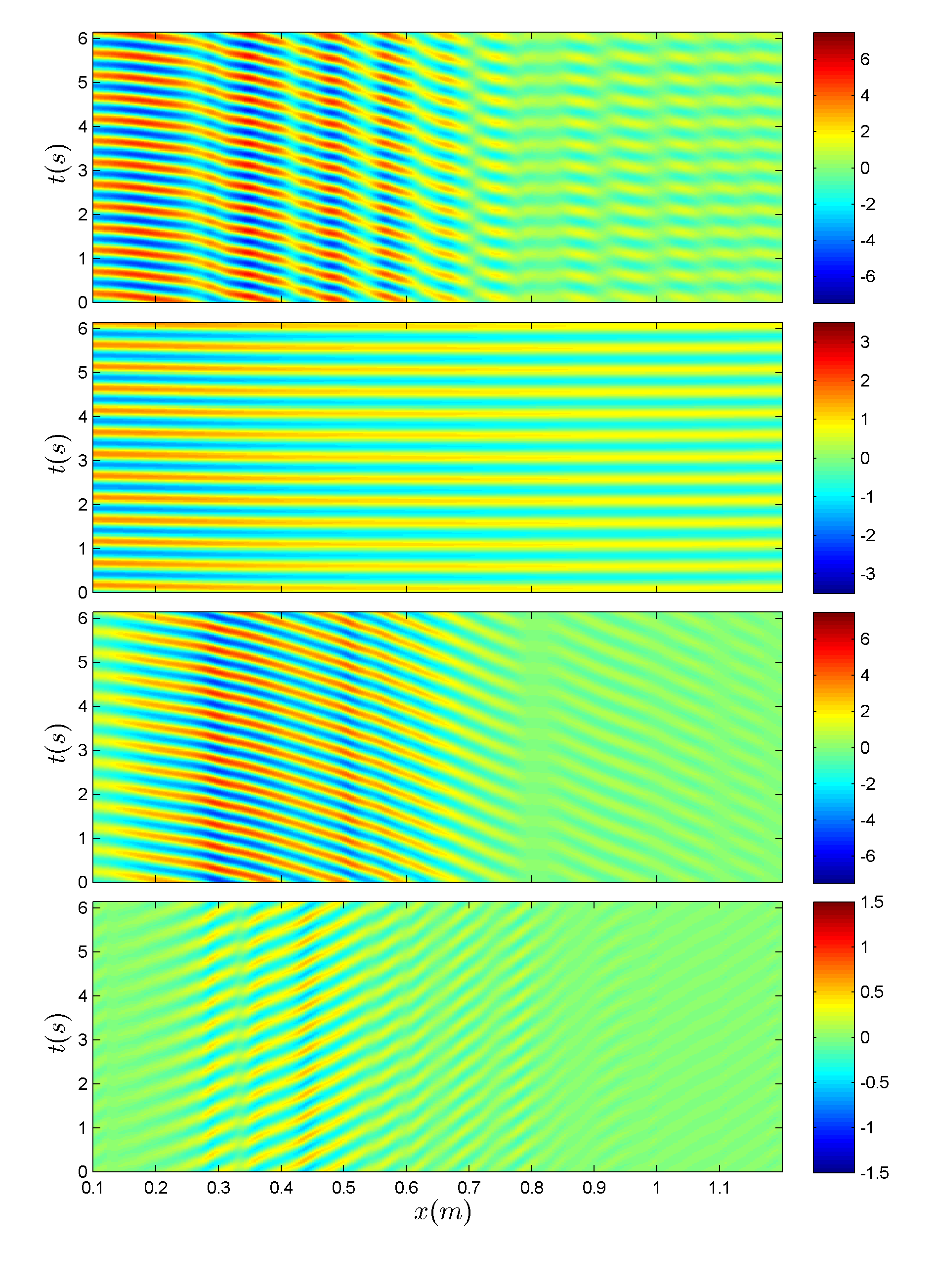}
\caption{Space-time diagrams filtered at the wave-maker angular frequency $ \omega_I = 3.14 \rm {Hz}$ for an asymptotic incoming wave amplitude $ a_I = 0.8 mm$: (a) a zoom of the Figure ~\ref{xvst050}; (b) incoming mode $k_I$; (c) blue-shifted mode $k_B$; (d) negative mode $k_N$. The blue-shifted and negative modes disappear in the vicinity of $x=0.8m$ because of their conversions towards free harmonics and not because of viscous dissipation as was assumed so far...}
\label{xvst2}
\end{figure}

The space-time diagrams of Fig.~\ref{xvst050} (explaining the process to remove the zero mode and the harmonics) and Fig.~\ref{xvst2} (explaining the process to separate the stimulated mode from the converted modes) permit to confirm the superposition of the modes at the wave-maker frequency decomposed as, on the one hand, the incoming mode I at $\omega$, the blue-shifted mode B at $\omega$ corresponding to Hawking radiation and the negative mode N at $\omega $ as observed in the Vancouver experiment because of the filtering of their results at $\omega$ and, on the other hand, the additional harmonics modes for the converted modes $B_{2\omega}$, $N_{2 \omega}$, transverse modes like $t_{3 \omega}$ and so on...

\begin{figure}[!htbp]
\includegraphics[width=8cm,height=5cm]{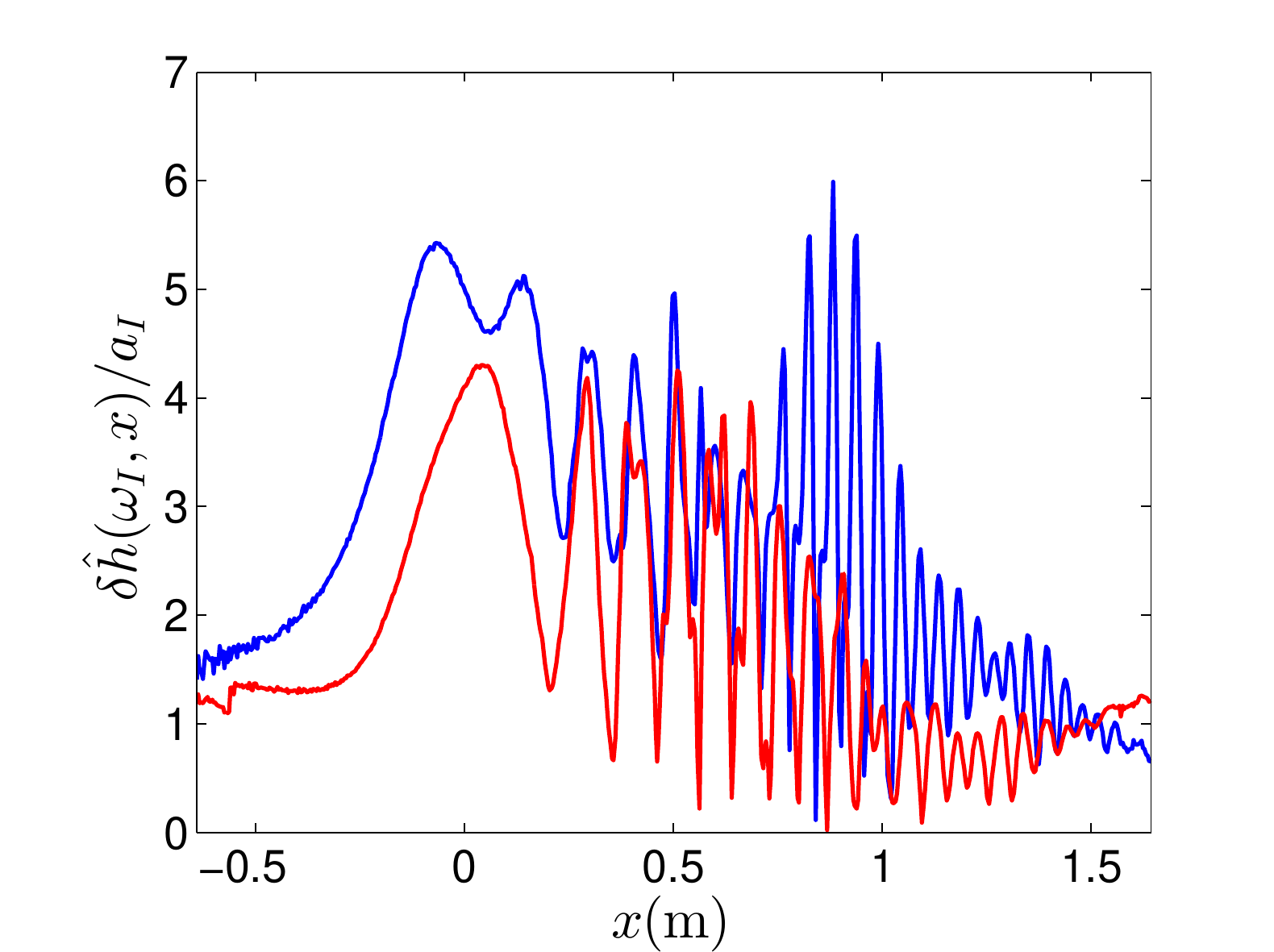}
\includegraphics[width=8cm,height=5cm]{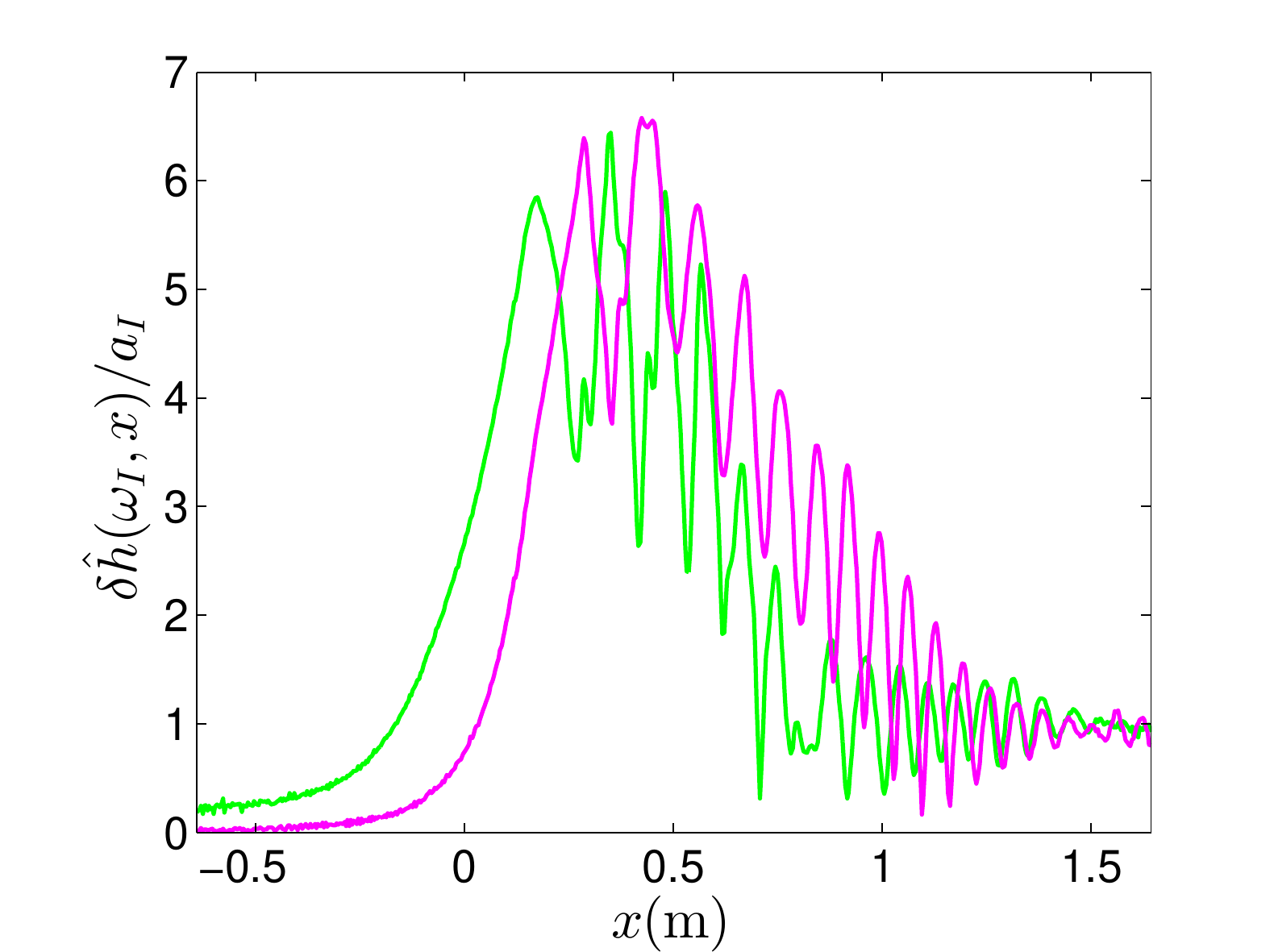}
\caption{Wave envelope of the free surface fluctuations at the filtered incident frequency made dimensionless by the amplitude of the asymptotic incoming wave amplitude $ a_I = 0.8 mm$. The color of the curves corresponds to different angular frequencies sent by the wave-maker $ \omega _I $: (Top) $1.13 \rm {Hz}$ (blue), $2.14 \rm {Hz}$ (red); (Bottom) $3.14 \rm {Hz}$ (green), $4.21 \rm {Hz}$ ( purple). The decreasing part is an exponential function of space provided that wave-blocking takes place and is a signature of evanescent waves that are present inside the forbidden region in the geometrical optics approximation.}
\label{EnvelopeTransmission}
\end{figure}

\begin{figure}[!htbp]
\includegraphics[width=8cm,height=5cm]{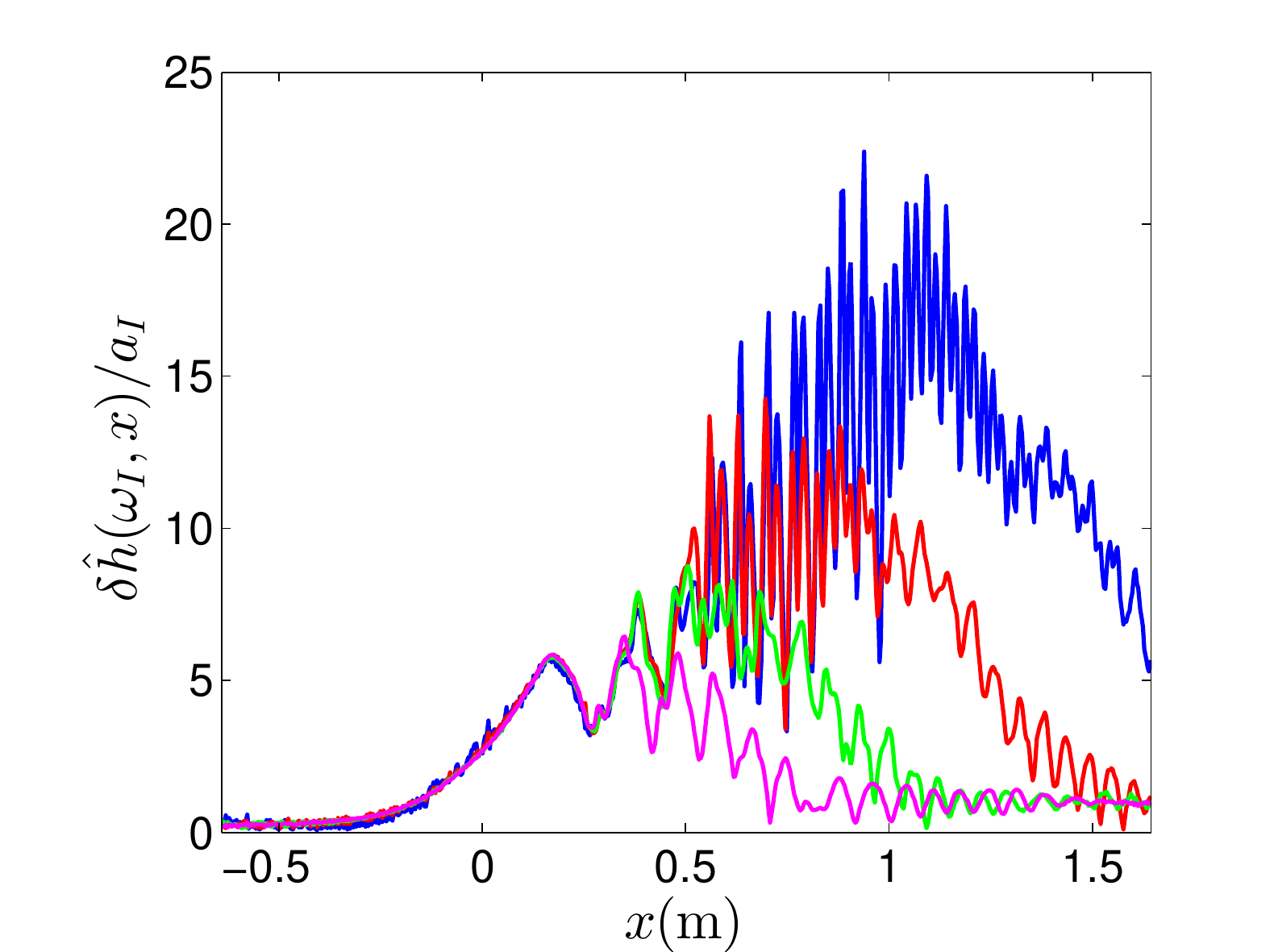}
\caption{Wave envelope as a function of the distance for several asymptotic incoming amplitudes of the stimulating mode in a blocking case at the filtered angular frequency $\omega_I = 3.14 \rm {Hz}$: $a_I=0.05 mm$ (blue), $0.15 mm$ (red), $0.35 mm$ (green), $0.8 mm$ (pink). The decreasing part is an exponential function of space.}
\label{Envelope}
\end{figure}

\begin{figure}[!htbp]
\includegraphics[width=8cm,height=6cm]{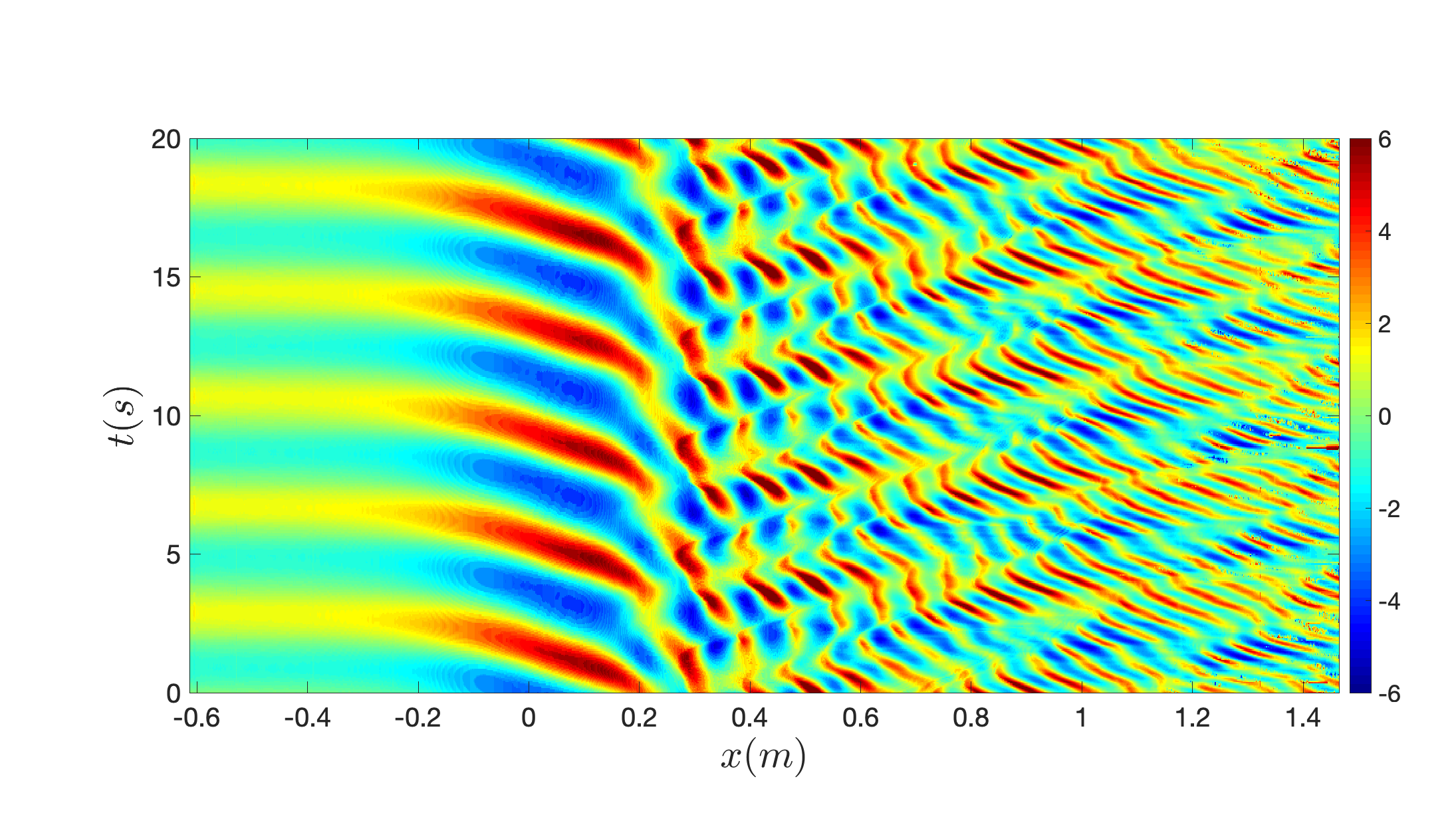}
\includegraphics[width=8cm,height=6cm]{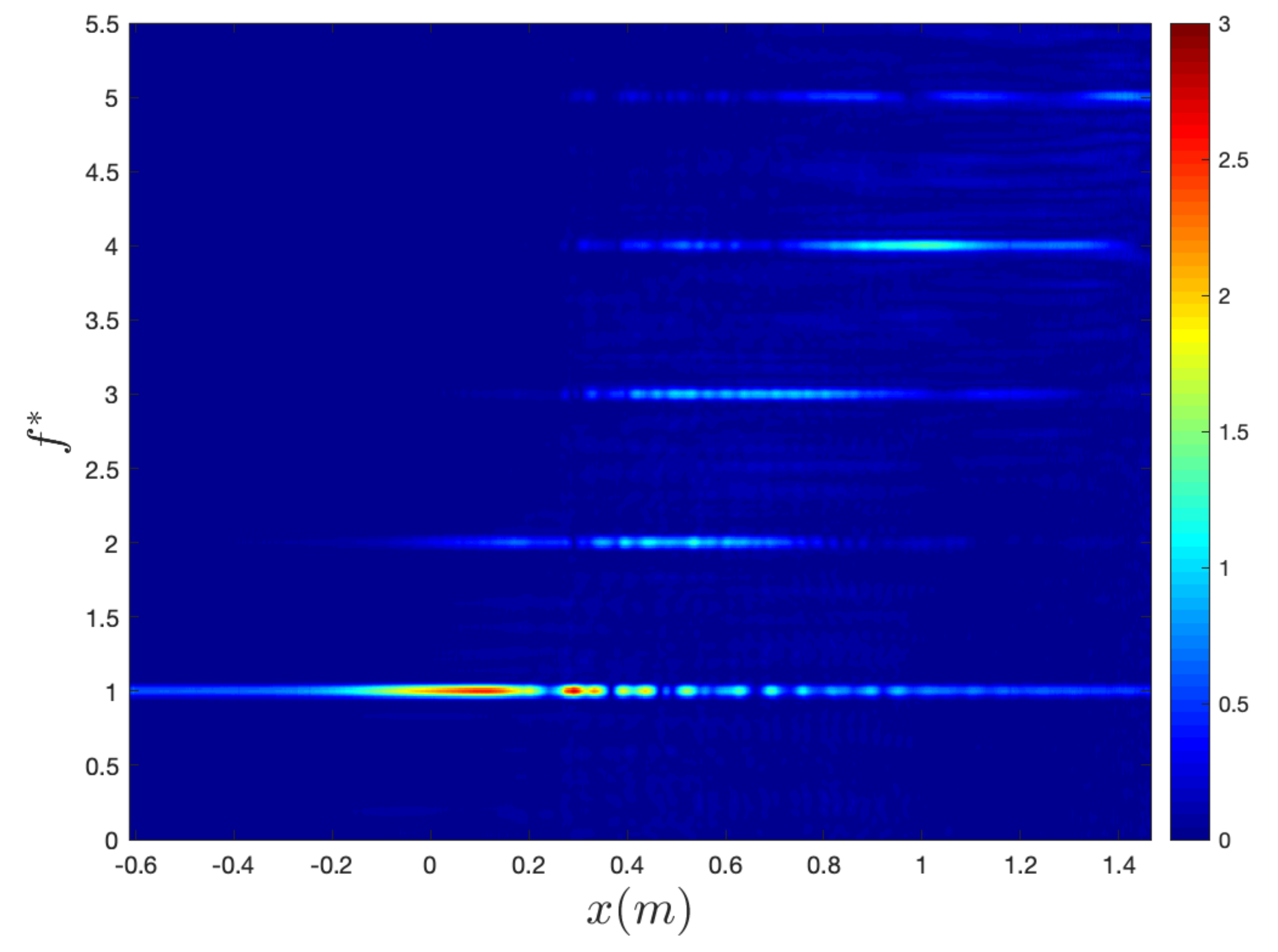}
\caption{Free Harmonics Generation in the Poitiers Experiments. (Top) Mode conversion in a non-blocking case (without wave breaking) towards free harmonics $B^{2\omega}$, $N^{2\omega}$, etc (BUT without $I^{2\omega}$ free harmonics as in the 2006-2010 Nice experiments, see below). Both converted modes are propagating in opposite directions as far group velocities are concerned on the top of the incoming waves $I^\omega$ ($\omega _I=1.63 Hz$ and $a_I= 0.8mm$); (Bottom) The spatial spectrogram $\tilde{\delta h}(x,\omega)$ in the asymptotic downstream region with a concentration of the signal at integer values of the dimensionless frequency $f^*=f/f_I=\omega/\omega_I$ namely 1, 2, 3, 4 or 5. The color bars for the water fluctuations are in mm.}
\label{B2}
\end{figure}

Concerning the wave-current interaction process, one notices both the influence of the simulating frequency for a given asymptotic wave amplitude (see Fig.~\ref{EnvelopeTransmission}) as well as the influence of the asymptotic wave amplitude for a given stimulating frequency on the presence of wave blocking or not (Fig.~\ref{Envelope}). Hence, we confirm the transmission in most of the frequencies used in \cite{PRL2011} but not reported as discussed in \cite{PRD2015}. This does not prevent neither mode conversion or free harmonics generation as shown for instance at a peculiar frequency in Fig.~\ref{B2} in accordance with the observations of \cite{NJP2008, NJP2010, Chaline}, a kind of generalized Hawking radiation for some harmonics in absence of a non-dispersive white hole horizon strictly speaking...

\subsection{A revisit of the Nice experiments}

Earlier, one of us had performed some experiments in 2006 \cite{NJP2008, NJP2010} and in 2010 \cite{Chaline} in the private company ACRI featuring a wave-tank 30 m long, 1.8 m wide and 1.8 m deep. The wave-maker was of a piston-type and generated waves with periods ranging from 0.35 s to 2.5s with typical wave heights ($2\times a$) around $5-10cm$ from crest to through as usual in coastal engineering.  For comparison, the value of the wave amplitudes in the Vancouver experiment was 1-2mm with a range of periods from 1.49s to 50s \cite{PRL2011}. We have shown conclusively that the Vancouver experiments were performed in a non-linear regime for the converted modes whereas the stimulating incoming mode did not suffer of either a Benjamin-Feir instability (namely a frequency downshifting due to a strong non-linear behavior \cite{Benjamin1967, Shugan2014}) or of free harmonics generation into a $I^{2\omega}$ mode (a weaker non-linear effect). Thanks to this, we are in position to finally understand part of the observations made in the 2006 Nice experiments \cite{NJP2008, NJP2010} by analyzing the improved 2010 Nice experiments \cite{Chaline} avoiding flow recirculations by decreasing the downstream angle of the bottom obstacle in the light of both Vancouver experiments \cite{PRL2011} and its present Poitiers reproduction with interface extraction and Fourier analysis. 

\begin{figure}[!htbp]
\includegraphics[width=8cm,height=4cm]{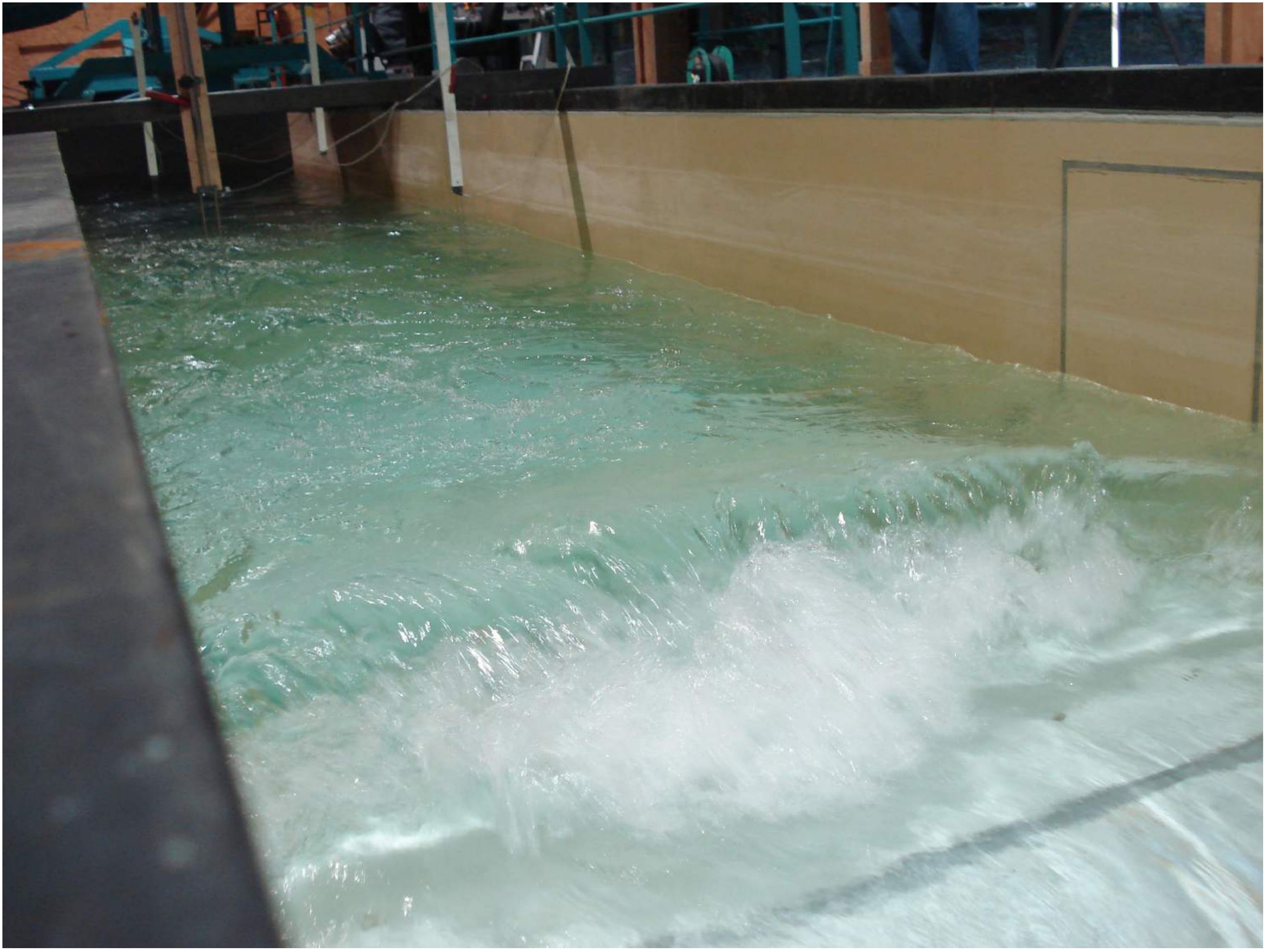}
\caption{The turbulent hydraulic jump in the original setup of the Nice 2008 experiments \cite{NJP2008} at high Froude numbers seen downstream just after the flat part of the obstacle whose boundary corresponds to the inclined grey band in the lower part of the image. The wave-maker with its cyan painting is located in the background of the picture: the flow points towards the picture from the lower right corner to the upper left corner. This transcritical regime was replaced by a subcritical one with a slower speed and an almost flat free surface or with a small depression in order to propagate counter-current waves on a smoother free surface from downstream the bottom obstacle.}
\label{Jump}
\end{figure}

\begin{figure}[!htbp]
\includegraphics[width=7cm,height=4cm]{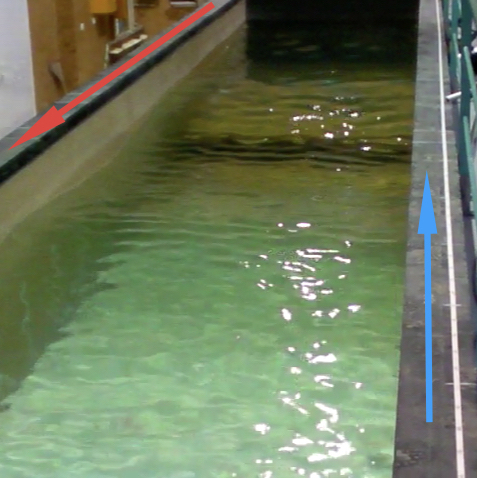}
\includegraphics[width=8cm,height=4cm]{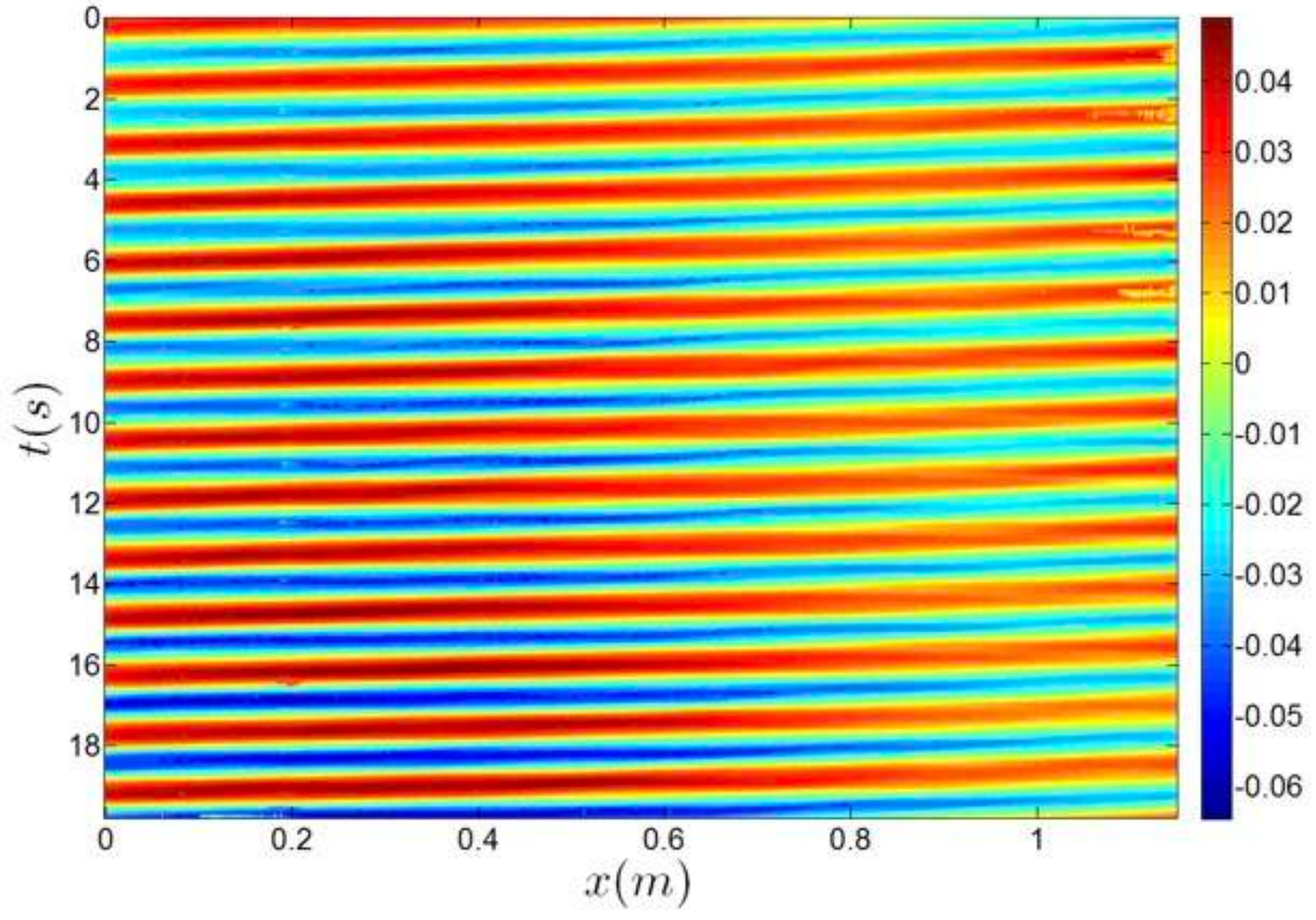}
\includegraphics[width=8cm,height=4cm]{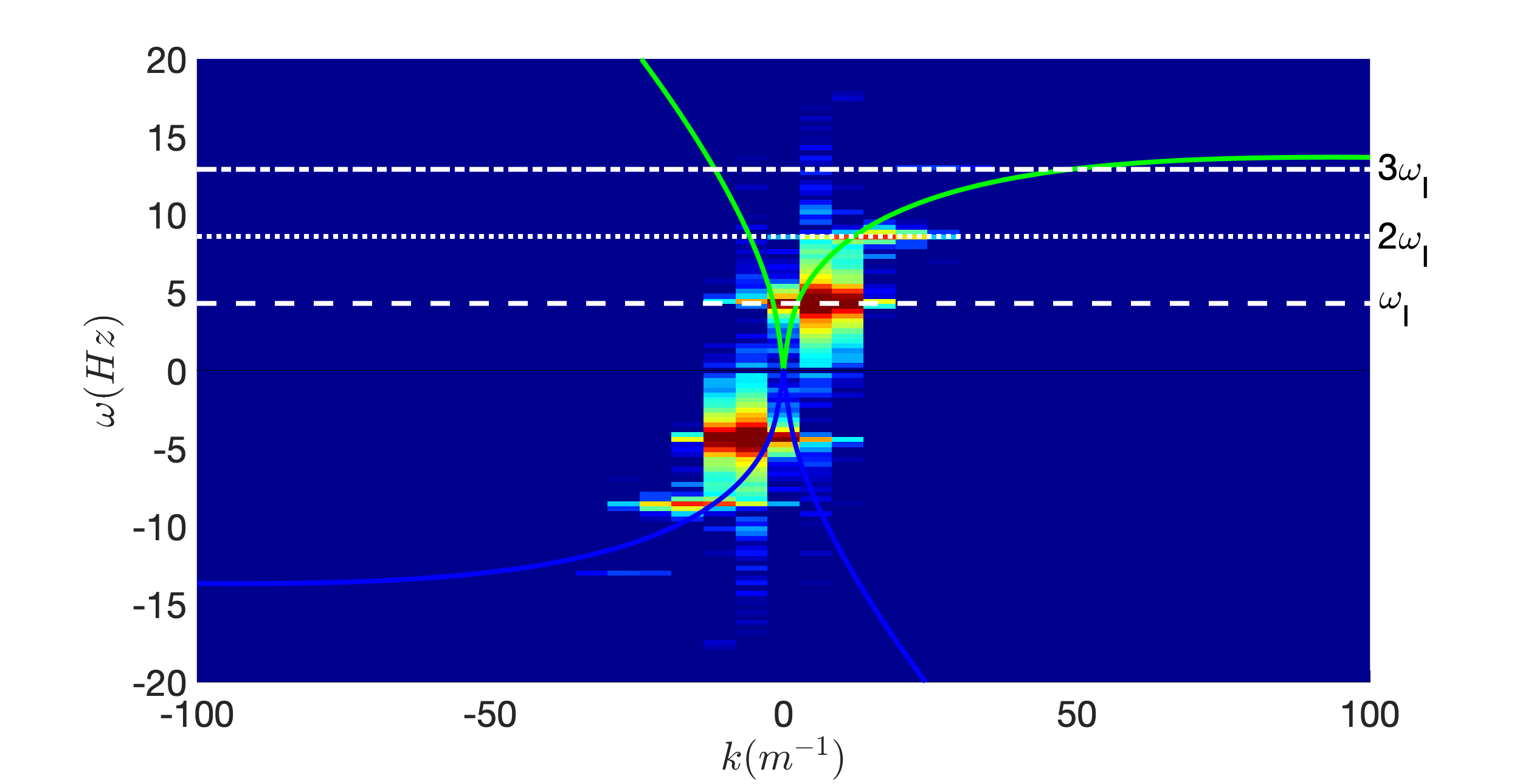}
\caption{Free harmonics generation in the 2010 Nice experiments corresponding to the reported mysterious (a) case of Figures 7 and 8 of \cite{NJP2008}: (Top) Mode conversion in a non-blocking case (without wave breaking) towards free harmonics $I^{2\omega}$ which are propagating in the same direction (red arrow) on the top of the incoming waves $I^\omega$ (wave-maker period $T_I = 1.5s$ or $\omega _I=4/3\pi \rm {Hz}$, asymptotic incoming wave amplitude $a_I= 5cm$, counter-current speed (blue arrow) $U(x)=0.274-0.03x$ m/s ($0<x<10m$, variator position=3)); (Middle) The space-time diagram without wave blocking in the region downstream on the obstacle (the color bar is in m); (Bottom) The dispersion relation in the descending part of the obstacle featuring both incoming $I^{\omega}$ and harmonics $I^{2\omega}$ modes contrary to the Vancouver experiment \cite{PRL2011} and its present Poitiers reproduction that display free harmonics of the converted modes only and not of the incoming modes as in the Nice experiment ($U_{theory}=0.188m/s$ and $h_{theory}=0.95m$ on average in the region of interest to compute the green (positive branches) and blue (negative branches) curves).}
\label{I2}
\end{figure}

We report in the Fig.~\ref{I2} an observation of the generation of free harmonics of the incoming mode without free harmonics generation of converted modes since the flow velocity does not allow the appearance of these converted modes. Hence, we confirm that there was positive energy modes generation of the type $I^{2\omega}$ (see Figures 7 and 8 case (a) of \cite{NJP2008}) without wave blocking due to non-linear effects: the latter do not conserved the energy of the incoming mode that keeps its frequency constant during the propagation despite the fact that a free harmonics appears that redistributes the energy in this new conversion channel. The temporal Fourier transform (not reported in \cite{NJP2008}) of the resistive sensor measuring the water depth as a function of time only displayed the appearance of a peak at twice the frequency which was wrongly interpreted as a bounded harmonics \`a la Stokes but whose relative amplitude with respect to the fundamental was considered as negligible: the authors of \cite{NJP2008} were unaware of the existence of free harmonics, solutions of the dispersion relation. The bounded harmonics would be naturally dismissed based on temporal measurements only since they are understood as just a non-linear deformation of the incoming mode propagating at the same phase speed contrary to the free harmonics which are propagating at different speeds with respect to the incoming mode. In Fig.~\ref{I2}, one observed the absence of wave blocking in the space-time diagram but the presence of free harmonics generation at twice the frequency of the incoming mode thanks to the spatial measurements of the free surface trace with cameras on the side of the water channel opposite to them. The nonlinear free harmonics $I^{2\omega}$ mode is superposed to the linear incoming mode $I^{\omega}$ and propagates with a speed smaller than the fast incoming mode in the same direction if one looks to the crests. When the wave-maker is stopped, the linear incoming mode vanishes rapidly in the water channel since it is dissipated by an absorbing beach at the end of the water channel after the bottom obstacle whereas the non-linear free harmonics mode are still propagating in a restricted region of space over the obstacle as can be seen thanks to optical caustics on the bottom of the water channel. This region of existence is dictated by the range of the velocity speeds compatible with free harmonics generation as in the Poitiers experiments for the converted modes only. In the case reported here, most of the flow speed downstream of the obstacle is inferior to the Landau speed of $23.1cm/s$ and partly inferior to the minimum group velocity of $17.8cm/s$ (see \cite{NJP2010}), hence the extra short waves on the top of the incoming waves are more likely to be considered as harmonics than mode converted at the frequency of the wave-maker. The Froude number which is the ratio between the flow speed $U$ and the long wave speed $\sqrt{gh}$ remains small around $Fr=0.13$ for the case treated. The water depth is quasi-constant along the water channel with a small depression of a few cm at most above the flat part of the bottom obstacle in the 2006-2010 Nice experiments (case D (for depression) in the hydraulic diagram of \cite{PTRSA2020}). The free harmonics mode $I_{2\omega}$ mimicked the converted mode $B_{\omega}$ since they propagate in the same direction and since they have similar wavelengths of medium sizes. Hence, one cannot discriminate visually between the two in permanent regimes by a basic visual inspection: one must rely on a free surface detection procedure with a spatial Fourier transform as first introduced by the Vancouver team and applied in this work to the 2010 Nice experiments for the first time.

The 2006-2010 Nice experiments were seminal in the sense that they raised numerous questions: do we need wave blocking in order to have mode conversion ? The answer is negative. The theoretical work \cite{NJP2010} shows that modes can be converted linearly without being blocked (see the wave phase diagram with the many regimes in the Figure 7 of the \cite{NJP2010} in deep water with surface tension and its generalization in \cite{Como, PTRSA2020} whatever the water depth). The Nice and Vancouver experiments were performed without blocking (a fact acknowledged in Nice and done on purpose to reduce a stronger non-linear effect namely wave-breaking due to the incoming amplitudes, a fact not acknowledged in the Vancouver work and revealed in the Poitiers reproduction \cite{PRD2015}) or with blocking and in two distinct non-linear regimes: free harmonics generation of the incoming modes was observed in Nice whereas free harmonics generation of the converted modes takes place in the Vancouver experiments but was filtered and displayed in the present Poitiers reproduction. Hence, what do we call Hawking radiation ? Do we need a (dispersive \cite{NJP2010} or even complex \cite{CS2018}) horizon ? Should the frequency stay conserved ? Both answers are negative since the stimulated Hawking radiation extends to free harmonics even in absence of wave blocking. $k_I^\omega$ and $k_B^\omega$ are equal at the blocking line corresponding to a dispersive group velocity white hole horizon in the case of wave-blocking \cite{PRL2009}; partial wave breaking is also observed because of the growing of the wave camber close to the horizon for the highest amplitudes and shortest periods.  To conclude, the Vancouver experiments feature the same drawbacks compared to the Nice experiments but the non-linear effects were smaller due to the range of wave amplitudes probed: {\it ``negative frequency waves with wrong frequency (indicating non-linear effects); negative frequency modes with and without acoustic horizons ; separation of flow"} \cite{PIRSA2009, PIRSA2011}.

\end{document}